\documentclass[preprint,preprintnumbers,amsmath,amssymb]{revtex4}
\usepackage{graphicx}
\usepackage{epsfig}
\usepackage{dcolumn}
\usepackage{amsmath}
\usepackage{amsfonts}
\usepackage{epsfig}
\usepackage{bm}
\usepackage{import} 
\usepackage{bm}
\usepackage{subfigure}
\usepackage{epstopdf}
\usepackage{rotating}
\usepackage{natbib}
\usepackage[dvipsnames]{xcolor}
\usepackage{color}
\usepackage{tikz}
\usetikzlibrary{decorations.pathreplacing,arrows}
\usepackage{placeins}
\usepackage{morefloats}
\definecolor{bro}{RGB}{255,0,255}
\definecolor{brown}{RGB}{222,125,0}
\definecolor{gray}{RGB}{10,36,106}

\newcommand{\be}{\begin {equation}}
\newcommand{\ee}{\end {equation}}

\newlength\figureheight 
\newlength\figurewidth 
\usepackage{float}
\usepackage{import}
\usepackage{array}
\usepackage{wasysym}
\makeatletter
\newcommand{\thickhline}{ \noalign {\ifnum 0=`}\fi \hrule height 1pt \futurelet \reserved@a \@xhline}

\newcommand{\argmin}{\mathop{\mathrm{argmin}}}

\newsavebox{\astrutbox}
\sbox{\astrutbox}{\rule[-5pt]{0pt}{20pt}}

\begin{document}
\title{ Turbulent Boundary Layer Features via Lagrangian Coherent Structures, Proper Orthogonal Decomposition and Dynamic Mode Decomposition}
\author{Naseem Ali $^1$}
\author{Murat Tutkun $^{2,3}$}
\author{Ra\'{u}l Bayo\'{a}n Cal $^1$}
\affiliation{$^1$Department of Mechanical Engineering, Portland State University, Oregon, USA}
\affiliation{$^2$Department of Process and Fluid Flow Technology, Institute for Energy Technology, 
NO-2027 Kjeller, Norway}
\affiliation{$^3$Universit\'{e} Lille Nord de France, \'{E}cole Centrale de Lille, 59655 Villeneuve d'Ascq, France}
\date{\today}
\begin{abstract}

High-speed stereo PIV-measurements have been performed in a turbulent boundary layer at Re$_{\theta}$ of 9800 in order to elucidate the coherent structures. Snapshot proper orthogonal decomposition (POD) and dynamic mode decomposition (DMD) are used to visualize the flow structure depending on the turbulent kinetic energy and frequency content. The first six POD and DMD modes show the largest and the lowest amount of energy and frequency, respectively. Lagrangian coherent structure (LCS) based on the algorithm developed using the variational theory is also applied to track the flow {\it{via}} attracting and repelling trajectories. The shapes and the length of the trajectories show variation with increasing advection time. LCS trajectories are overlayed with the individual POD and DMD modes. Repelling and attracting lines cover the structure of these modes. Reconstructed flow fields from individual POD modes are also used to generate new LCS trajectories. The energy and frequency content have a direct impact on the length of the trajectories, where the longest reconstructed trajectories associate with the higher energy and lower frequency modes, and vise verse. The multiple intersection points between the repelling and attracting lines marked the low momentum regions.  
 
\end{abstract}
\maketitle
\clearpage

\section{Introduction}\label{intro}
The turbulent boundary layer is a complex flow possessing dynamical behavior over wide ranges of scales both spatially and temporally. The interaction between the boundary layer regions and the diversity of the scales make the flow physics difficult to uncover. Coherent structures within a turbulent flow are extensively studied as the amalgamation of chaotic velocity fluctuations inherently define the flow behavior. As \citet{T} posed, coherent structures exist in the shape of what was coined as horseshoes vortices. This was later supported experimentally and numerically with direct numerical simulation (DNS) data  \cite{R1991}. Further experimental and numerical studies have enabled to look into details of the complex organization. \citet{AMT} described the vortex packet based on the two dimensional measurements further supported by \citet{gana} and \citet{tomkins2}. In what was called a hairpin vortex as well as low speed streaks existing in the boundary layer, examples have been provided in \cite{kahler20,Adrian2007,herpin2,schlatter,marusic2010,schroder,dennis201}. \citet{jacobi2013} investigated the very-large-scale motion (VLSM) and their influence on smaller scales using cospectral density. Different techniques have allowed to visualize these structures all rooted in an Eulerian frame of reference and also employing the velocity gradient tensor such as Q-criterion \cite{HWM}, $\delta$-criterion \cite{CPC}, $\lambda_{2}$-criterion \cite{JH}, and swirling strength \cite{ZABK}. However, these criteria lack objectivity due to their dependence on the chosen frames \cite{haller2015}.

Low-order descriptors are also used to extract the dominant flow features based on energy content thus decomposing the flow field into coherent and less coherent parts. One of the most common modal decompositions is the proper orthogonal decomposition (POD) also known as singular value decomposition or principle component analysis. \citet{lumley1967} presented proper orthogonal decomposition in the context of turbulent flow and \citet{sirovich1987turbulence} introduced snapshot POD that conforms to the difficulties of the classical orthogonal decomposition. \citet{herzog1986l} performed three dimensional POD analysis for low Reynolds number pipe flow. \citet{moin1989} used proper orthogonal decomposition to analyze the channel flow data obtained by direct numerical simulation. \citet{liberzon200l} used vorticity-based POD to extract the coherent structures of the DNS field. \citet{wu2010spatial} used orthogonal decomposition modes to distinguish between the large scales and small scales in the turbulent boundary layer over smooth and rough surfaces. Large scales indicated the general features of the hairpin vortex packet. \citet{baltzer2011st} showed that the hairpin vortex packets in the wall turbulence are strongly related to the dominant modes. \citet{shah2014very} analyzed VLSM of atmospheric boundary layer performed by wall-modelled large eddy simulation (LES). Different POD modes of the vertical velocity field were extracted to then be correlated with the horizontal velocity and temperature. Furthermore, momentum and heat flux modes yielded from the analysis. POD follows a linearity of procedures of Fourier analysis and the diagonalizing of the temporal-spatial correlation matrix, thus extracting orthogonal structures \cite{berkooz1993proper}. However, the POD mode loses the phase information and might not be temporally independent due to the averaging process of the correlation tensor.  

In the need to include the temporal, dynamic mode decomposition (DMD) allows to reveal this information as the energetic modes are described in terms of their frequency content. \citet{schmid2008de} introduced DMD built on the theory of Krylov subspaces as a variant of the Arnoldi algorithm. The computation requires elements of the time series to be linearly independent and have distinct eigenvalues. \citet{rowley2009spectral} introduced DMD through the theory of Koopman spectral analysis, presented firstly by \citet{mezic2005spectral} in the model reduction. Based on Koopman theory, the modes completely characterize the dynamics of the nonlinear flow and associate the particular eigenvalues to constitute a Koopman decomposition. Hence, DMD is considered as a particular numerical algorithm that calculates an approximate Koopman decomposition. The connection between the DMD and Koopman theory is preserved if and only if the dataset columns are linearly independent. \citet{schmid2010dynamic} presented DMD as a linear stability analysis and showed its validity for nonlinear flow {\it{via}} analyzing a linear tangent approximation to the underlying flow. Furthermore, the relation between POD and DMD through an singular value decomposition, SVD, based algorithm was explored. \citet{schmid2011application} used the dynamic mode decomposition to investigate the dominant features of flame, jet and flow within a cavity. In each of these instances, the temporal content was extracted and visualized. \citet{muld2012flow} used DMD to extract the  coherent structure aspects of a simulated flow in the wake of the high speed train. \citet{pan2011dynamical} analyzed the wake of an airfoil, which yielded a  correlation  between the frequency of each mode with respect to the wavelength. \citet{tang2012dynamic} extracted periodic dynamical behavior of the hairpin vortices generated by a hemisphere protuberance. In a cylinder wake, \citet{zhang2014identification} distinguished the structure in the spatial and frequency domain; concluding that, POD modes are contaminated by uncorrelated structures.

In a Lagrangian frame of reference, a Lagrangian coherent structure (LCS) was used to detect features about the flow pertaining to stability as the trajectory evolves based on the expansion rates computed. Hyperbolic LCSs reveal the surface of ordered fluid trajectories and describes the most repelling, attracting and shearing at these surfaces. Repelling material lines (stable manifolds) are responsible for the stretching of passive tracer groups normal to the manifold, whereas attracting material lines (unstable manifolds) for stretching tangent to the manifold and organize the motion of nearby fluid tracers as pointed by \citet{haller2000finding}. \citet{H} showed an LCS event should appear as a ridge of the Finite Time Lyapunov Exponent, FTLE. These ridges represent stable or unstable manifolds in the flow. \citet{wang2003closed} presented that a fluid blob is deformed when a fluid tracer group along a stable manifold is incident upon an unstable manifold. \citet{SLM} suggested an alternative definition of the LCS as ridge of FTLE, thus showing a negligible flux across the well-defined LCS. However, \citet{Ha} proved the discrepancy of ridge theory, where LCS is not necessary to be the ridge of the FTLE  and vice versa. A remedy was attained {\it{via}} variational theory based on the maximization of the normal repulsion rate of material lines at specific time interval. \citet{FH2} developed an improved computational approach based on variational theory to extract hyperbolic LCSs as smooth curves. \citet{farazmand2013attracting} relaxed the inconsistency of the calculation of repelling and attracting at the same time by calculating from a single forward or backward cascade. \citet{farazmand2015hyperbolic} showed that the hyperbolic LCSs align with the time average scalar field (zero level curves) as a consequence to  forward or backward advection.

Lagrangian coherent structures have been applied to obtain an understanding of fuel mixing for a flame thrower \cite{wang2003closed} and ocean currents \cite{inanc2005optimal}. \citet{SLM} studied the attracting LCS on an oscillating flow over an airfoil. \citet{GRH} found hairpin-like LCSs using direct Lyapunov exponents, DLE, of a DNS turbulent channel flow. \citet{PD} used LCSs to investigate predator-prey behavior of jellyfish. \citet{pan2009identification} identified Lagrangian coherent structures in a low Reynolds number turbulent boundary layer by computing FTLE on a wall-normal-streamwise plane. \citet{beron2010invariant,beron2012zonal} observe in the spatially periodic geophysical flows that the shearless invariant tori (zonal jet cores) can be characterized when the trenches coincide in forward and backward FTLE calculations. \citet{WTR} identified LCSs on a spanwise-streamwise plane of a high Reynolds number turbulent boundary layer. A more recent review  can be found in \citet{haller2015}. The present work analyzes high-speed PIV data obtained on a wall parallel plane in turbulent boundary layer flow at high Reynolds number. Coherent structures in the turbulent boundary layer are investigated {\it{via}} implementing  snapshot POD and DMD. Furthermore, obtained results are compared and contrasted to the found LCS to observe the coinciding features amongst these three techniques in {\it{lieu}} of their coherence. Finally, LCSs are constructed {\it{via}} POD modes and analyzed as the mode basis.

\section{Theory}

In this study, theoretical frameworks for proper orthogonal decomposition, dynamic mode decomposition and Lagrangian coherent structures are described as detailed below.

\subsection{\label{theory1} Snapshot Proper Orthogonal Decomposition}

Proper orthogonal decomposition is a statistical technique that identifies the optimal deterministic function and is presented as eigenvalues and eigenfunctions. POD gives the opportunity to decompose the flow into its large and small scales depending on the turbulent kinetic energy associated with these scales. The fluctuating velocity is obtained by subtracting the individual snapshots from the mean velocity and can be represented as,

\begin{equation}
{\vec{u}}={{u}}({\vec{x}},{t^{n}}),  
\end{equation}
where ${\vec{x}}$ and {\textit{t}} refer to the spatial coordinates and time, respectively. {\textit{n}} denotes the sample number. The matrix of the velocity components is processed as following,
 \begin{equation}
{\vec{u}}({\vec{x}},{t^{n}}) =
\begin{pmatrix}
  { u_{1}^ 1}& {u_{1}^2} & {u_{1}^3} & \dots  & {u_{1}^N} \\
\vdots & \vdots & \vdots & \vdots & \vdots \\
    {u_{M} ^1}& {u_{M}^2} & {u_{M}^3} & \dots  & {u_{M}^N}\\
    {v_{1}^ 1}& {v_{1}^2} & {v_{1}^3} & \dots  & {v_{1}^N }\\
\vdots & \vdots & \vdots & \vdots & \vdots \\
    {v_{M} ^1}& {v_{M}^2} & {v_{M}^3} & \dots  & {v_{M}^N}\\
    {w_{1}^ 1}& {w_{1}^2} & {w_{1}^3} & \dots  & {w_{1}^N }\\
\vdots & \vdots & \vdots & \vdots & \vdots \\
    {w_{M} ^1}& {w_{M}^2} & {w_{M}^3} & \dots  & {w_{M}^N}
\end{pmatrix},
\end{equation}
where {\textit{u}}, {\textit{v}}, {\textit{w}}, {\textit{N}} and {\textit{M}} is the streamwise, wall-normal, spanwise fluctuation velocity components, total number of the snapshots, and spatial grid points, respectively. A set of the orthonormal basis functions, ${\vec{\phi}}$, can be presented as,
\begin{equation}
 {\vec{\phi}=\sum\limits_{n=1}^N A(t^{n}){\vec{u}}({\vec{x}},{t^{n}})}.  
\end{equation}

\noindent Mathematically, the optimal functions, ${\vec{\phi}}$, require the average error between the field data and its projection onto ${\vec{\phi}}$ to be minimized, thus the averaged projection is maximized in mean square sense. The two-point correlation tensor is required to obtain the maximization of the projection onto the flow field and can be determined as follows,
 \begin{equation}
 {R(\vec{x},\vec{x}^{'})=\frac{1}{N}\sum\limits_{n=1}^N {\vec{u}^{T}}({\vec{x}},{t^{n}}) {\vec{u}} (\vec{x}^{'},t^{n})},  
\end{equation}
where ${R(\vec{x},\vec{x}^{'})}$ is a spatial correlation between two points ${\vec{x}}$ and ${\vec{x}^{'}}$ and {\textit{T}} denotes the transpose of the matrix. Using the calculus of the variations, the maximization of the projection is given by the following Fredholm integral equation,

\begin{equation}
{\int_\Omega R(\vec{x},\vec{x}^{'}) \phi(x^{'}) dx^{'}=\lambda  \phi(x)},  
\end{equation}
 where ${\Omega}$ and ${\lambda}$ are the domain and the eigenvalue, respectively. The optimal deterministic function is reduced to the eigenvalue problem and takes the following form  
\begin{equation}
{C G=\lambda G},
\end{equation}

\noindent where {\textit{C}} and {\textit{G}} are the correlation tensor and basis of eigenvector, respectively. Each eigenvalue corresponds to the distinct eigenfunction and both  are ordered in optimal sense. The eigenvalues are positive and real and represent the contained energy of the eigenfunction as well as arranged in decreasing order as,

\begin{equation}
{\lambda_{1}\geq \lambda_{2}\geq \lambda_{3}\geq .\ .\ .\geq \lambda_{N-1}>0}.  
\end{equation}

\noindent The matrix of orthonormal eigenvectors is normalized to represent the basis vectors or modes that describe the spatial information of the turbulence structures. Thus, summation of the eigenvalues is equivalent to the total turbulence kinetic energy, ${E}$, in a finite domain:
\begin{equation}
{E}={\sum_{n=1}^{N} \lambda_{n} }.
\end{equation}

\noindent The normalized energy content of each mode, ${\eta_{n}}$, and the fraction of the cumulative energy, ${\zeta_{n}}$, can be represented as, 
\begin{equation}
{\eta_{n}}={\frac{\lambda_{n}}{\sum_{n=1}^{N} \lambda_{n} }},
\end{equation}

\begin{equation}
{\zeta_{n}}={\frac{\sum_{n=1}^{n} \lambda_{n} }{\sum_{n=1}^{N} \lambda_{n} }}.
\end{equation}

One of the aspects of the POD is the ability to reconstruct the stochastic velocity using the eigenfunctions. The contained energy  of the reconstructed velocity depends on the number of modes that is used to reconstruct the flow field. The reconstructed velocity can be described by,
 \begin{equation}
{\vec{u}(\vec{x},t^{n})=\sum\limits_{n=1}^N a_{n} {\vec{\phi}}^{ \ n}(x)},  
\end{equation}

\noindent where ${a_{n}}$ is a back projecting coefficient of eigenfunction onto stochastic velocity and can be determined by,

\begin{equation}
 { a_{n}(t^{n})} = {\int_\Omega \vec{u}}({\vec{x}},{{t}^{n})} {\phi^{n}} {(x^{'}) dx^{'}.}  
\end{equation}

\subsection{\label{theory2} Dynamic Mode Decomposition}

The dynamic modes of the field data can be extracted  through to Arnoldi algorithm which start from a Krylov sequence that classifies a set of successive vectors {\it{via}} spanned spaces and that corresponding to the Krylov subspaces, $\mathcal{K}$, \cite{trefethen1997numerical,tirunagari2012analysis}. Let an instantaneous velocity vector, ${\vec{\tilde{u}}}$, organized into an ensemble snapshot  as,  

\begin{equation} 
{\vec{\tilde{u}}^{N}_{1}}={[\vec{\tilde{u}}_{1}, \ \vec{\tilde{u}}_{2}, \ \vec{\tilde{u}}_{3},\ ....\ , \vec{\tilde{u}}_{N}]}. 
\end{equation}

\noindent A linear mapping between the snapshots given by
\begin{equation}
{\vec{\tilde{u}}^{N}_{1}}={[\vec{\tilde{u}}_{1}, \ B \vec{\tilde{u}}_{1}, \ B^{2} \vec{\tilde{u}}_{1},\ ....\ , B^{N-1} \vec{\tilde{u}}_{1}]},  
\end{equation}
 where {\textit{B}}  is a constant of the linear mapping over the data and related to the standard Arnoldi iteration problem as following, 
\begin{equation}
{B\ \vec {\tilde{u}}^{N-1}_{1}} \approx  {\vec {\tilde{u}}^{N-1}_{1}\ S}, 
 \end{equation}
where {\textit{S}} is a companion matrix that linearly combines the last snapshot {\textit{N}} to the previous one, {\textit{N-1}}. In a least square sense, the companion matrix is optimally determined as following,

\begin{equation} 
{S}= {\argmin_{S}  \lVert { \vec {\tilde{u}}^{N}_{2} - \vec {\tilde{u}}^{N-1}_{1} S }  \rVert },
\end{equation}

\noindent where {argmin} is the argument of the minimum. Following the algorithm proposed by \cite{schmid2011application}, the dynamic mode decomposition is presented briefly. The snapshot matrix is decomposed into an orthogonal matrix and an upper triangular matrix {\it{via}} $QR$ decomposition,
 \begin{equation}
{[Q, \ R]}={qr(\vec{\tilde{u}}^{N-1}_{1}, \ 0)}. 
 \end{equation}
 Companion matrix, {\textit{S}}, can be computed as,
 \begin{equation}
 {S}={R^{-1} \ Q^{H}\  \vec{\tilde{u}}^{N}_{2}},  
  \end{equation}
where ${Q^{H}}$ is the complex conjugate transpose of the matrix {\textit{Q}}. The resulting eigenvalue problem is solved to get the eigenvalues, {\textit{L}}, and eigenvectors, {\textit{X}}, of the {\textit{S}} matrix,
 \begin{equation}
{[X,\ L]}={eig(S)}.  
 \end{equation}
 
\noindent The eigenvalue of the companion matrix is mapped onto the complex planes to acquire the dynamic spectrum of the modes, 
 \begin{equation}
{\Pi}={log(L)/ \delta t}, 
 \end{equation}

\noindent where ${\Pi}$ and ${\delta t}$ are the logarithmic spectrum and time step between snapshots, respectively.
Finally, the dynamic mode, {\textit{DM}}, can be extracted as,
 \begin{equation}
{DM}={\vec{\tilde{u}}^{N-1}_{1} \ X}. 
 \end{equation}

\noindent The DMD algorithm can be used to reconstruct the stochastic flow field as,

 \begin{equation}
{\vec{u}(\vec{x},t)=\sum\limits_{n=1}^N \omega_{n} \ \mathrm{exp} (L_{n} t) \ {DM}_{  n}(\vec{x})},  
\end{equation}

\noindent where $\omega_{n}$ is the amplitude of the modes.

\
\subsection{\label{theory3} Lagrangian Coherent Structures}

Consider the differential equation:
 \begin{equation}
 \dot{{x}}={\vec{u}}({\vec{x},t)}, 
 \end{equation}

\noindent where ${\vec{u}(\vec{x},t)}$ is a velocity field defined in an open domain, {\textit{D}} $\subset \mathbb{R}^2$, for a determined interval of time. Flow map can be defined as,
 \begin{equation}
 {F_{t_{0}}^{t}:D} \longrightarrow {D} \hspace{10pt} \mathrm{and} \hspace{10pt} {F_{t_{0}}^{t}}({\vec{x}_{0}}) = {\vec{x}( t, t_{0}, {\vec{x}_{0}})}, 
 \end{equation}
\noindent where ${\vec{x}_{0}}$, and ${t_{0}}$ is the initial position and initial time, respectively. 
If ${M(t_0) \subset D}$ is a region occupied by a mass of fluid at time ${t_0}$, then, 
 \begin{equation}
{F_{t_{0}}^{t}(M(t_0)) = M(t)}. 
 \end{equation}

\noindent For any smooth velocity field, $\nabla {F}_{t_0}^{t}({\vec{x}_0})$ is the invertible tensor and denote the derivative of the flow map at a point ${\vec{x}}_0 \in D$. The Cauchy-Green strain tensor is obtained as follows:
 \begin{equation}
{C_{t_0}^{t}({\vec{x}_0})} {:=} \nabla {F_{t_0}^{t}}({\vec{x}_0})^{{\mathrm{T}}} \ \nabla {F_{t_0}^{t}}({\vec{x}_0}). 
 \end{equation}

\noindent The tensor has two real eigenvalues, ${\Gamma_1({\vec{x}}_0,t_0,t)} \; \mathrm{and} \; {\Gamma_2({\vec{x}}_0,t_0,t)}$, and two orthogonal eigenvectors, ${\xi}_1({\vec{x}_0,t_0,t)}$ and ${\xi}_2({\vec{x}_0,t_0,t)}$. Since ${C}_{t_0}^{t}({\vec{x}_0})$ is positive definite, its eigenvalues satisfy
 \begin{equation}
{ 0 < \Gamma_1}({\vec{x}_0,t_0,t)} { \leq \Gamma_2({\vec{x}}_0,t_0,t)}. 
 \end{equation}
 
\noindent  The repulsion rate is defined at each point ${\vec{x}}_0 {\in M(t_0)}$ by 
  \begin{equation}
{\rho_{t_0}^{t}(}{\vec{x}_0,{n}_0)} {=} {\langle} {n}_t, \nabla {F_{t_0}^{t}(}{\vec{x}_0)}{n_0 \rangle},
 \end{equation}
\noindent where ${n_0}$ is a unit normal to ${M(t_0)}$ at ${\vec{x}_0}$, ${n_t}$ is a unit normal to ${M(t)}$ at $\nabla {F_{t_0}^{t}}{(\vec{x}_0)}$, and ${\langle\: \cdot \;, \: \cdot \: \rangle}$ is the standard inner product. If ${\rho_{t_0}^{t}(}{\vec{x}_0,}{n_0) > 1}$, then the normal perturbation to ${M}$ is repelling between ${t_{0}}$ and ${t}$. Similarly, if ${\rho_{t_0}^{t}(}{x_0,}{n_0) < 1}$, then the normal perturbations to ${M(t_0)}$ decreases (attracting). The magnitude of $\nabla {F_{t_0}^{t}(}{\vec{x}_0) }{e_0}$ where ${e_0}$ is a unit tangent to ${M(t_0)}$ at ${\vec{x}_{0}}$  measures the tangential growth in ${M}$. 

\
For planar field, the true path of trajectories, depending on in-plane path, should satisfy certain requirements. As described in \citet{mathur2007uncovering}, the Lagrangian divergence measures the strength of cross-plane motion with respect to the in-plane motion and should be checked at each time step. In addition, the meander away from the plane should be small relative to the normal length scale of the plane. These have been previously tested and verified by \citet{WTR} for the same dataset. Furthermore, the velocity field is extended beyond the measurement domain in streamwise direction using the Taylor's frozen field hypothesis. \citet{WTR} describes the application of Taylor's hypothesis which is therefore not repeated here. To compute hyperbolic LCS,  four conditions are used as following \cite{FH2},
\[ 
\begin{array}{l}
1. \; {\Gamma_1(}{\vec{x}_0,t_0,t)} {\ne \Gamma_2(}{\vec{x}_0,t_0,t) > 1}, \\
2. \; {\langle}  {\xi_2, \nabla^2\Gamma_2}{\xi_2 \rangle \le 0}, \\
3.  \; {\xi_{1}}(\vec{x}_0,t_0,t) \| {T_{\vec{x}_0}}{M(t_0)},  \\
4. \;  {\overline{\Gamma_{2}}}\text{ is maximal} , 
 \end{array} 
 \]
 where ${T_{\vec{x}_0}}$ is the tangent space to ${M(t_0)}$ at ${\vec{x}_{0}}$, the over bar indicates the average value. The first condition keeps the hyperbolic LCS away from the degenerate points. The second condition ensures the ${\Gamma_{2}}$ is a local maximum near each point $({\vec{x}_{0}})$ on the LCS. The third and fourth conditions guarantee that the normal vector at each point $({\vec{x}_{0}})$ is aligned with the direction of the most repelling and attracting trajectories, and the LCS is more repelling and attracting than any close material line. 

\section{Description of Data and Methods}\label{experim}

The planar (streamwise-spanwise) turbulent boundary layer data were obtained at the Laboratoire de M\'{e}canique de Lille (LML) wind tunnel using a time resolved stereo particle image velocimetry. Dimensions of the test section are 20.6 m in length, 2 m in width and 1 m in height. The wind tunnel is designed for high Reynolds numbers and large boundary layer thickness at low speeds. Figure 1 illustrates the experimental setup, where the axis $x$ is parallel to the wall in streamwise direction, the axis $y$ is normal to the wall and the axis $z$ is the transversal. 

The free stream velocity, $U_{\infty}$, was 5 ms$^{-1}$ and corresponding Reynolds number based momentum thickness, $\mathrm{Re}_{\theta}$, became 9800. Boundary layer thickness at this Reynolds number was approximately 30 cm. The dimension of the measurement plane was 5.9 cm in streamwise direction and 4.8 cm in spanwise direction. To facilitate these measurements, two Phantom cameras located in symmetrical forward-scattering conditions with 1 mm laser thickness sheet were used. Poly-Ethylene Glycol particles of size 1 $\mu$m were seeded within the test section. The Scheimpflug condition was used to ensure uniform image focus in both cameras across the field. 

The spatial coordinates are nondimensionalized by viscous length scale which equal to $\delta_v=\nu/u_{\tau}$, where $\nu$ is kinematic viscosity and $u_{\tau}$ is the friction velocity. In this study, viscous length scale and the friction velocity are 0.09 mm and 0.174 ms$^{-1}$, respectively. The $(x-z)$ plane is parallel to the wall and located at $ y=4.5$ cm, which equals to 50 wall units. A total of 1200 packets of 40 time resolved data were recorded with resolutions of 0.67 ms in time and 0.5 mm in space. Further information on the experimental set up can be found in \citet{FOUCAUTetal2007,WTR}.

\begin{figure}
\begin{center}
\subfigure[ Side view]{\label{exp1} \includegraphics[width=0.47\textwidth,height=2.2 in]{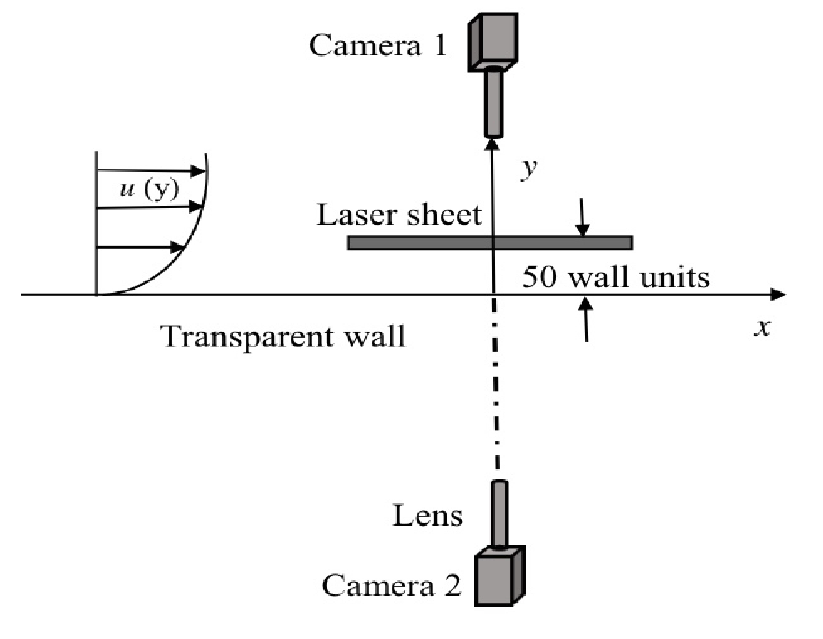}}
\subfigure[Upstream view]{\label{exp2} \includegraphics[width=0.48\textwidth,height=2.4 in]{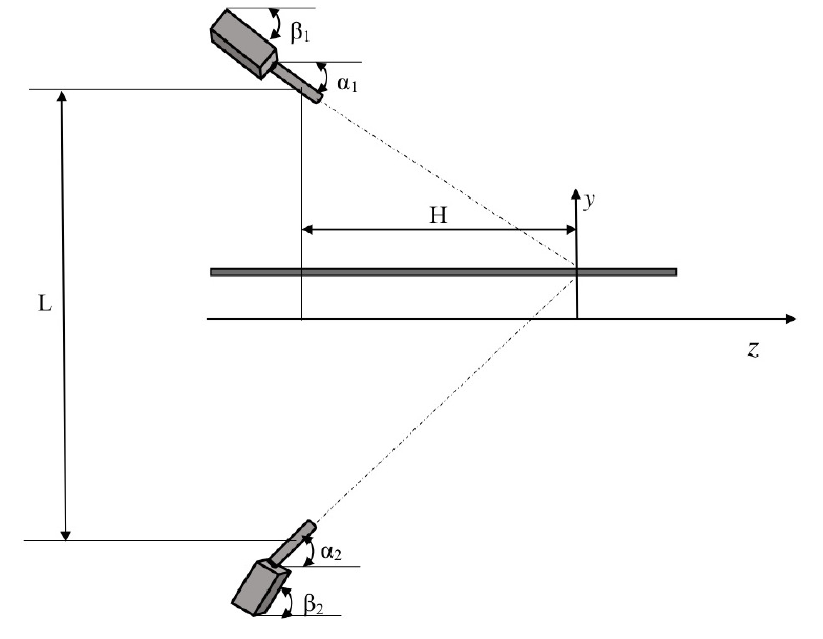}}
\end{center}
\caption{\label{1}Experimental setup: (a) side view; (b) upstream view.}
\end{figure}

\section{Results}\label{resu}
\subsection{\label{Results1} Proper Orthogonal Decomposition Analysis}

A total  of 40 successive snapshots are analyzed {\it{via}} proper orthogonal decomposition. The eigenvalues are normalized by total energy, or summation of the eigenvalues, as shown in equation  2.9. Figure 2(a) presents the percentage of the energy carried by individual POD modes. As the mode number increases, the eigenvalues decreases. 
The first mode contains 32\% of the total turbulence kinetic energy. The second and third modes possess half of the energy, 15\% and 11\%, respectively. Modes 4-6 carry energy of approximately 5\% to 7\%. Finally, the modes from 11 to 40 do not exceed 1\%, summing up to less than 30\% of the total turbulence kinetic energy. Consequently, the first six modes carry most of the energy content in the flow. Correspondingly, figure 2(b) displays the convergence of eigenvalues, following equation (2.10).  The cumulative energy of the first three, first six and first ten modes is 50\%, 75\% and 92\% of the total energy, respectively and a small percentage contribution is observed thereafter. 

\begin{figure}
\begin{center}
\subfigure[ Normalized eigenvalue distribution]{\label{sum energy} \includegraphics[width=0.48\textwidth]{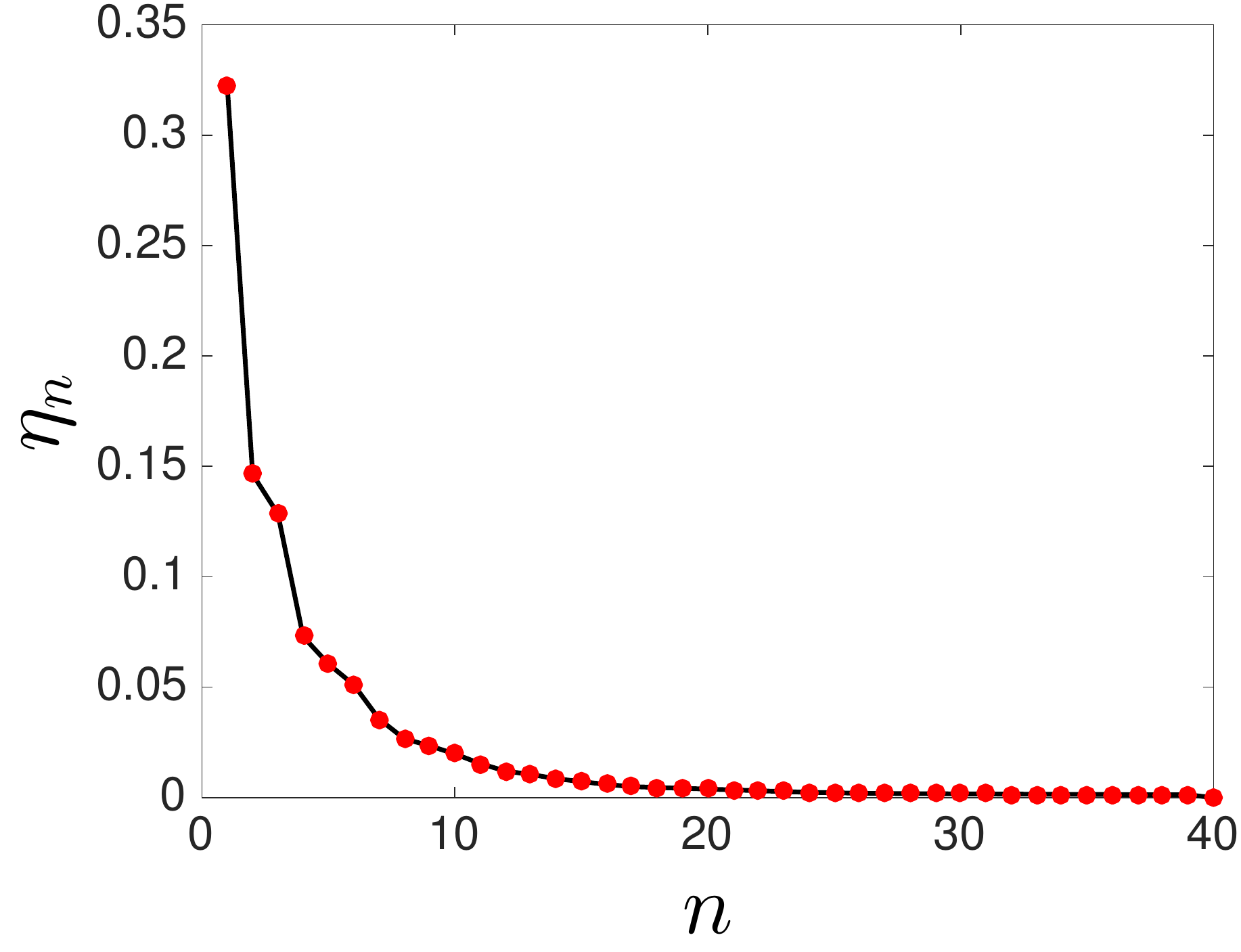}}
\subfigure[Normalized cumulative energy]{\label{single energy} \includegraphics[width=0.48\textwidth]{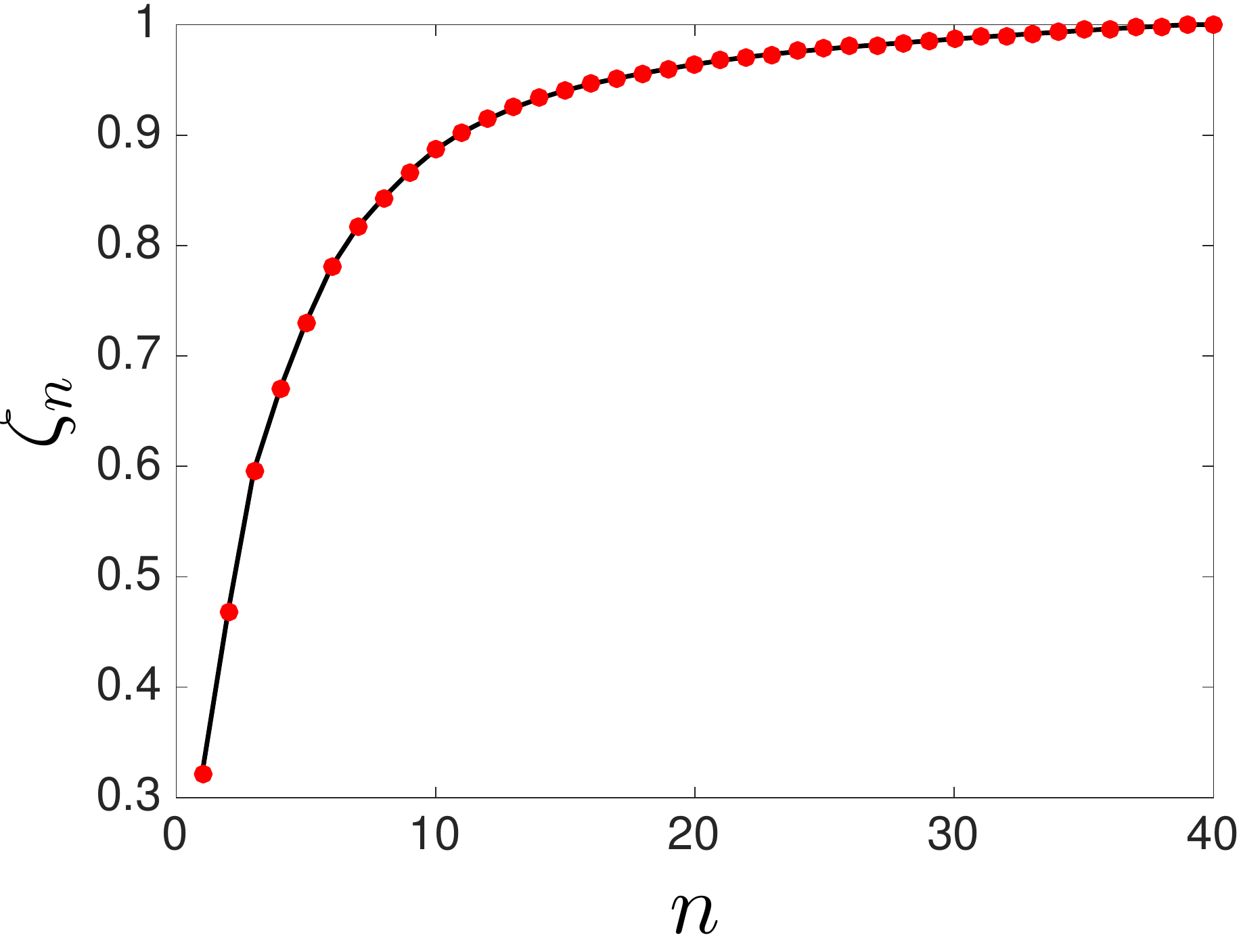}}
\end{center}
\caption{\label{EPOD.}Energy content of the of POD modes.}
\end{figure}

Six different POD modes associated with the streamwise direction are visualized in figure 3. For simplicity, $P_{U_n}$ denotes the $n$$^{th}$ POD mode associated with the streamwise component. Figure 3(a) depicts the first mode, $P_{U_1}$; the decomposition yields the mean flow in the streamwise direction. The mode also contains the greatest coherence in comparison with subsequent modes. Although the second mode, $P_{U_2}$, contains 50\% of the  energy carried by the first mode, large structures are still pronounced especially at the region at $x^{+} \in [-100, 100]$ and $z^{+}>0$. Although the coherence in the flow domain decreases per mode 3, structures within the flow are still recognized as shown in figure 3(c). After $P_{U_3}$, relatively small spatial structures appear which can be thought as a signature of higher mode numbers. In figures 3(d-e), modes 4 and 5 are presented to show the decrease in coherence where flow features are not as marked. Indeed, this is a trait of the framework as eigenvalues are ordered as the energy content is diminished. In stark contrast, $P_{U_{20}}$ highlights the small structures and their incoherence in comparison to the low mode numbers. 

 Vortical flow prevails where shear is present. The higher POD modes are related to small scale vortical motion. The features of the decomposed flow at low mode numbers resemble those as observed in streak-like phenomena as also observed in \citet{green1995fluid}. Wall streaks are initiated as a result to the penetration of the hairpin vortex leg from outer layer and contact with wall. Therefore, wall streaks  can be considered as the passive trail to the hairpin vortices convected closest to the wall. The streaks  have the repetition aspect or deterministic in the POD modes especially after $P_{U_4}$. As  mentioned in \citet{smith1991dynamics}, when the legs of the hairpin vortices are closer or when the vortex is further from the wall, the streaks merge and become single streak as shown in the $P_{U_2}$. The events that take place in the outer part of the boundary layer are exhibited as a footprint in the near-wall flow.

\begin{figure}
\begin{center}
\subfigure[$P_{U_1}$]{ \label{a} \includegraphics[
width=0.31\textwidth]{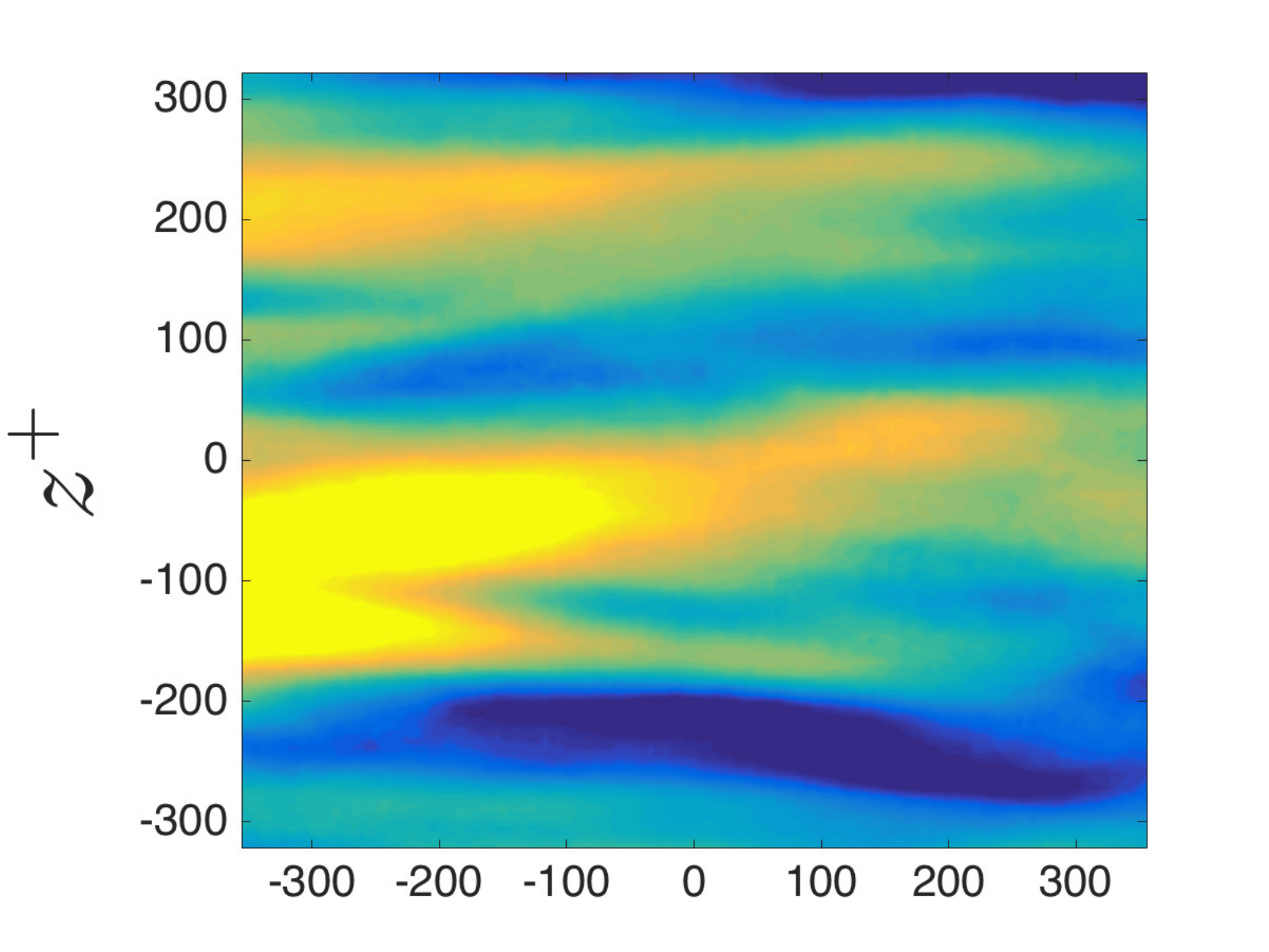} }
\subfigure[$P_{U_2}$]{ \label{b} \includegraphics[
width=0.31\textwidth]{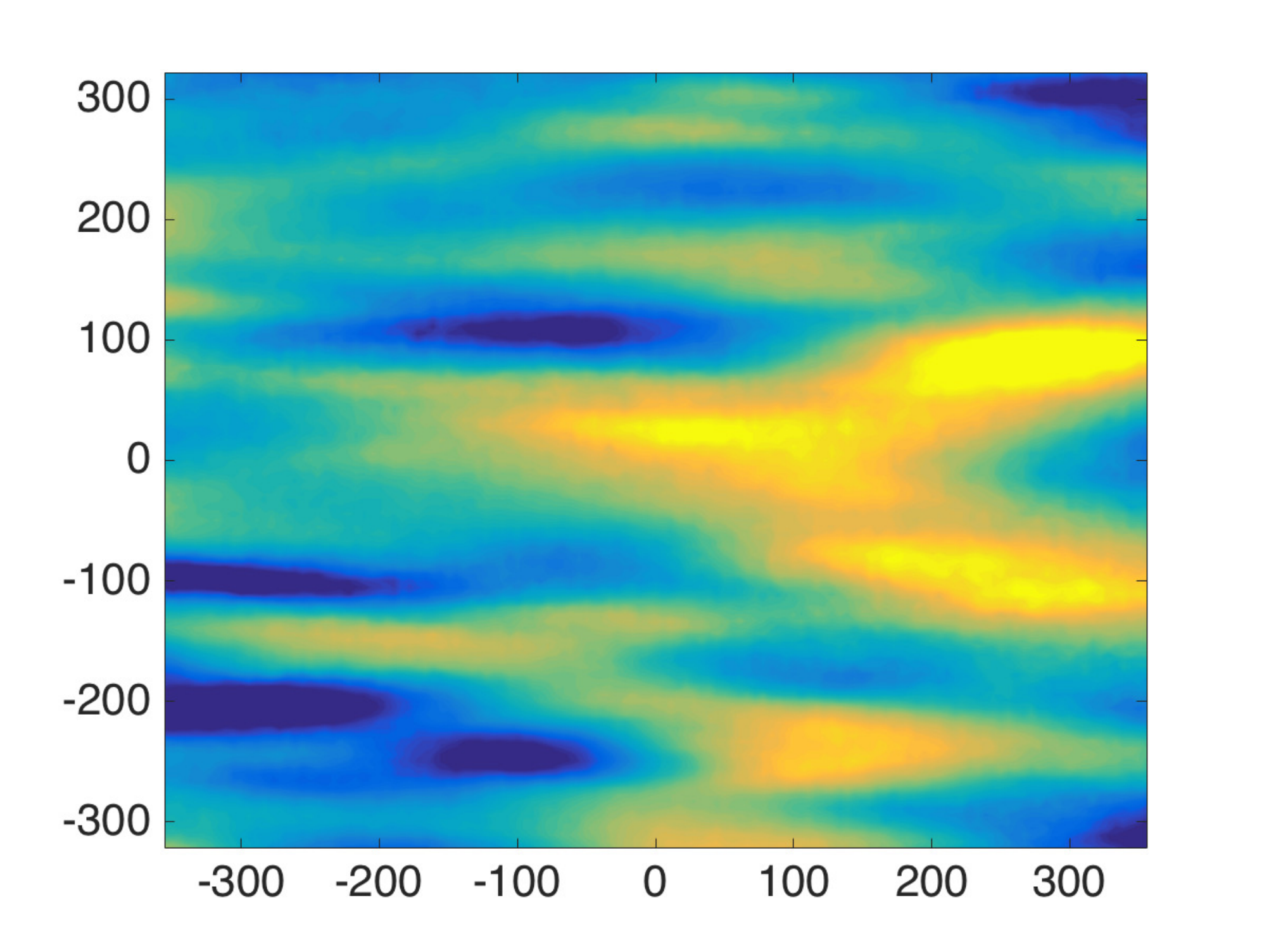} } 
\subfigure[$P_{U_3}$]{ \label{c} \includegraphics[
width=0.31\textwidth]{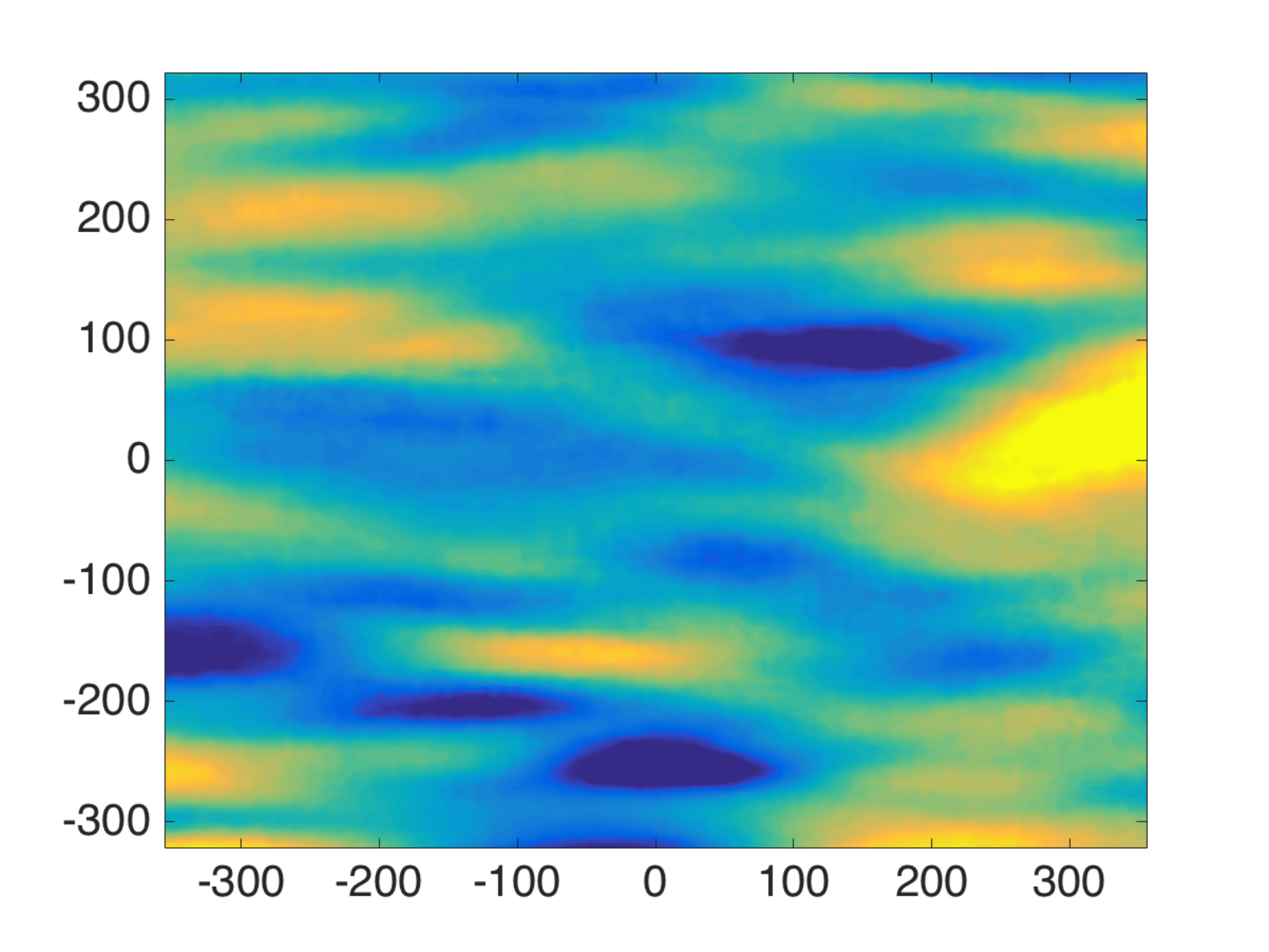} } 
\subfigure[$P_{U_4}$]{ \label{d} \includegraphics[
width=0.31\textwidth]{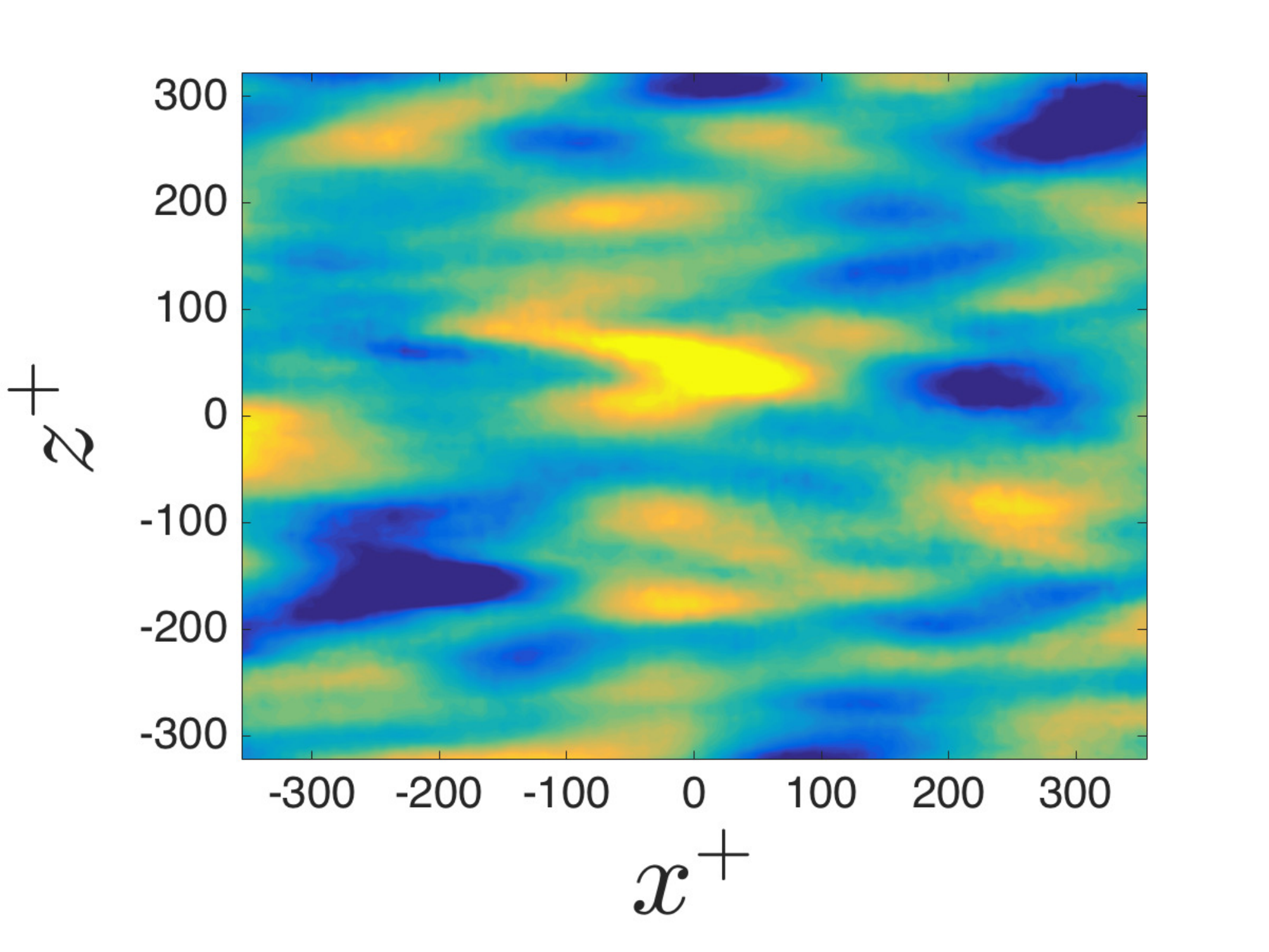} } 
\subfigure[$P_{U_5}$]{ \label{e} \includegraphics[
width=0.31\textwidth]{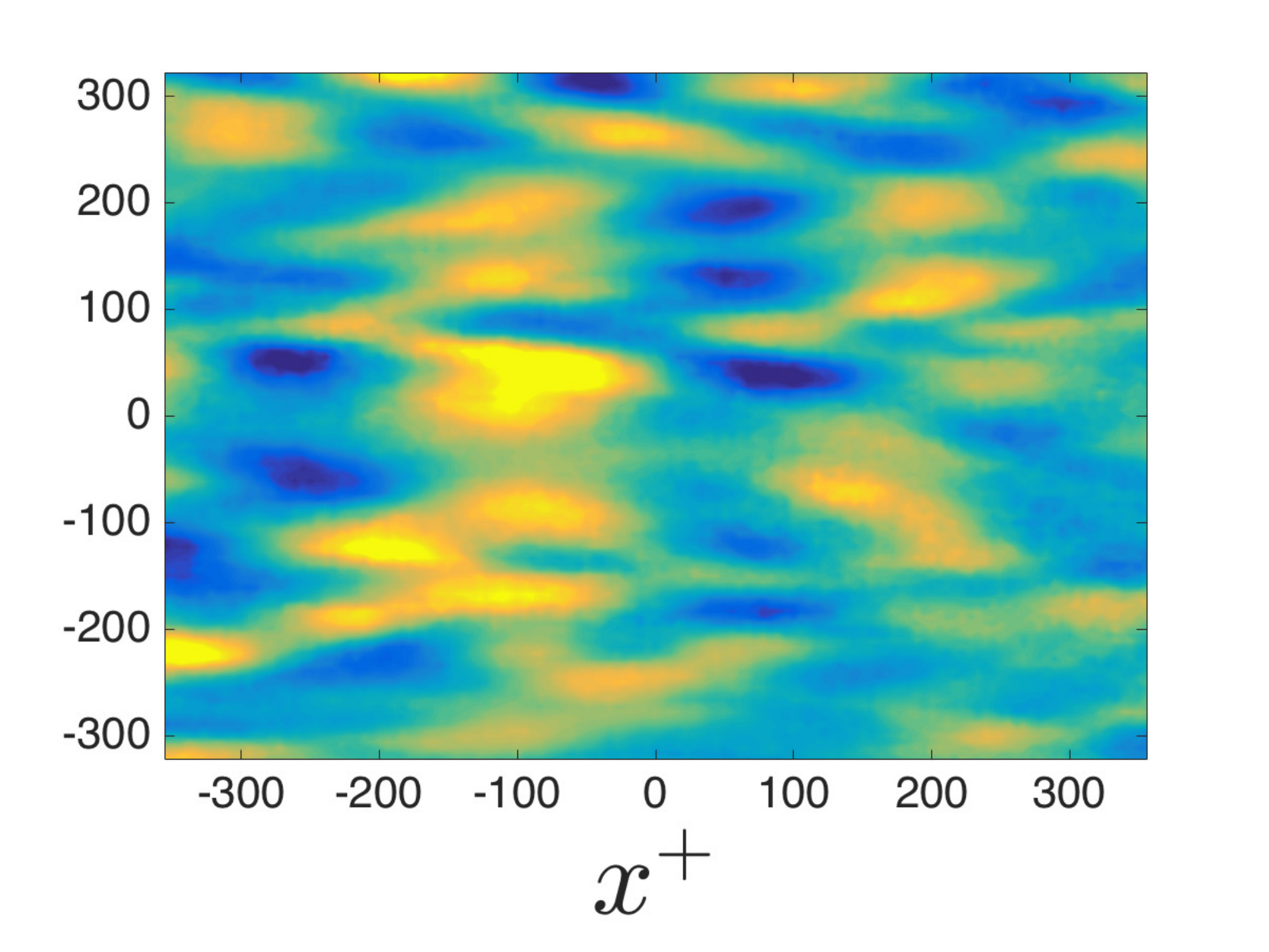} }
\subfigure[$P_{U_{20}}$]{ \label{f} \includegraphics[
width=0.31\textwidth]{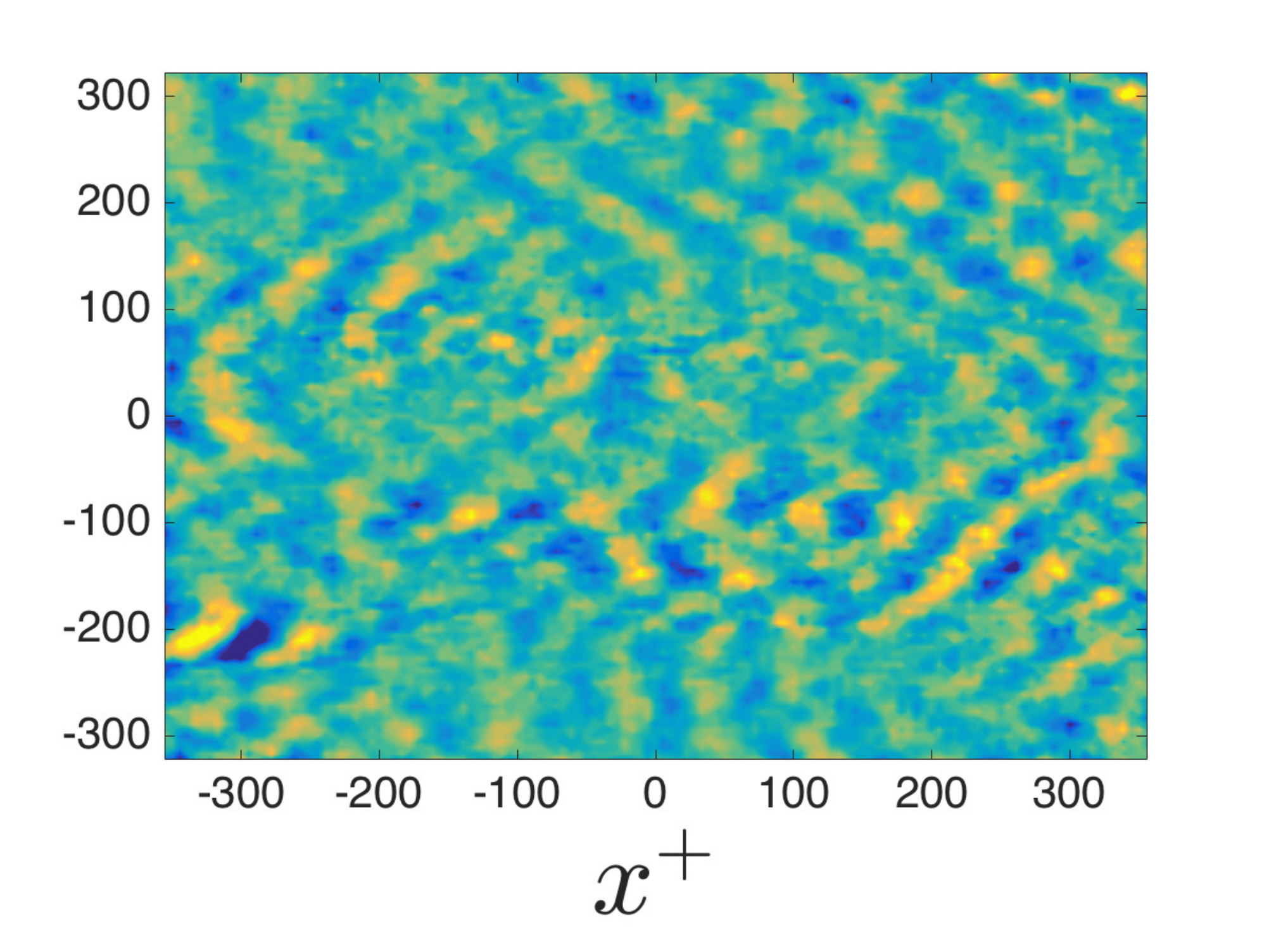} } 
%\\
\end{center}
\caption{\label{} Streamwise POD modes: (a) the first POD mode; (b) the second POD mode; (c) the third POD mode; (d) the fourth POD mode; (e) the fifth POD mode; (f) the twentieth POD mode.}
\end{figure}

\subsection{\label{Results2} Dynamic Mode Decomposition Analysis}

Projecting velocity data onto the associated eigenvector gives the spatial dynamic modes corresponding to the imaginary part of the companion matrix eigenvalues. Orthogonality in time is observed through the DMD modes. The logarithmic mapping of eigenvalues on the complex plane measures phase velocity of the dynamic modes as well as the growth/decay rates {\it{via}} the imaginary, $\Pi_{i}$, and real part, $\Pi_{r}$, respectively, as formulated in \cite{schmid2010dynamic}.

\begin{figure}
\begin{center}
\subfigure{ \includegraphics[scale=0.45]{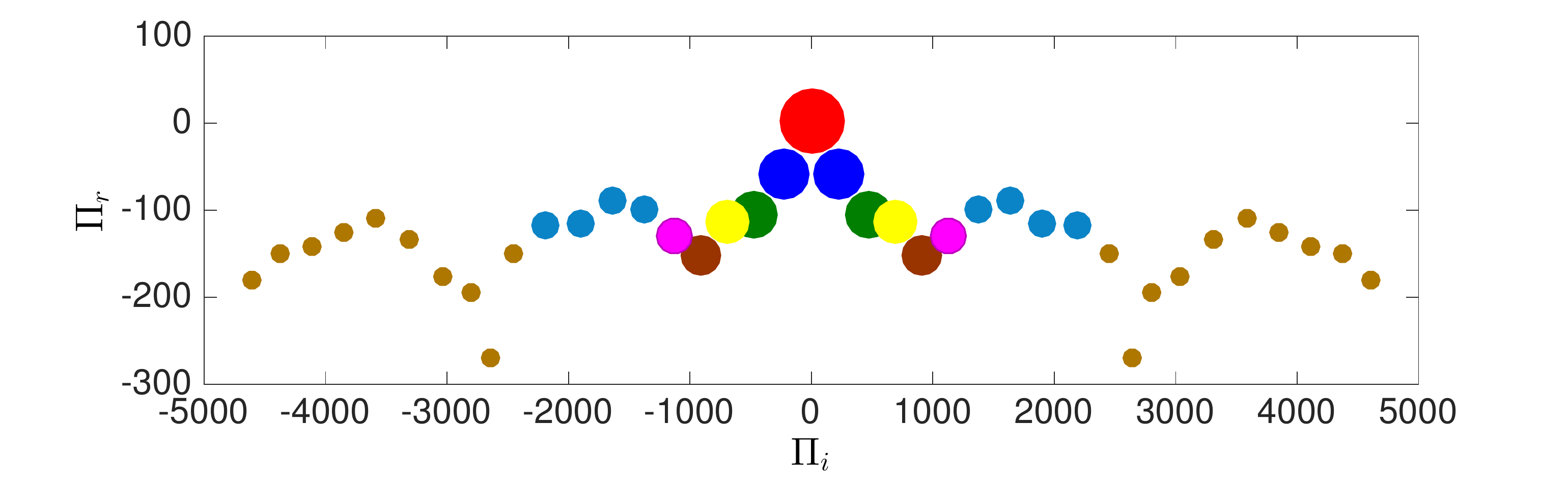} }
\end{center}
\caption{\label{}Spectrum of DMD modes. The marker size and color indicate the coherence of the DMD modes where the large and small circles identify the large and small scale structures, respectively.}
\end{figure}

\begin{figure}
\begin{center}
\subfigure[$D_{U_1}$]{ \label{a} \includegraphics[
width=0.31\textwidth]{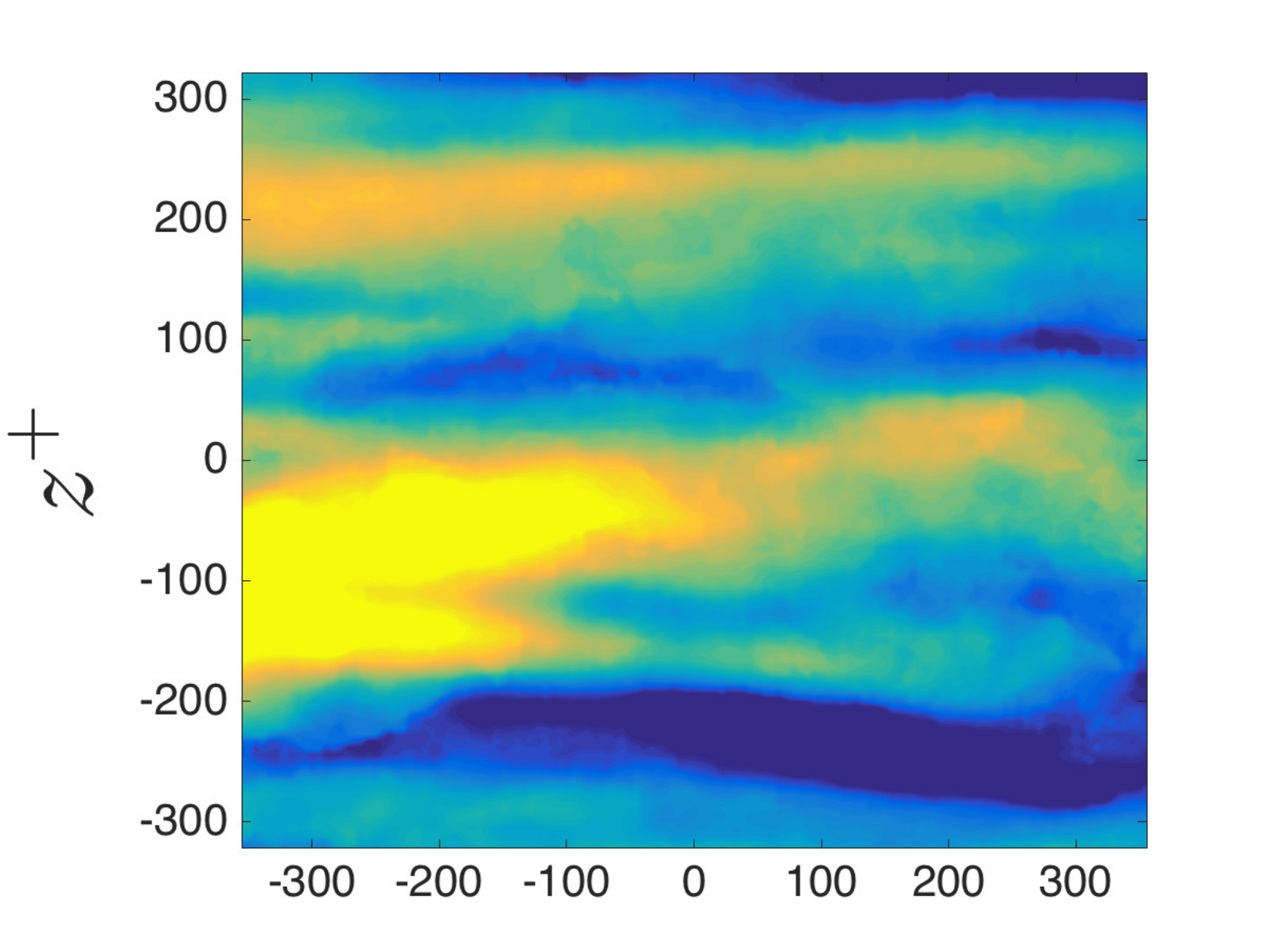} }
\subfigure[$D_{U_2}$]{ \label{b} \includegraphics[
width=0.31\textwidth]{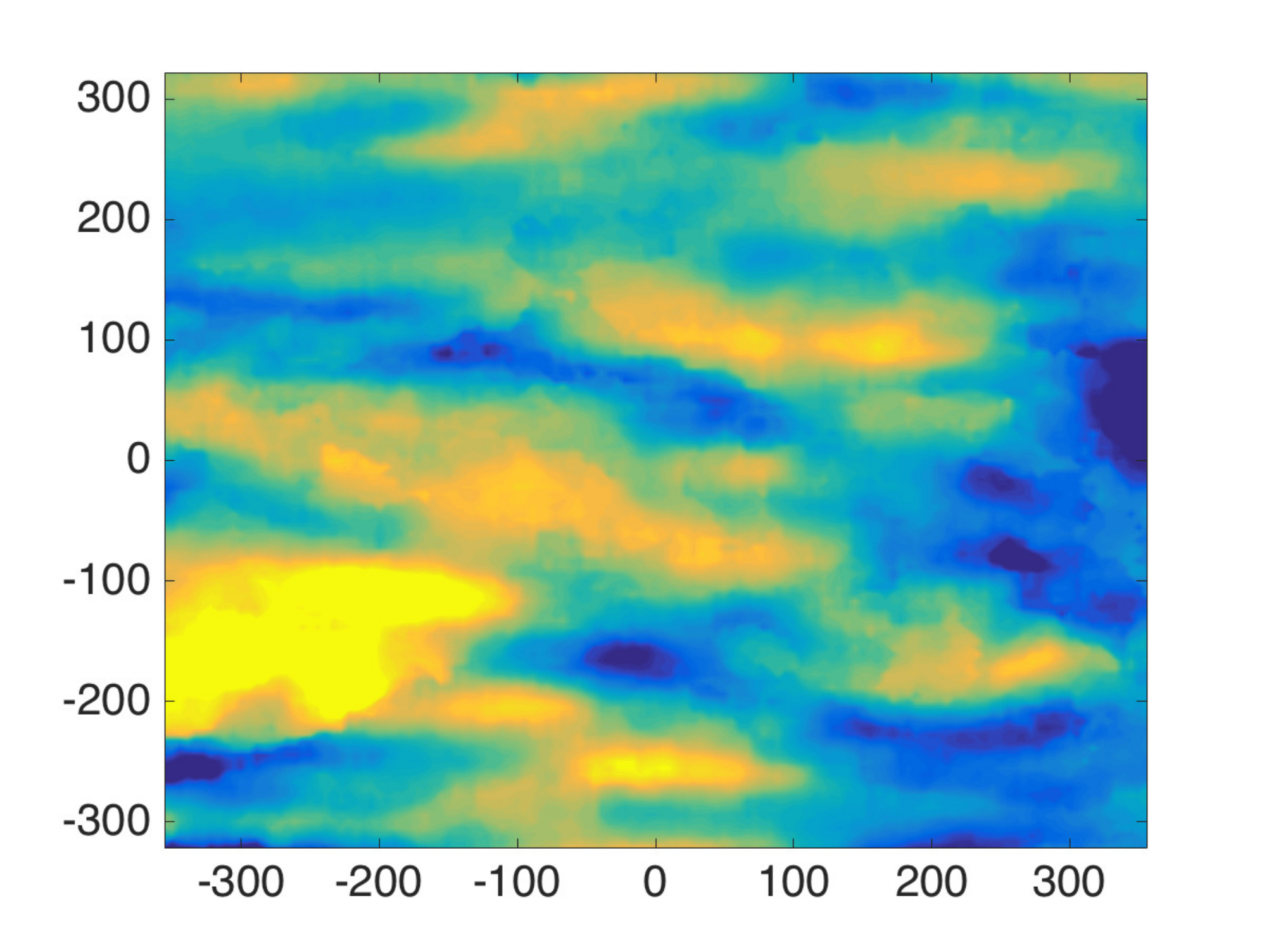} }
\subfigure[$D_{U_3}$]{ \label{c} \includegraphics[
width=0.31\textwidth]{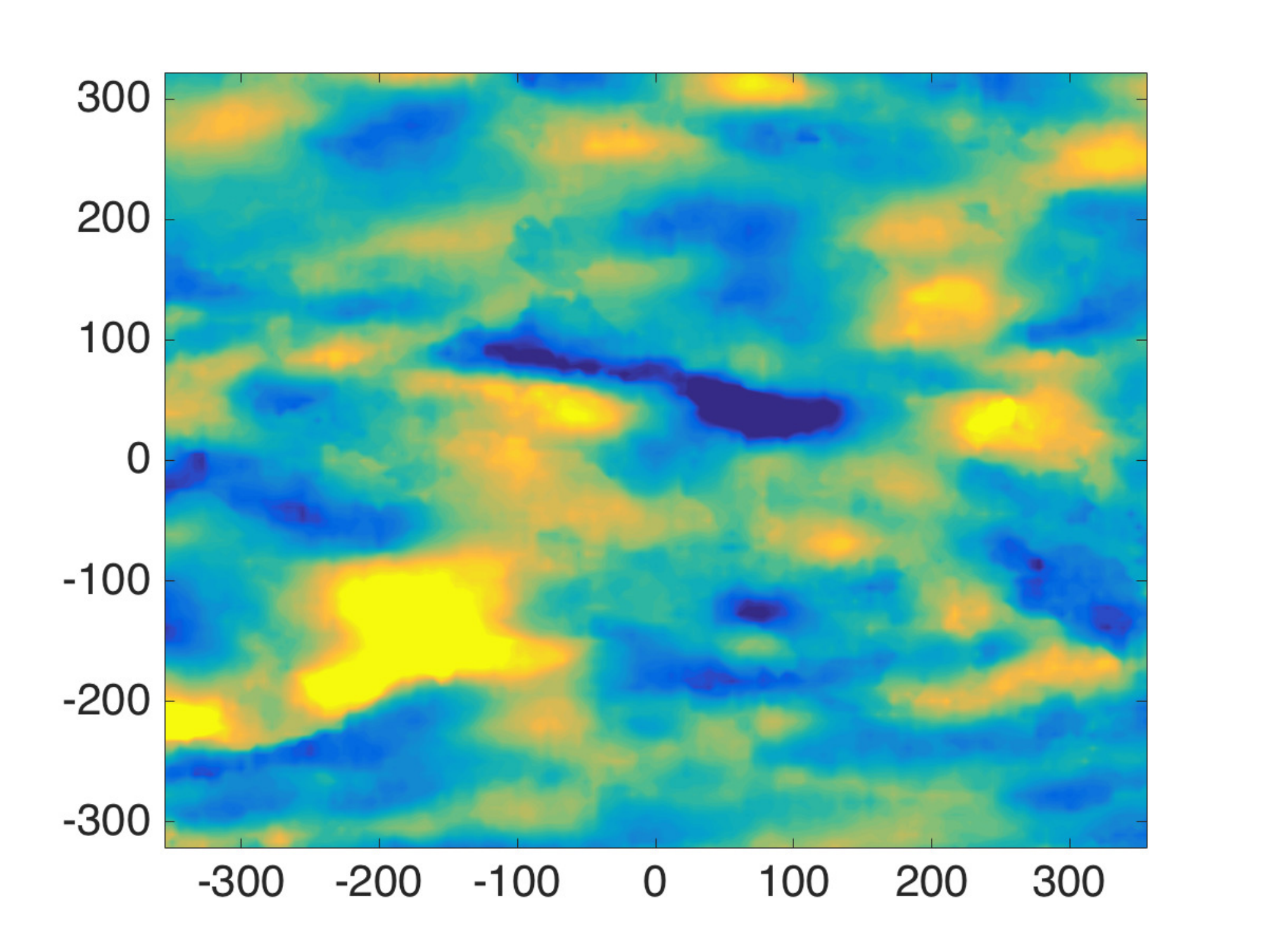} } 
\subfigure[$D_{U_4}$]{ \label{d} \includegraphics[
width=0.31\textwidth]{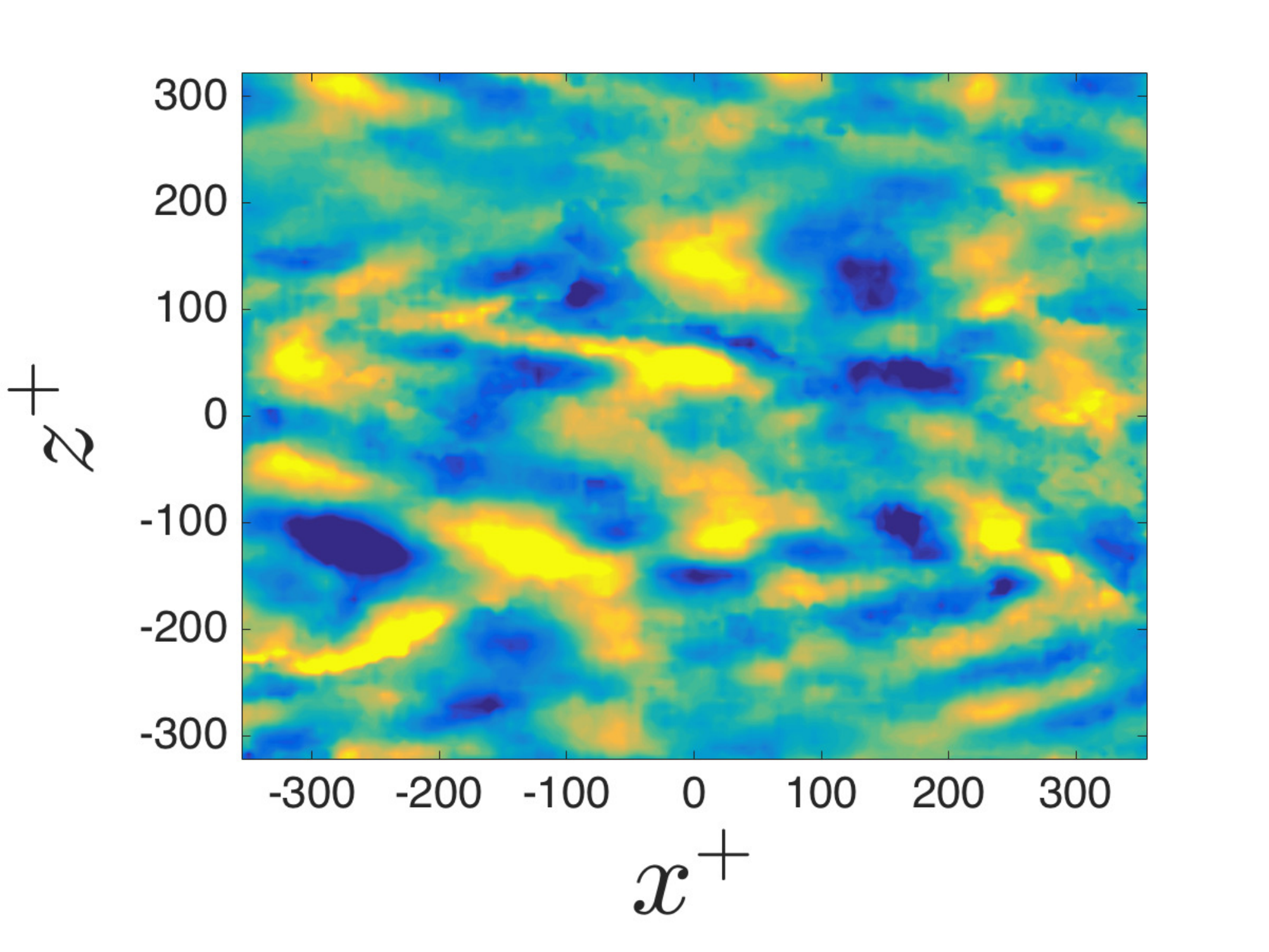} } 
\subfigure[$D_{U_5}$]{ \label{e} \includegraphics[
width=0.31\textwidth]{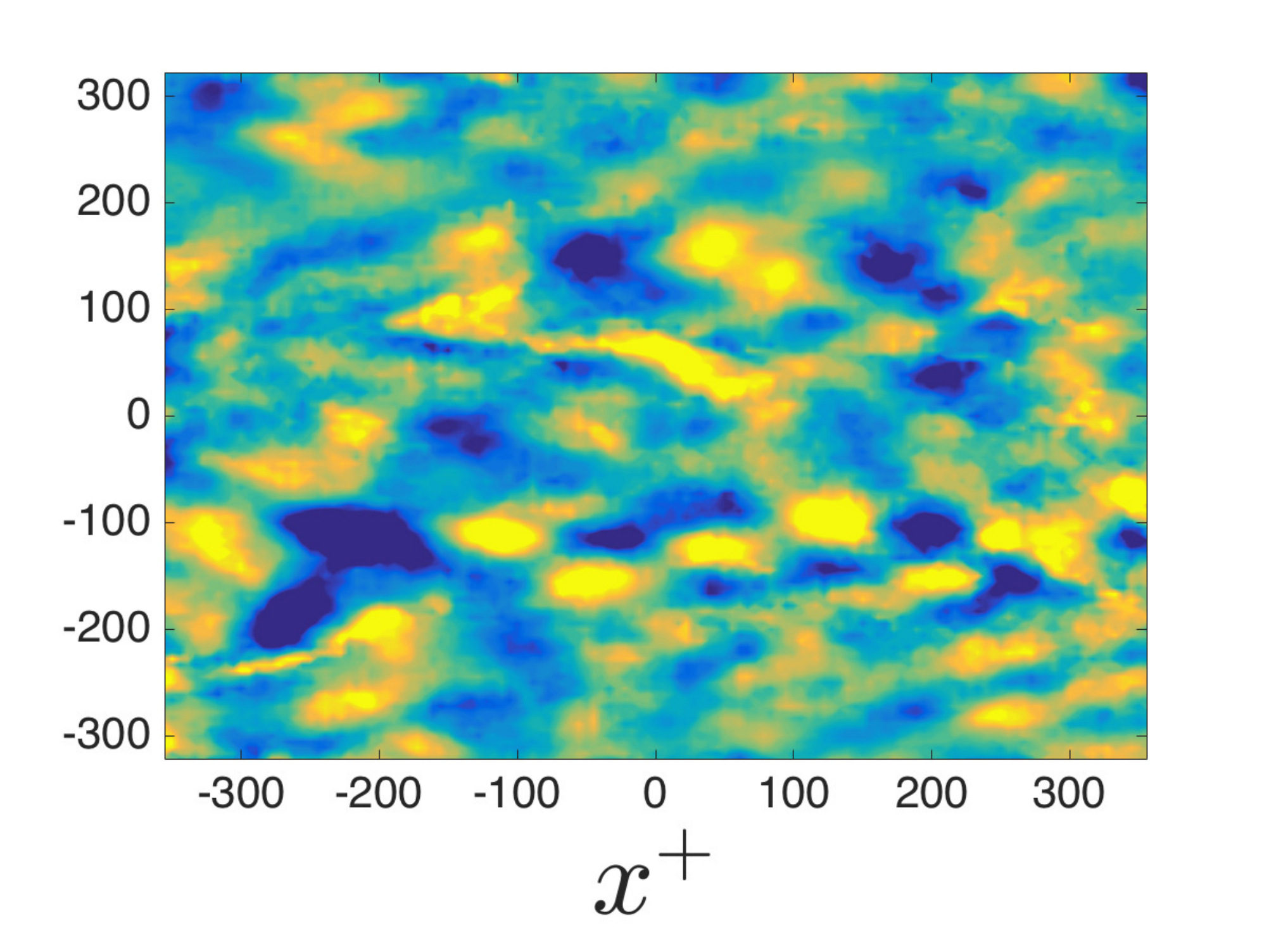} }
\subfigure[$D_{U_{20}}$]{ \label{f} \includegraphics[
width=0.31\textwidth]{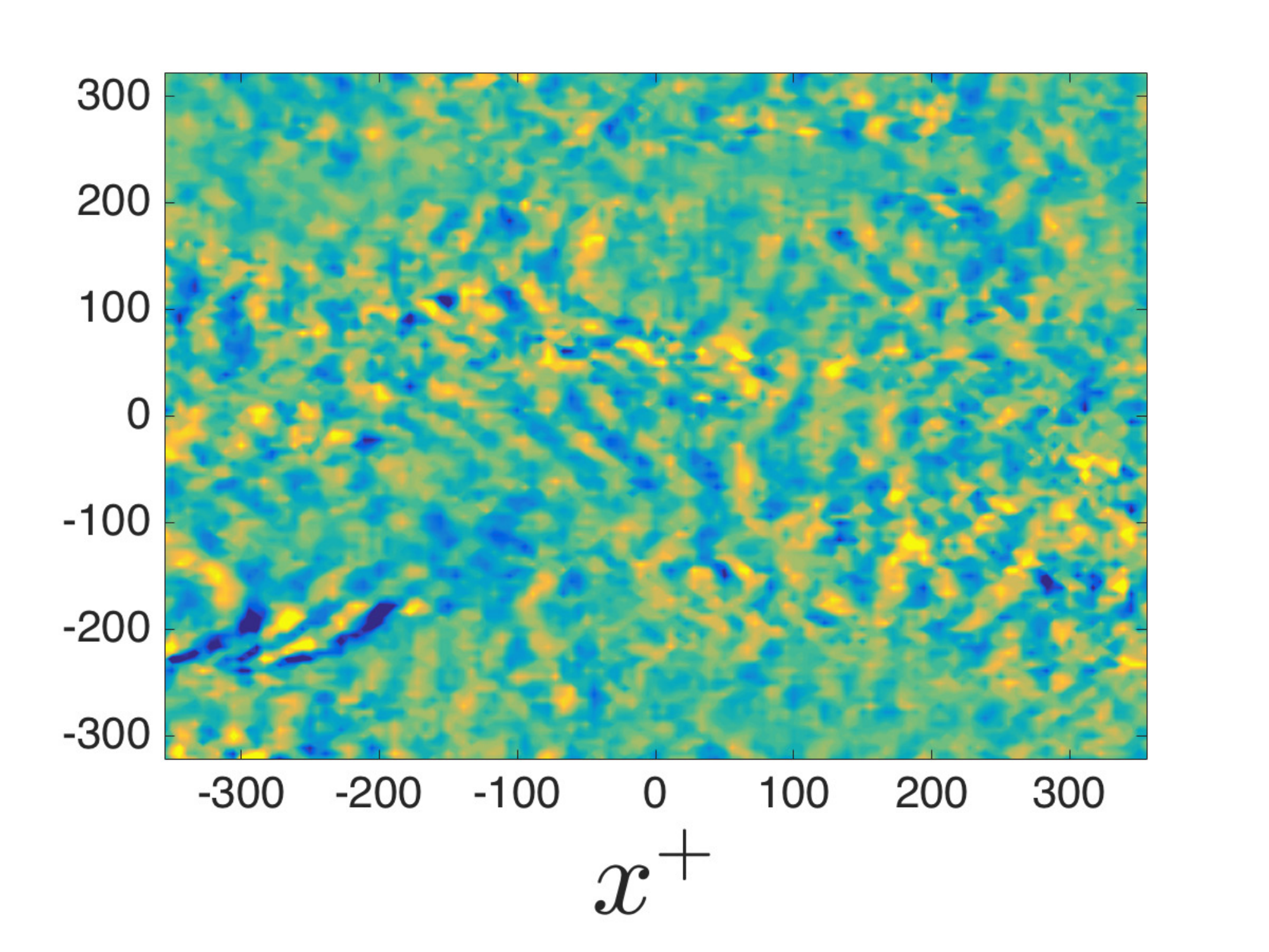} }
 %\\
\end{center}
\caption{\label{}Streamwise DMD modes: (a) the first DMD mode; (b) the second DMD mode; (c) the third DMD mode; (d) the fourth DMD mode; (e) the fifth DMD mode; (f) the twentieth DMD mode.}
\end{figure}

In figure 4, the logarithmic mapping is shown where $\Pi_r$ and $\Pi_i$ are plotted. The extracted spectrum in the complex plane is distinguished by different colors and sizes as these correspond to the structures and their dynamic content. Similar to POD, these are organized in descending value for both frequency and importance. The spectrum is symmetric and six dominant eigenvalues are zero-centered around $\Pi_i$. These correspond to the most dominant structures and are relatively large as their frequency is modest in comparison higher frequency modes, $\Pi_i < \lvert 1200\rvert$.  In observing the relative size of the circles, the first mode located at $\Pi_i=0$ is largest and shows a growth rate as $\Pi_r > 0$, thus corresponding to the mean flow. Thereafter, a wide range of frequencies are captured and characterized within each following DMD mode. Modes two through five (blue, green, yellow and brown circles) experience an increase in decay rate, although the rates are different from mode to mode. Interestingly, the decay rate doubles from modes one through three at $\Pi_r$ $\approx$  0, -50 and -100, respectively. The variation from mode three to four is minor.  The trend from modes five to six shifts to a lessened decay rate. Past mode six at $\Pi_i < \lvert 1200\rvert$, modes begin to oscillate and manifest a negligible influence in the flow field. This can be attributed to the influence of the incoherent structures which contain frequencies related to the small scales. 

In visualizing the particular DMD modes, streamwise modes are shown in figure 5 and similar to the POD are denoted as $D_{U_n}$ for simplicity. Figure 5 presents the first five modes as well as the twentieth mode to provide a relative comparison of the most energetic modes and an incoherent/high frequency mode. $D_{U_1}$, once again, represents the mean flow, but in this instance obtained {\it{via}} DMD. Coherence of the first mode is visible as compared with  following modes, where structures are elongated covering the entire interrogation area in some instances. As mode numbers increase, large scale features begin to break up and coherence disappears even for the second mode, $D_{U_2}$, see figure 5(b). Frequency of the structures increases as the size of the structure decreases. For example, a small difference between $D_{U_3}$  and $D_{U_4}$  exists when these are compared in their frequencies, 0.475 kHz for $D_{U_3}$, 0.694 kHz for $D_{U_4}$, respectively. $D_{U_5}$ displays much smaller scales which oscillate at frequency of 0.91 kHz. Incoherence is apparent at mode number 20 which also has a higher frequency,  4.61 kHz, figure 5(f). 

Proper orthogonal decomposition and dynamic mode decomposition analyses the data enabling to extract relevant flow features through dynamic based modes and energy based modes. Although these two decomposition techniques have differences in their approach, many similarities are found. For example, the first mode of the POD and DMD is very similar in particular. This also  means that the two decompositions extract the same structures, i.e., the mean flow.

\subsection{\label{Results3} Lagrangian Coherent Structures}

LCS can identify the flow structure {\it{via}} their distribution over the flow domain. Figure 6 illustrates the orientation of repelling and attracting LCSs at four different non-dimensionalized times, $t^{+}= t u_{\tau}/\delta_{\nu}=1.29, 18, 37.5$  and 49, and the advection of structures is clear within the domain as also shown in \cite{WTR}. LCSs demonstrate a variety of signatures in terms of the alignment, curvature and size. It is difficult to extract the trend of these trajectories due to the complexity of the turbulent boundary layer flow. In general, the characteristic shapes of the LCSs possess straight or inclined patterns, which bend in the middle and/or have an open ring at the ends. Some LCSs are visualized at different times indicating the advection through the domain of the structures at $y^{+}=50 \delta_{\nu}$. 

With increasing advection time, the shape and length of the LCSs change as the flow map evolves both in time and space. \cite{farazmand2015hyperbolic} remarked that the length of repelling and attracting lines rapidly shrinks with advection forward or backward a consequence of initial conditions. This is more pronounced for attracting lines (red trajectories) than repelling lines (black trajectories). When $t^{+}=1.29$ and $18$, the attracting trajectories are more organized and  less tangled than the repelling trajectories; this is reflected through the length of the repelling structures. However, this state is reversed with increasing time, where the repelling lines become more organized and longer compared to  attracting lines, in particular region between $z^{+}=-100$  and $z^{+}=-200 $ as shown in figure 6(c and d).

\begin{figure}
\begin{center}
\subfigure[$t^{+}=1.29$]{ \label{a} \includegraphics[
width=0.48\textwidth]{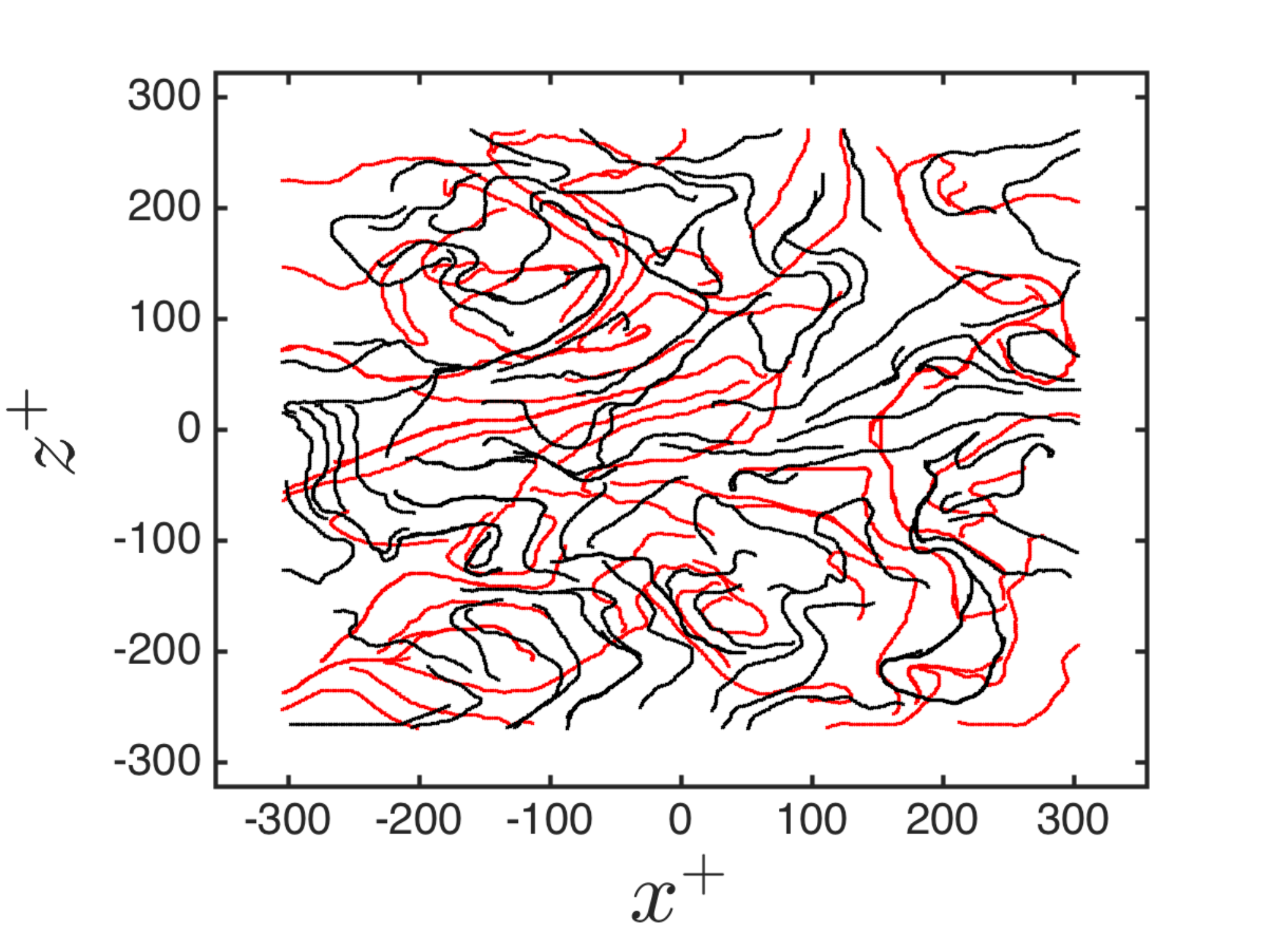} }
\subfigure[$t^{+}=18$]{ \label{b} \includegraphics[
width=0.48\textwidth]{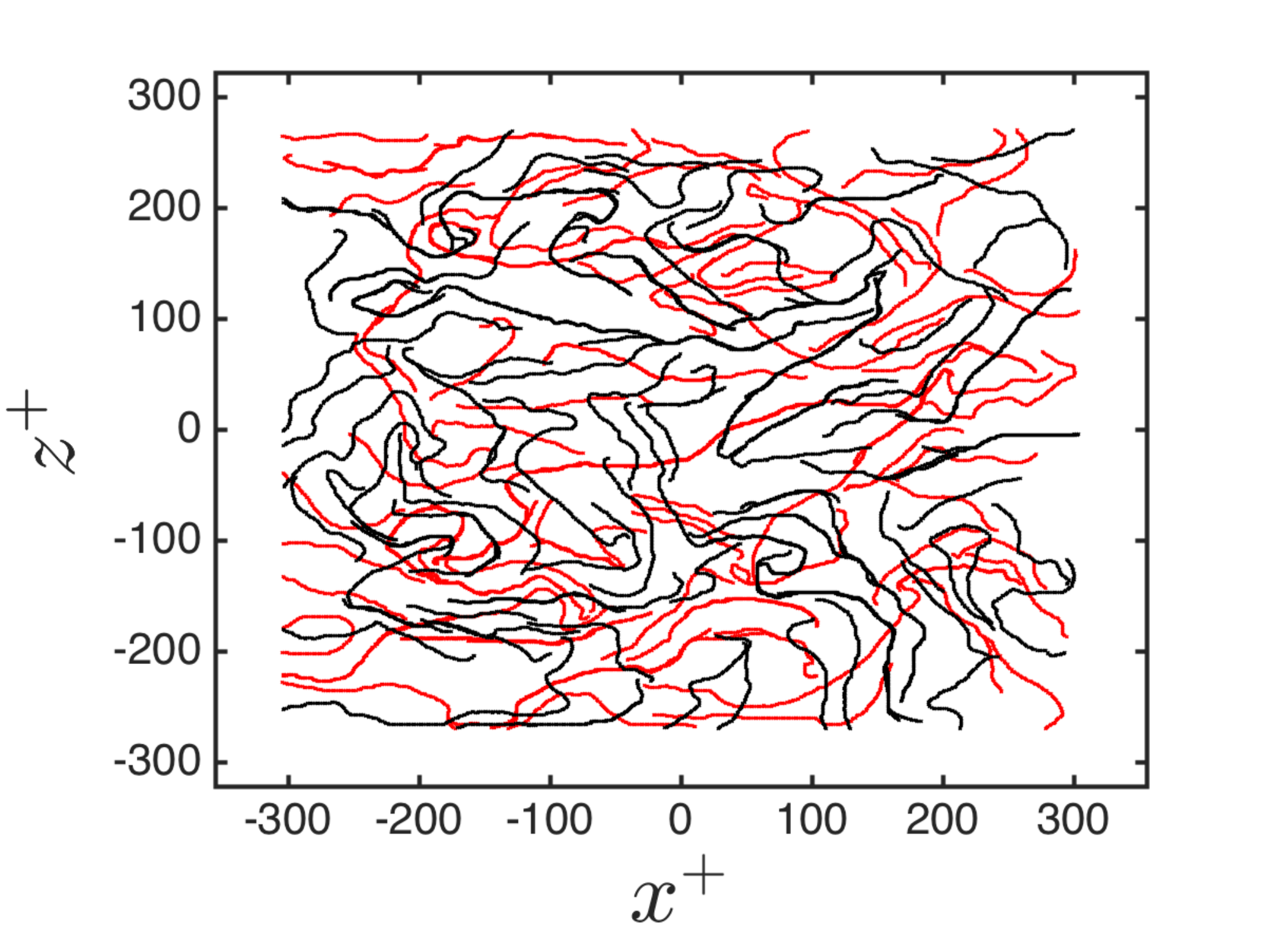} } 
\subfigure[$t^{+}=37.5$]{ \label{c} \includegraphics[
width=0.48\textwidth]{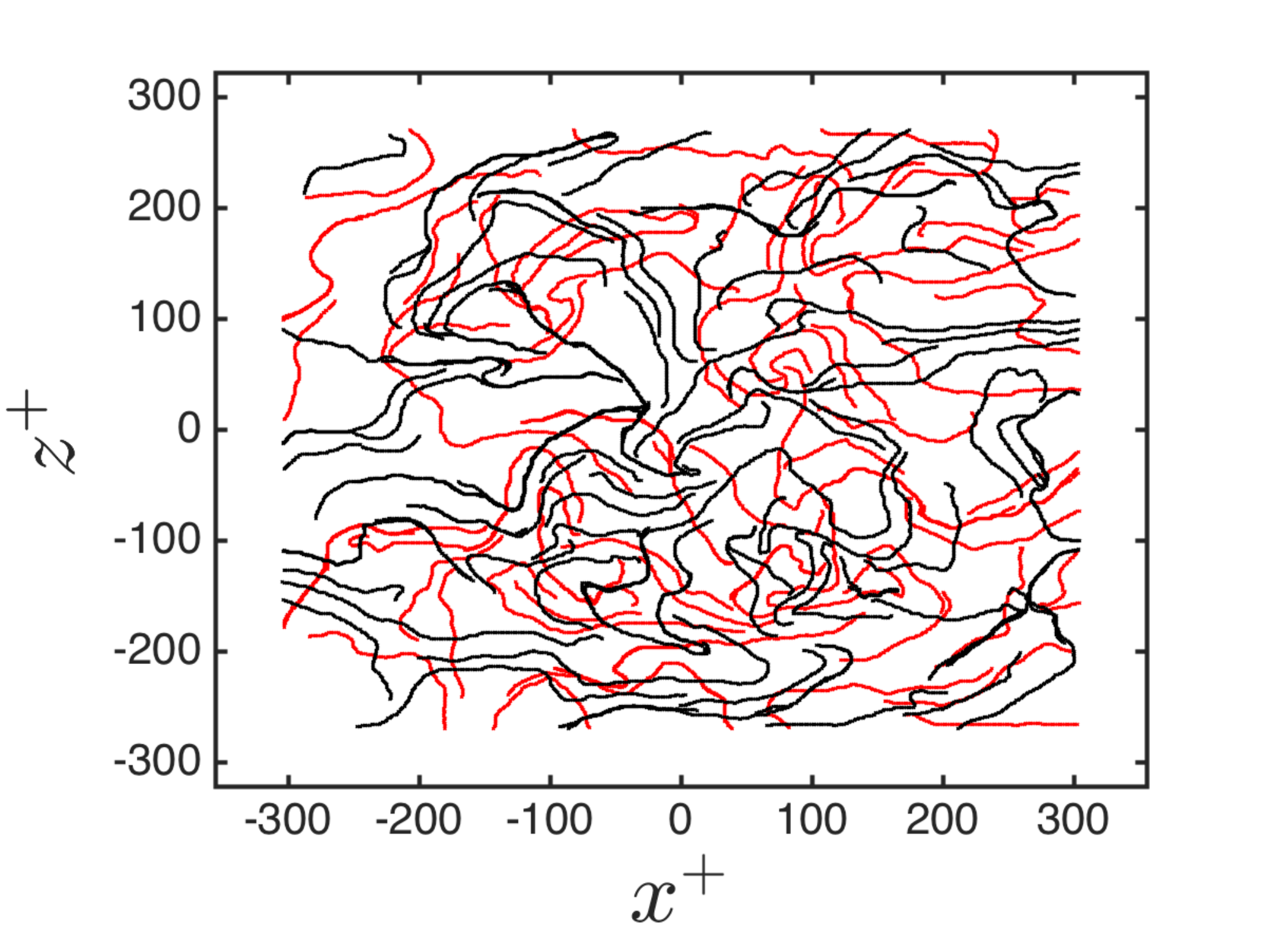} }
\subfigure[$t^{+}=49$]{ \label{c} \includegraphics[
width=0.48\textwidth]{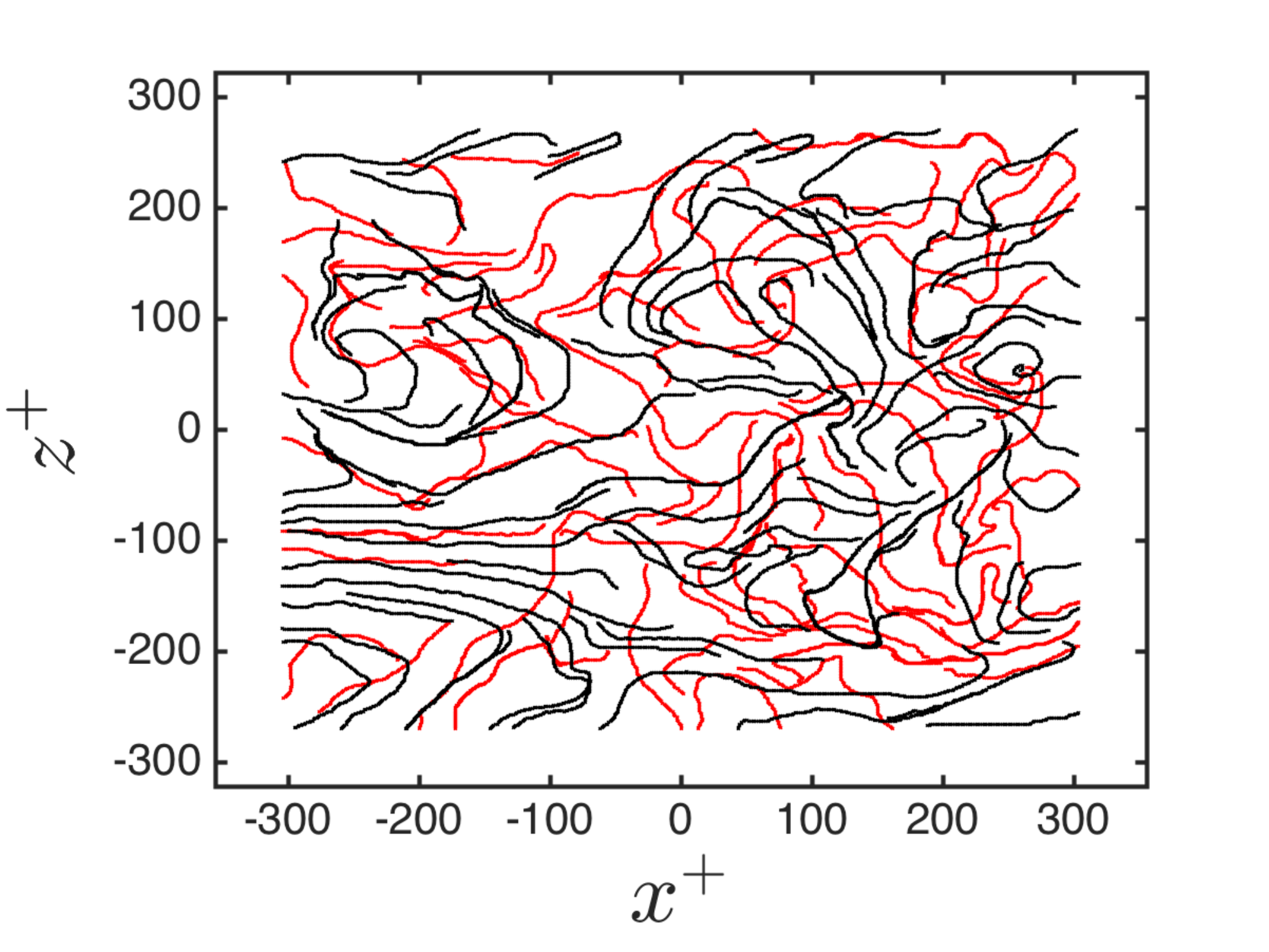} }
\end{center}
\caption{\label{} LCSs computed at four different times. Red lines (attracting) and black lines (repelling): (a) $t^{+}=1.29$; (b) $t^{+}=18$; (c) $t^{+}=37.5$; (d) $t^{+}=49$.}
\end{figure}

Repelling and attracting lines are observed in order to distinguish the dynamically most distinct structures withing the flow. Both LCS, POD and DMD capture structures within the flow while the latter two sort the structures according to their energy content or frequency, respectively. It is then also possible to compare and contrast how the computed LCS fare with the structures found {\it{via}} POD and DMD as shown in section \ref{Results1} and \ref{Results2}. In order to visualize this relationship, LCS are overlayed on contours of specific POD and DMD individual modes as shown in figures 7 and 8, respectively. Specifically, modes 1 through 5 and the twentieth mode are used. Unequivocally, the LCSs tend to delineate the most important features as captured by the modes and the trajectories tangent to the structure. These attracting and repelling LCSs touch the boundaries and are roughly perpendicular to the structure or arc to embrace the structures with maximum and minimum energy content. Although there are some differences between the POD and DMD modes, LCSs identifies all the structures of these modes and thus the techniques compliment each other well.
 
In order to visualize the relation between LCS trajectories and the POD, repelling and attracting LCSs are overlayed over the first POD mode at four different times, $t^{+}=1.29, 18, 37.5$ and 49 as demonstrated in figure 9. Although the shape and length of repelling and attracting LCS are changing with time, they still capture most structures identified by the POD. In addition, the morphing of LCS shape  comes from the variations of the structure temporally as these are advected downstream. At the beginning of time advection, $t^{+}=$1.29 and 18, attracting and repelling lines spread randomly over the features of the mode. With increasing time, LCSs begin to distort their orientation and moving as  straight lines over structures especially between $z^{+}$ of 0 and -200. Repelling  LCSs are arc-shaped lines surrounding the regions between $z^{+}$ of -200 and -300, and once again, change their orientation with time and achieve a more uniform distribution.

\begin{figure}
\begin{center}
\subfigure[$P_{U_1}+ LCS$]{ \label{a} \includegraphics[
width=0.48\textwidth]{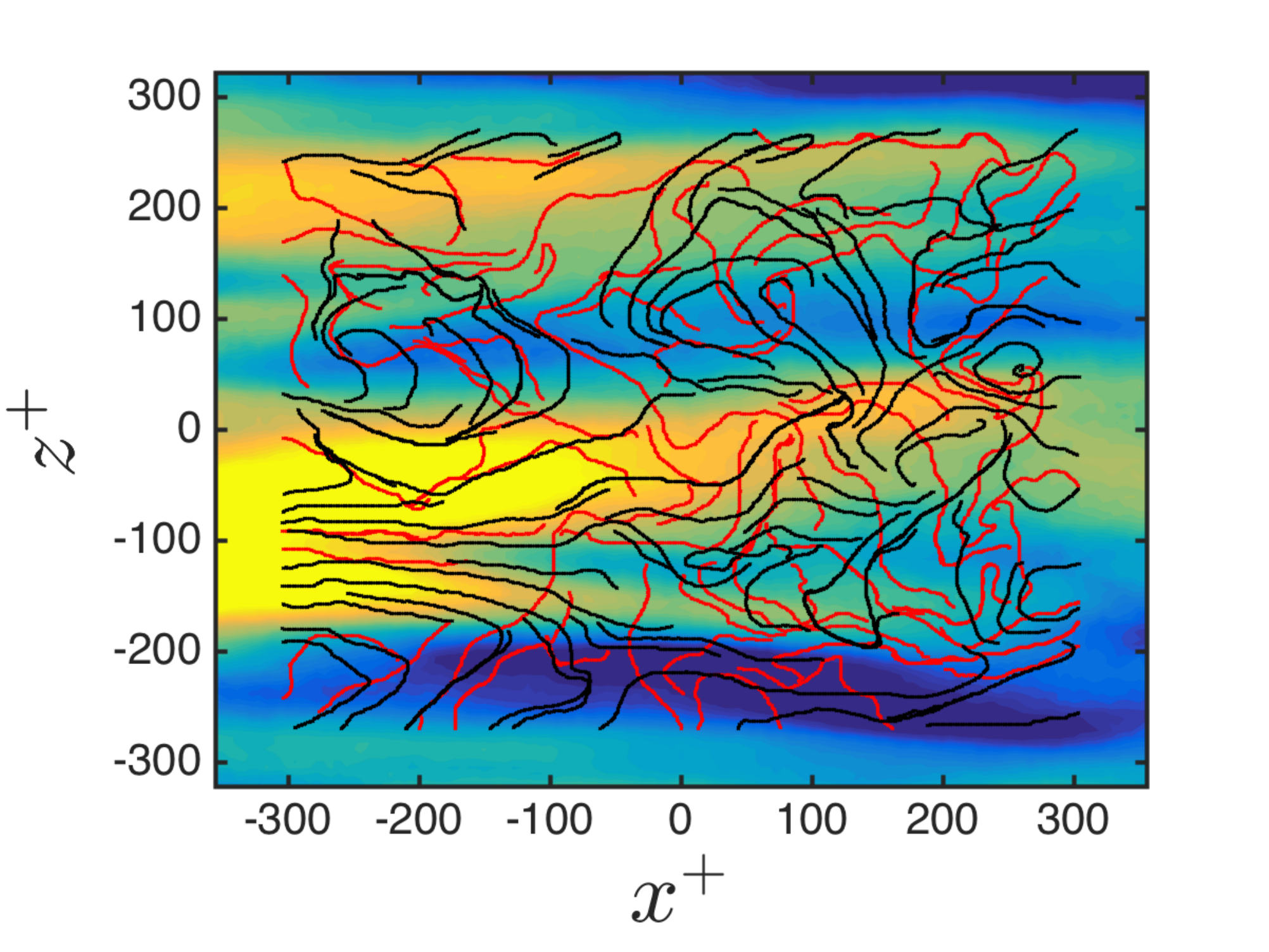} }
\subfigure[$P_{U_2}+LCS$]{ \label{b} \includegraphics[
width=0.48\textwidth]{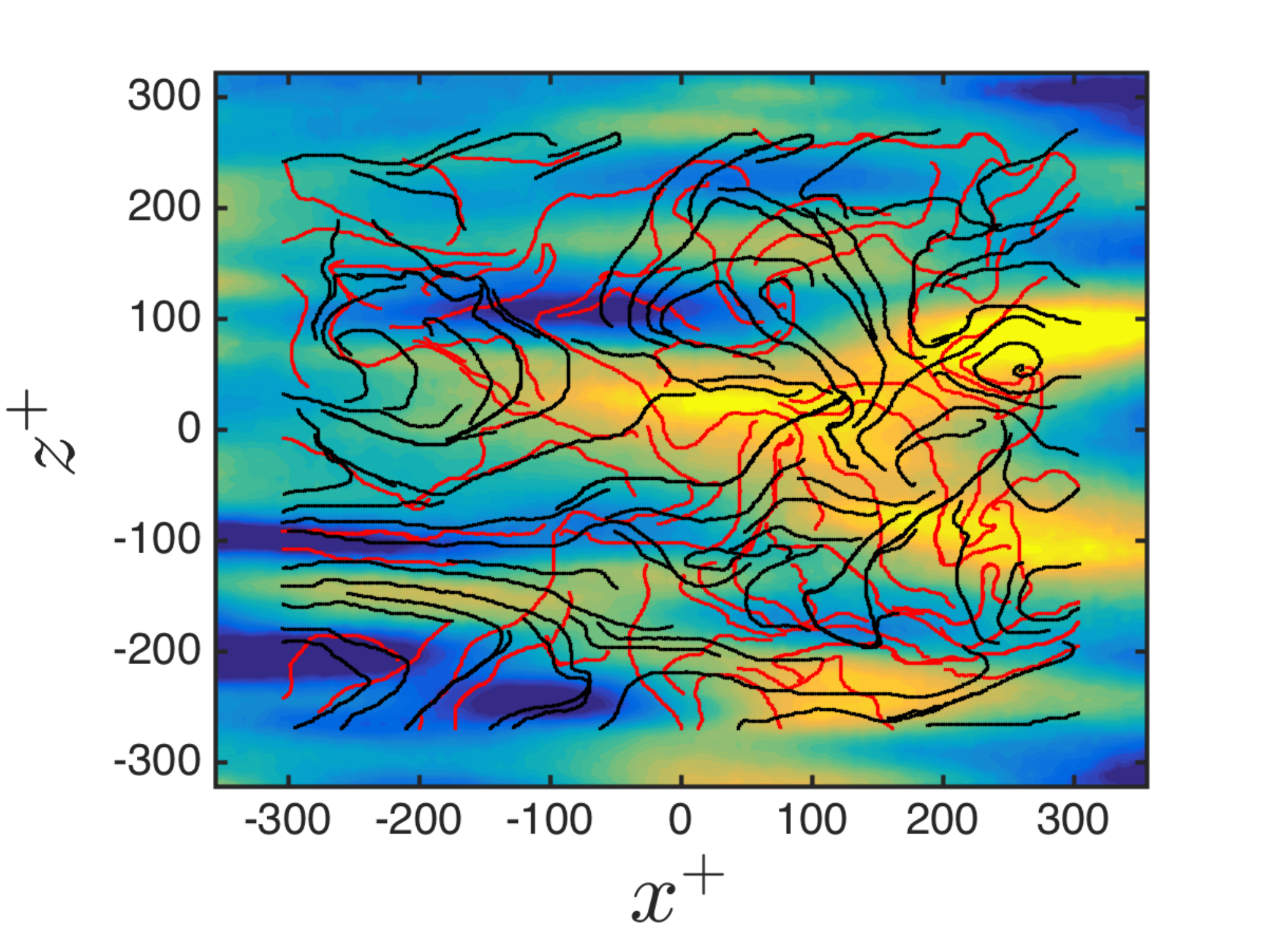} } 
\subfigure[$P_{U_3}+LCS$]{ \label{c} \includegraphics[
width=0.48\textwidth]{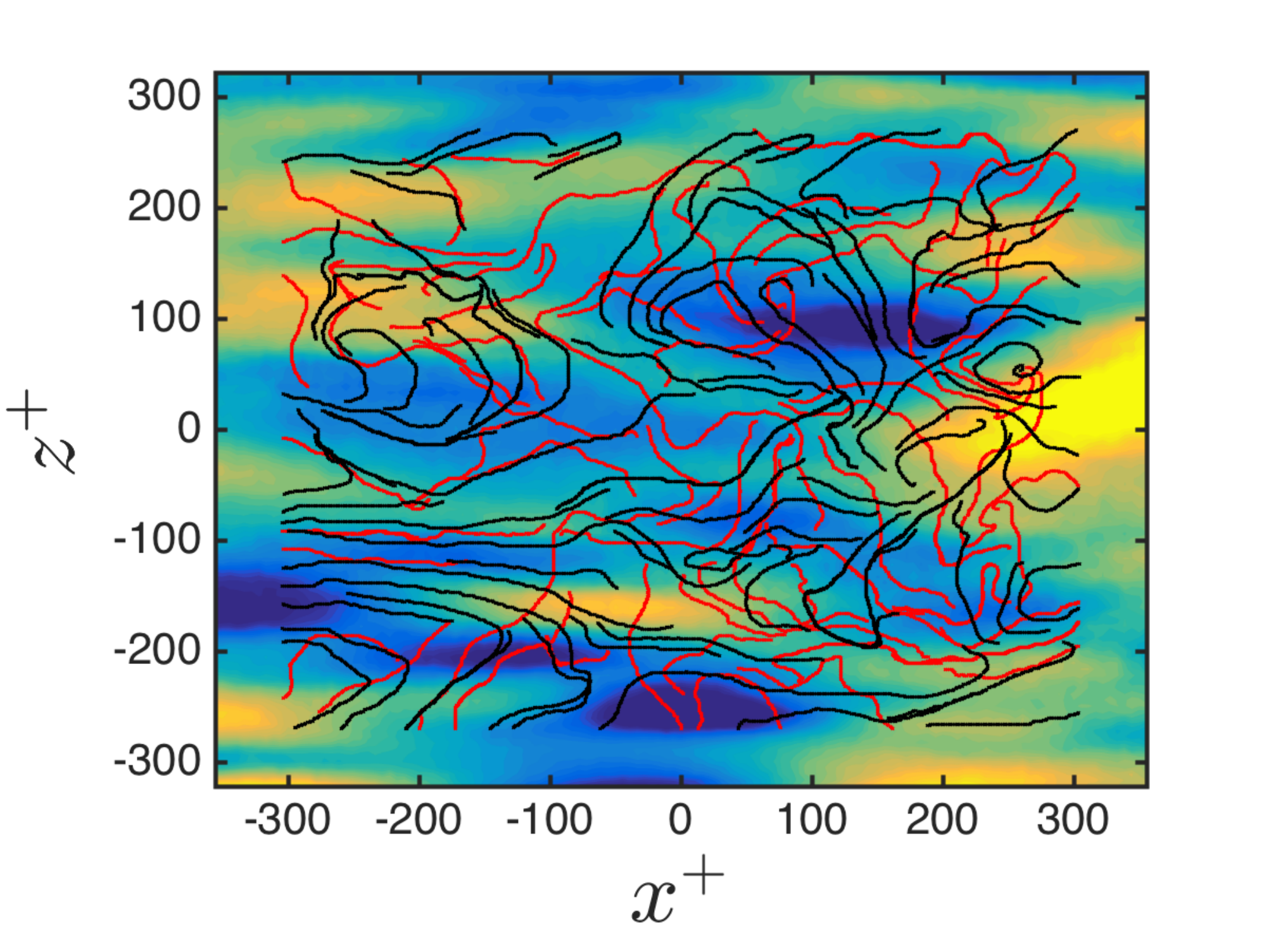} } 
\subfigure[$P_{U_4}+LCS$]{ \label{d} \includegraphics[
width=0.48\textwidth]{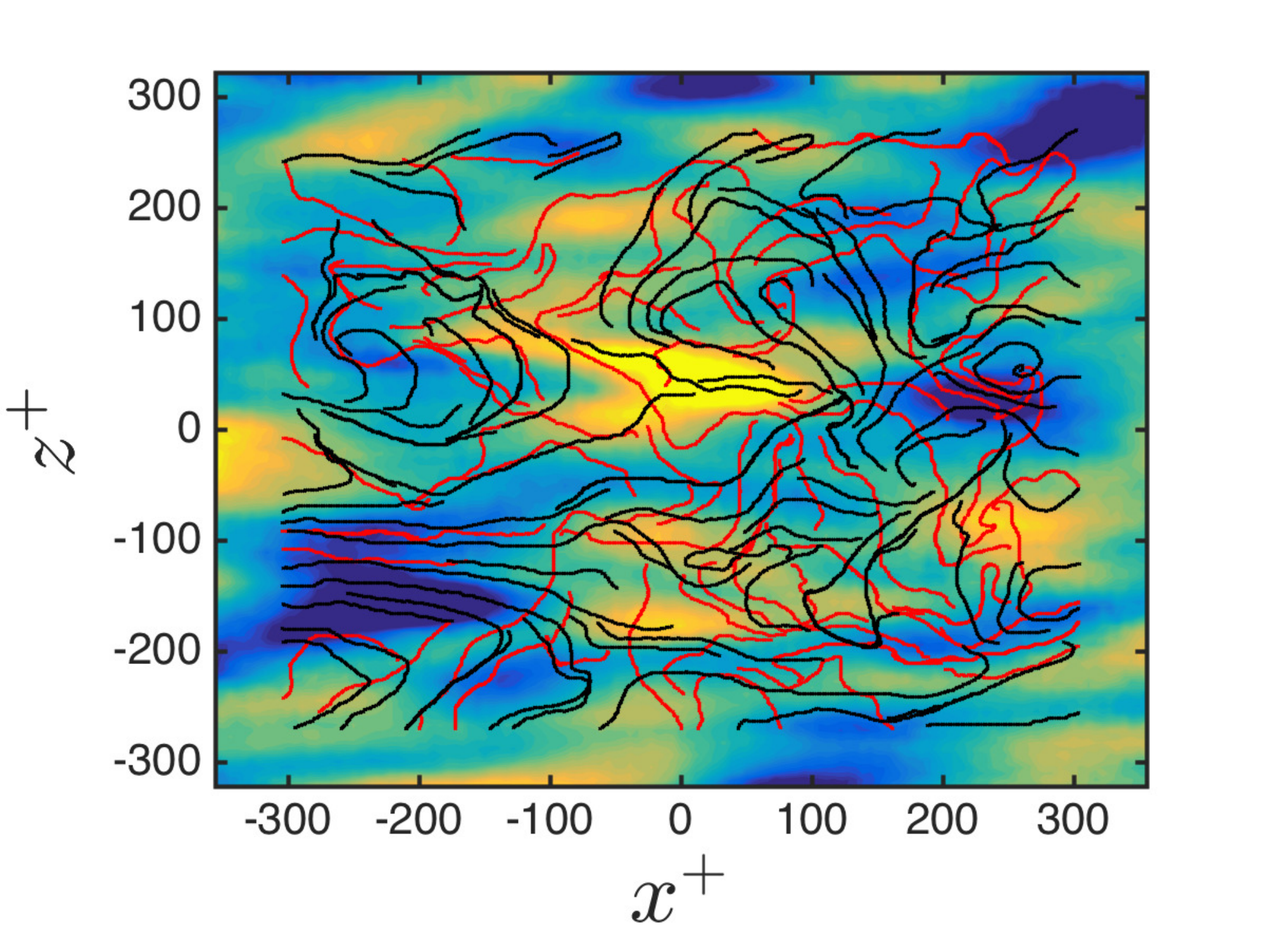} }
\subfigure[$P_{U_5}+LCS$]{ \label{e} \includegraphics[
width=0.48\textwidth]{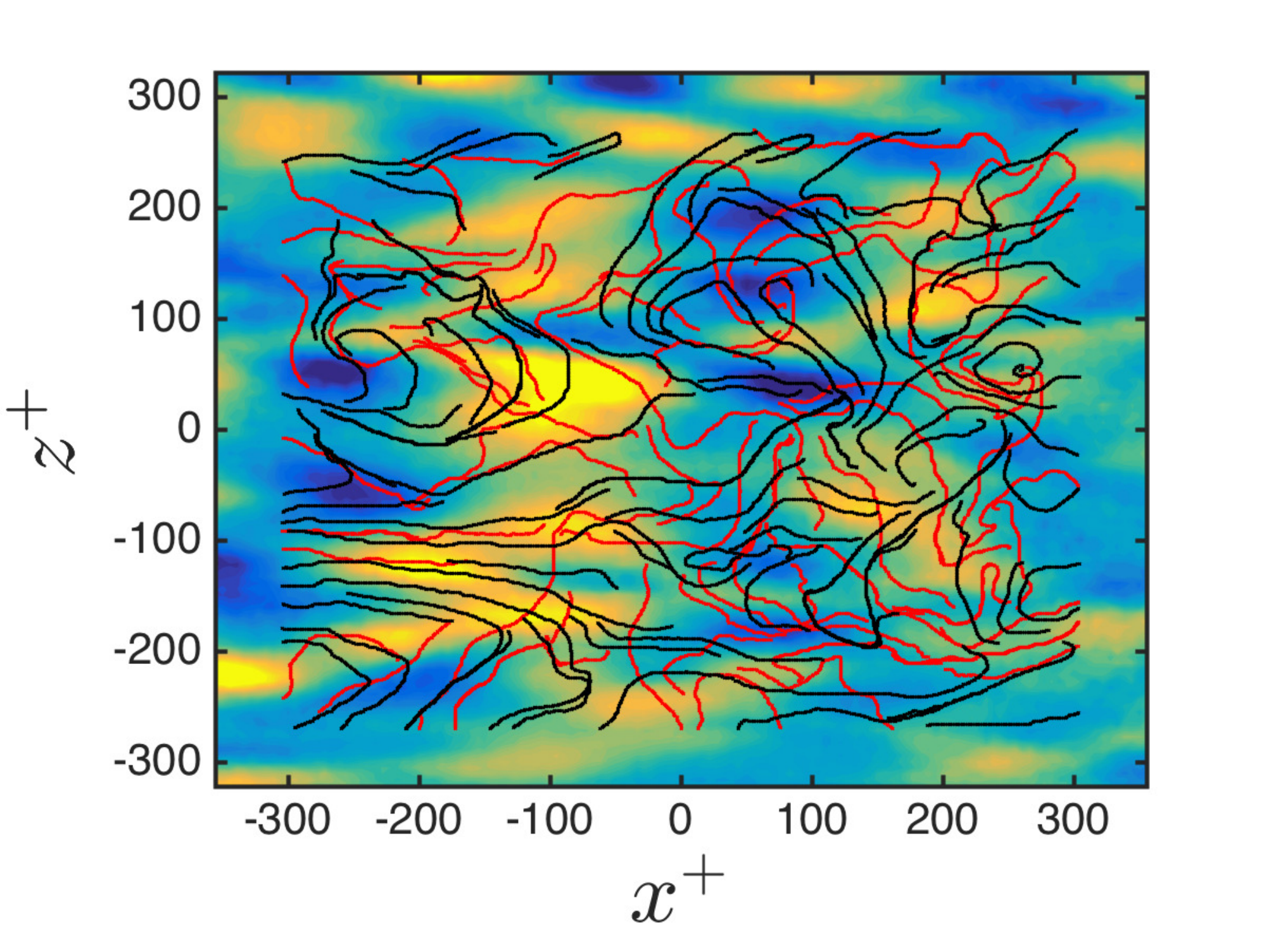} } 
\subfigure[$P_{U_{20}}+LCS$]{ \label{f} \includegraphics[
width=0.48\textwidth]{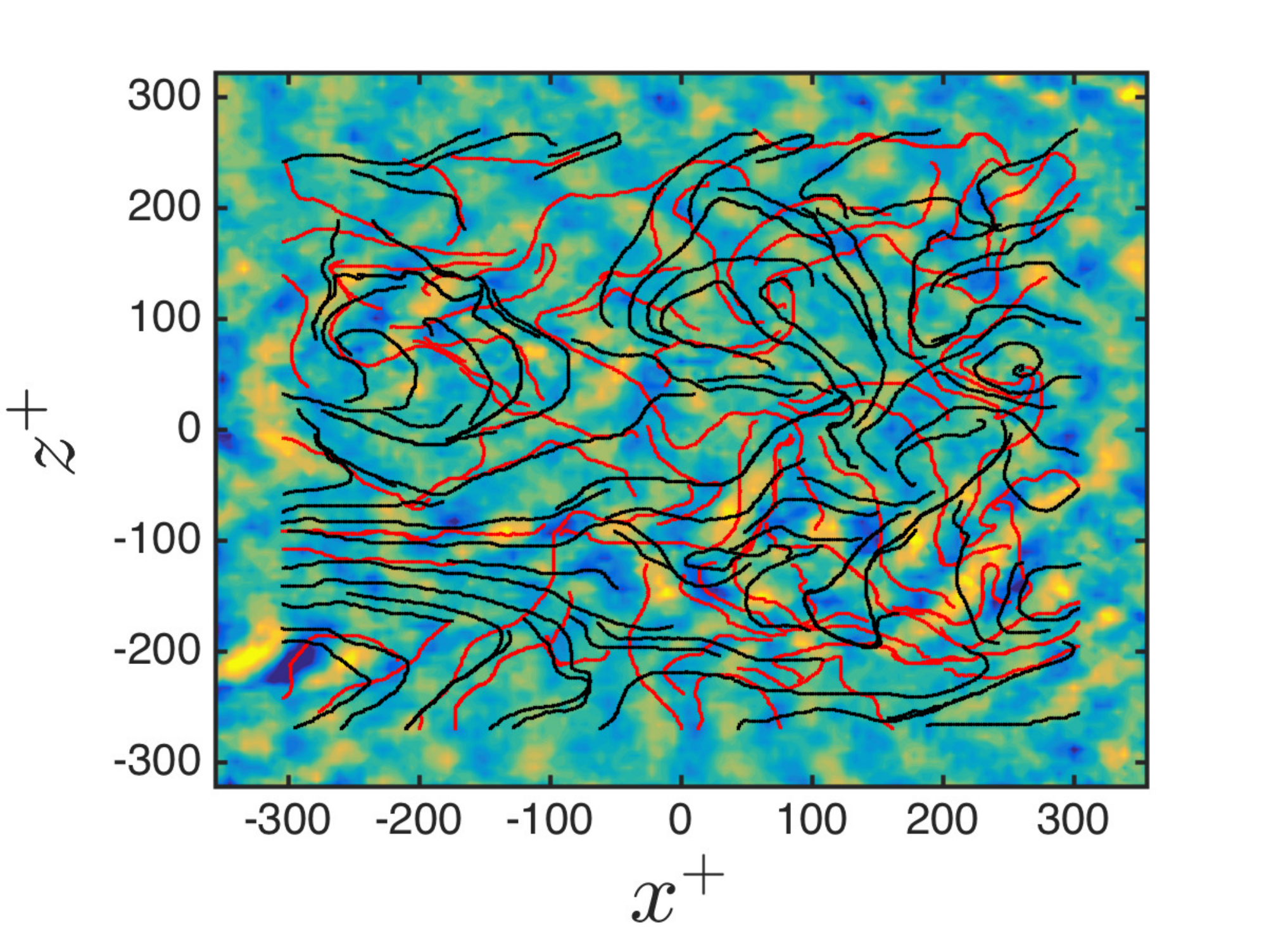} }
\end{center}
\caption{\label{}Repelling (black lines) and attracting (red lines) LCSs with: (a) the first POD mode; (b) the second POD mode; (c) the third POD mode; (d) the fourth POD mode; (e) the fifth POD mode; (f) the twentieth POD mode.}
\end{figure}

\begin{figure}
\begin{center}
\subfigure[$D_{U_1}+ LCS$]{ \label{a} \includegraphics[
width=0.48\textwidth]{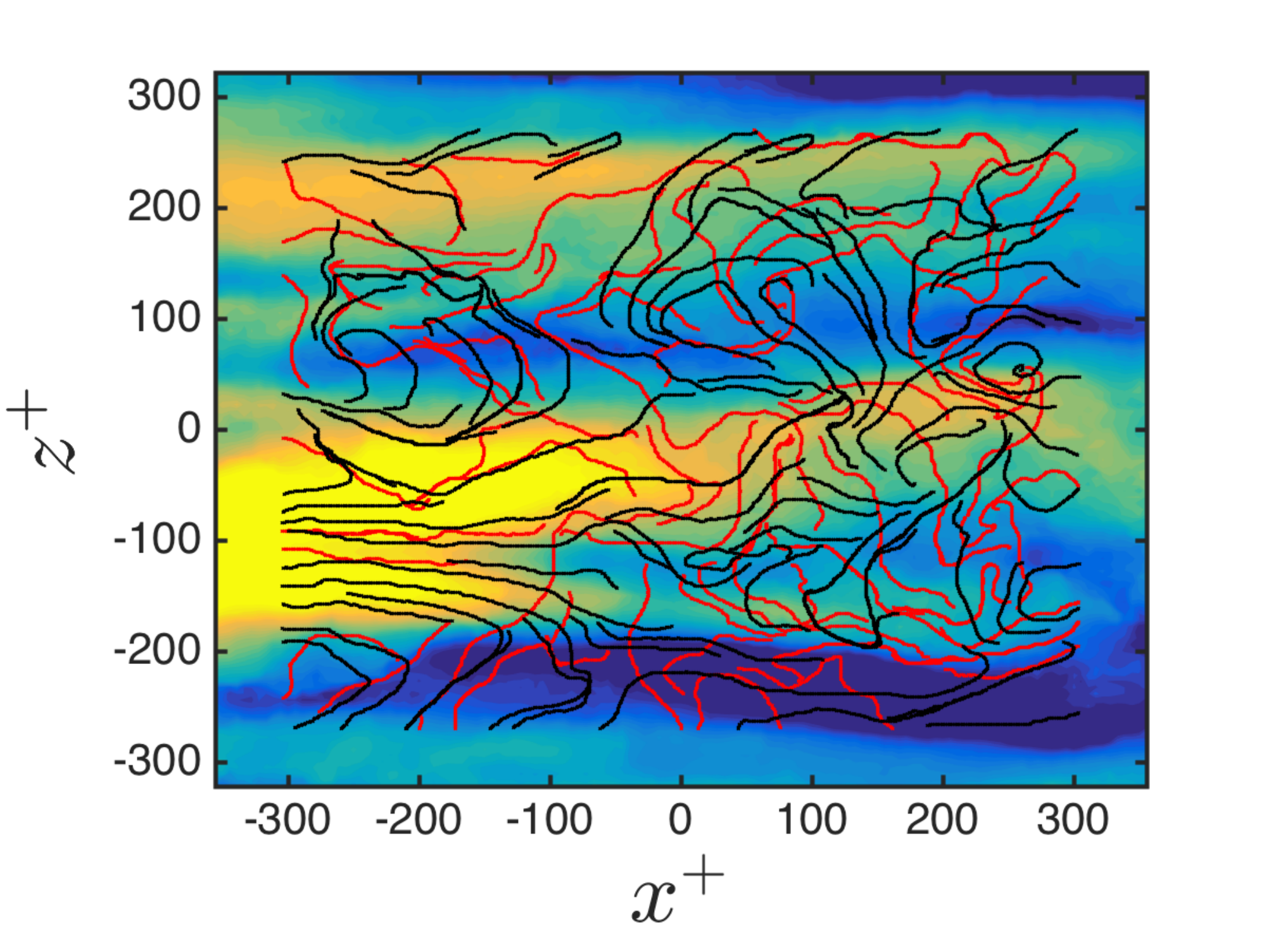} }
\subfigure[$D_{U_1}+LCS$]{ \label{b} \includegraphics[
width=0.48\textwidth]{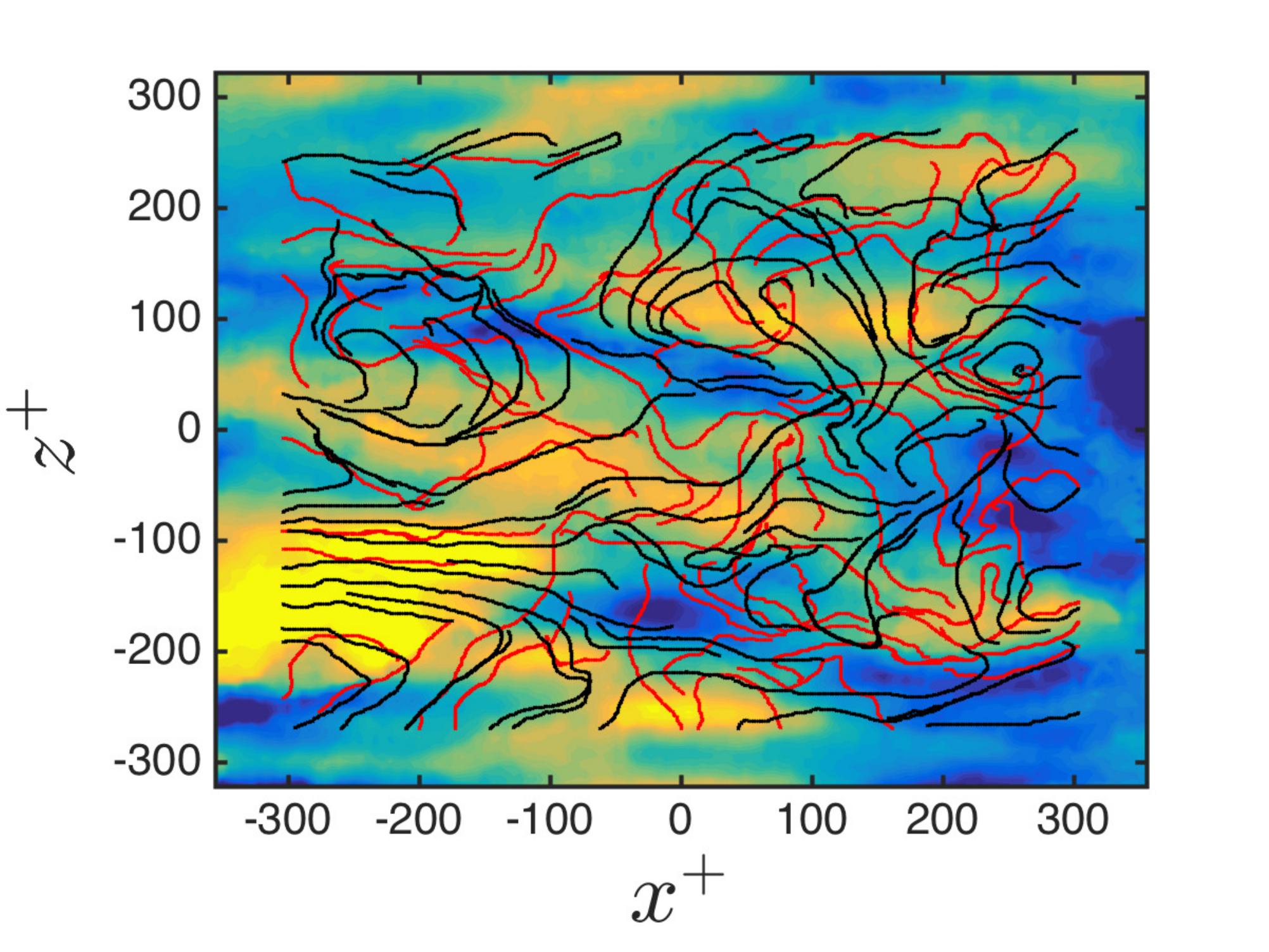} } 
\subfigure[$D_{U_1}+LCS$]{ \label{c} \includegraphics[
width=0.48\textwidth]{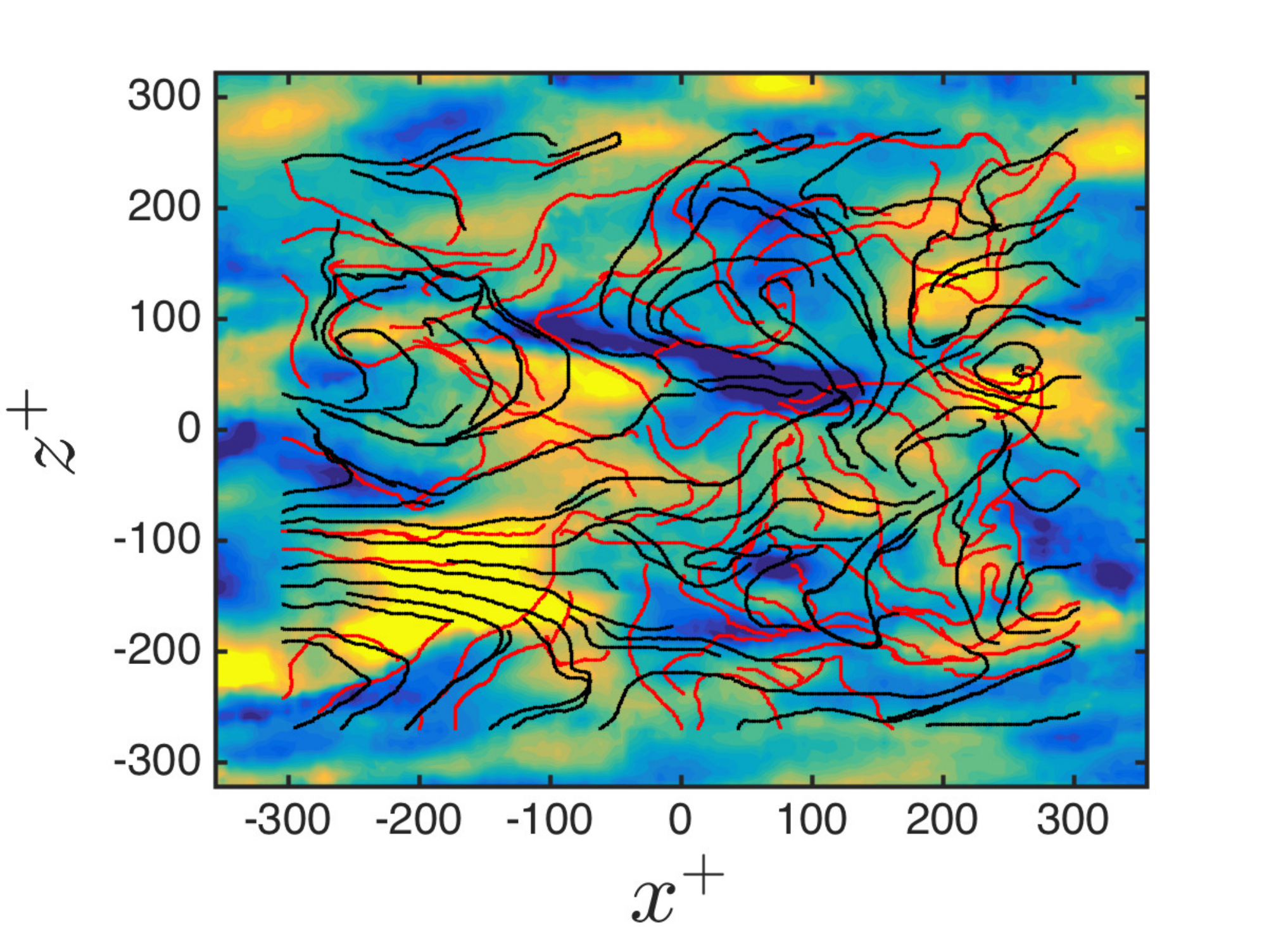} } 
\subfigure[$D_{U_1}+LCS$]{ \label{d} \includegraphics[
width=0.48\textwidth]{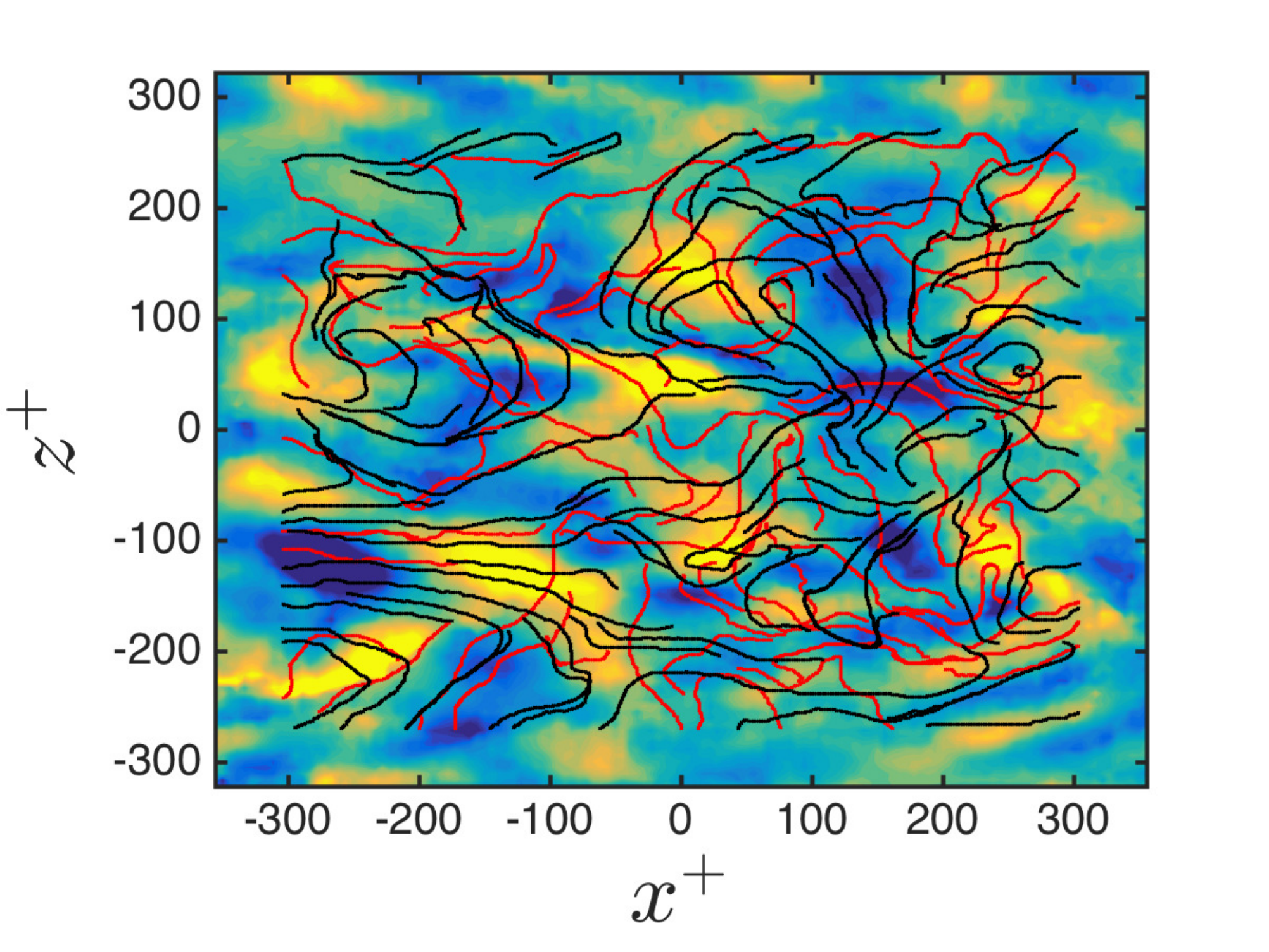} }
\subfigure[$D_{U_1}+LCS$]{ \label{e} \includegraphics[
width=0.48\textwidth]{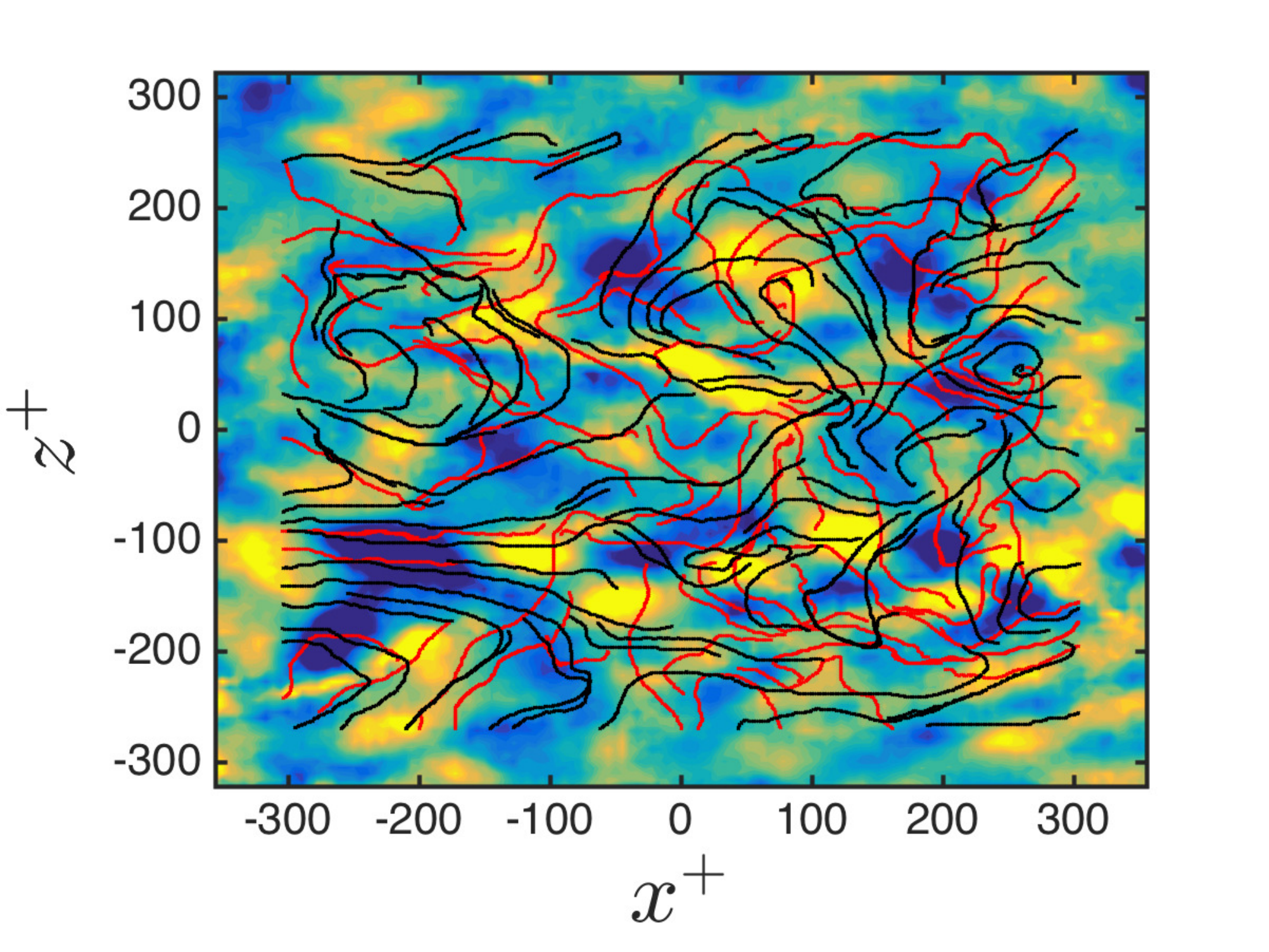} } 
\subfigure[$D_{U_{20}}+LCS$]{ \label{f} \includegraphics[
width=0.48\textwidth]{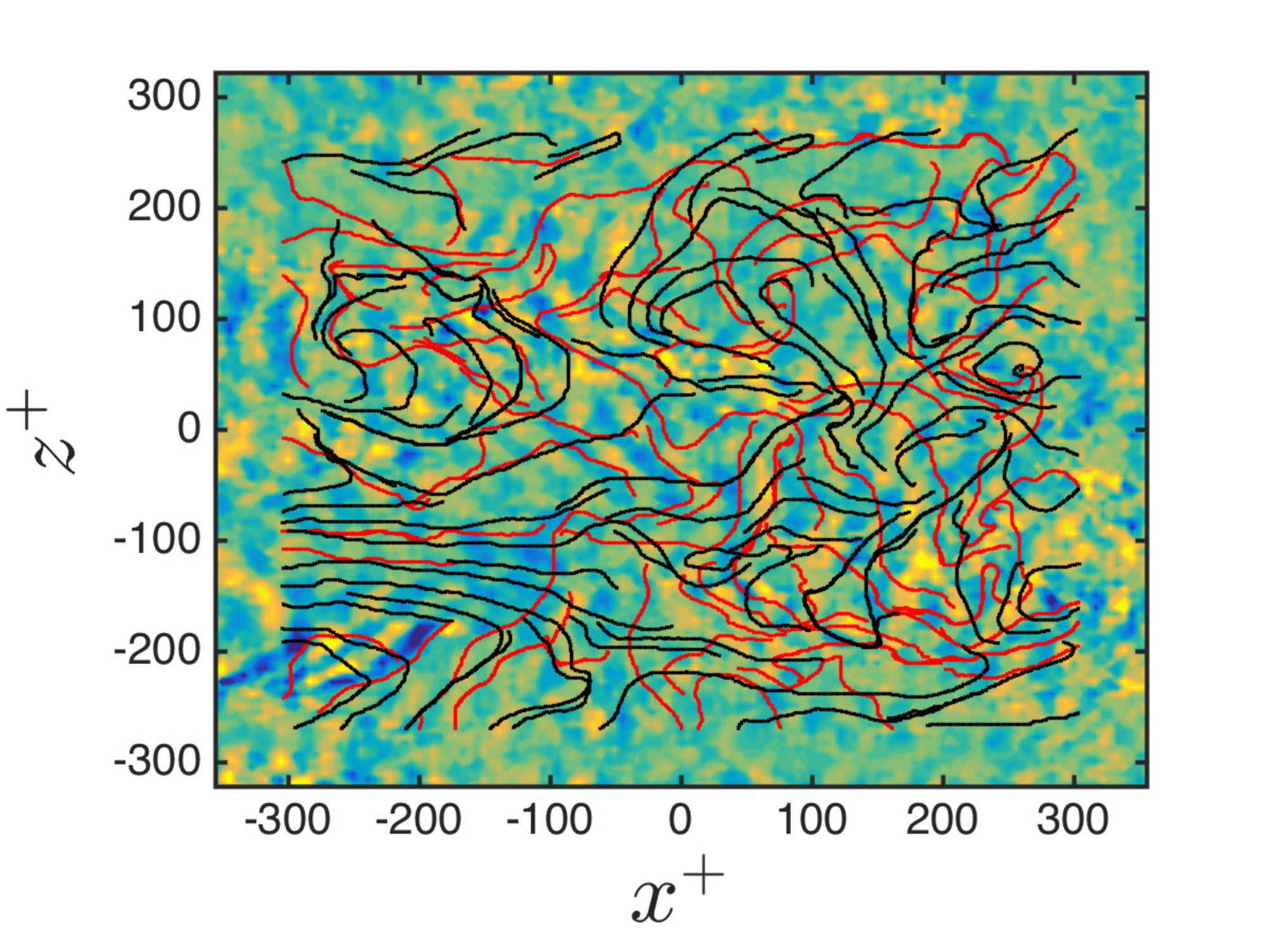} }
%\\
\end{center}
\caption{\label{}Repelling (black lines) and attracting (red lines) LCSs with: (a) the first DMD mode; (b) the second DMD mode; (c) the third DMD mode; (d) the fourth DMD mode; (e) the fifth DMD mode; (f) the twentieth DMD mode.}
\end{figure}

\subsection{\label{Results4} Lagrangian Coherent Structures of Reconstructed Fields}

The numerical computations of Lagrangian coherent structures is rather challenging due to numerical error that can lead to inaccurate hyperbolic LCSs. The numerical error is particularly sensitive in the computation of Cauchy Green eigenvectors especially near hyperbolic LCSs and near degenerate points as well as discontinuity of the eigenvector orientations. Numerical error is remedied by increasing the accuracy of the finite difference calculation. Adding auxiliary grids of four nodes to the each nodes in the original grid increases the accuracy significantly. The discontinuity of the eigenvector orientation can be identified when acquiring the reversing sign during the comparison between the eigenvectors at consequent integration time steps. This problem can be addressed by following the sign reversal of the eigenvectors \citep{FH2}. Turbulent boundary layers have a wide range of scales and it is therefore difficult  to avoid the `noise' associated with the small scales. This may indeed cause failure or underestimate the repelling and attracting LCSs. This problem is inevitable even with increased mesh resolution since the latter may be located on the same side of the LCS and experience less stretching than in the main mesh. 

Since the LCS are aligned with maxima and minima within the POD and DMD low mode numbers as shown in section \ref{Results3}, POD reconstructions are employed to obtain the fluctuating velocities associated with that particular mode and therefore compute the LCSs corresponding to the mode. Consequently, POD is an efficient filter commonly applied to exclude the small, otherwise incoherent, structures through the reconstruction technique.

\begin{figure}
\begin{center}
\subfigure[$P_{U_1}+LCS_{t^{+}=1.29}$]{ \label{a} \includegraphics[
width=0.48\textwidth]{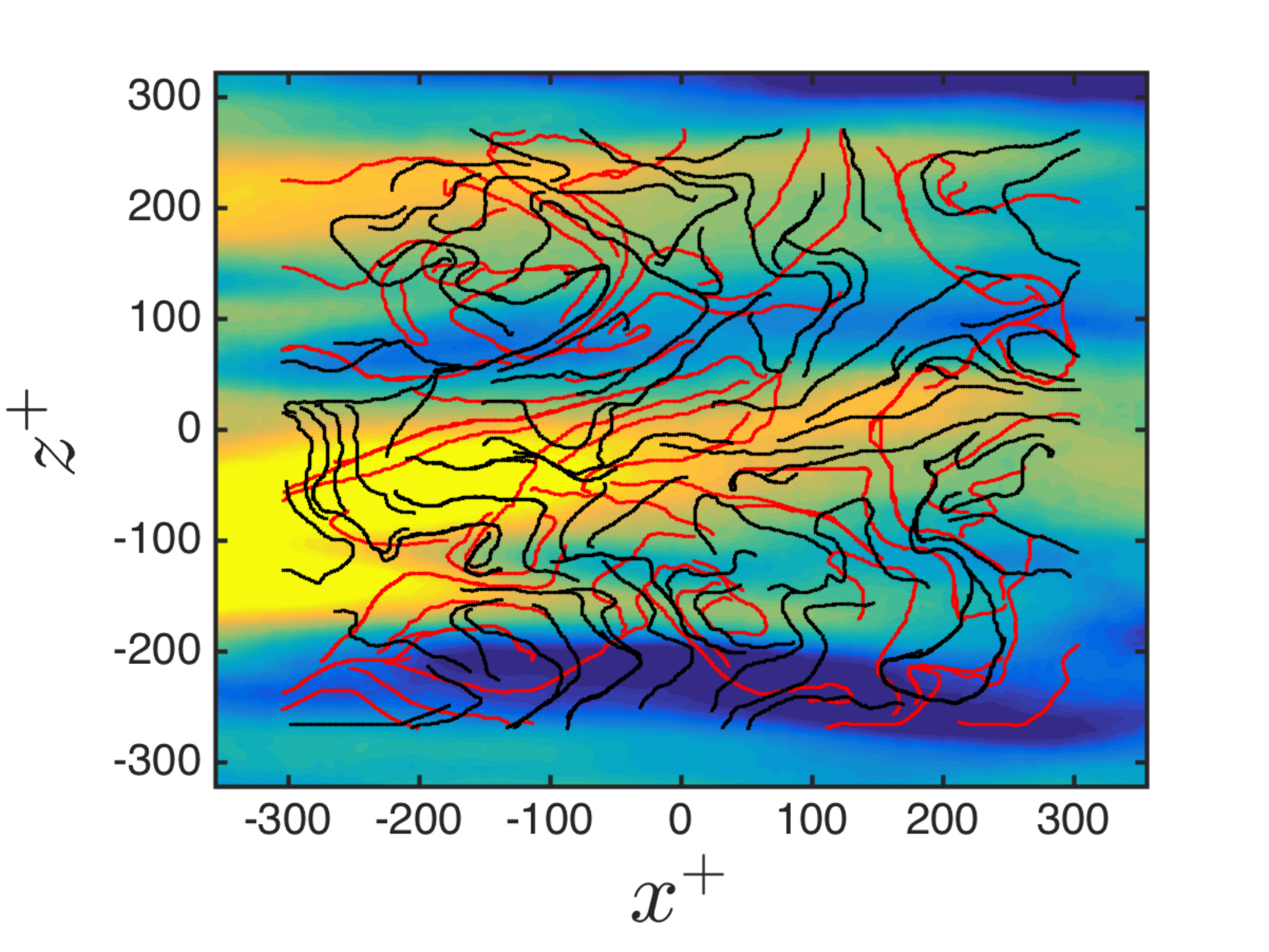} }
\subfigure[$P_{U_1}+LCS_{t^{+}=18}$]{ \label{b} \includegraphics[
width=0.48\textwidth]{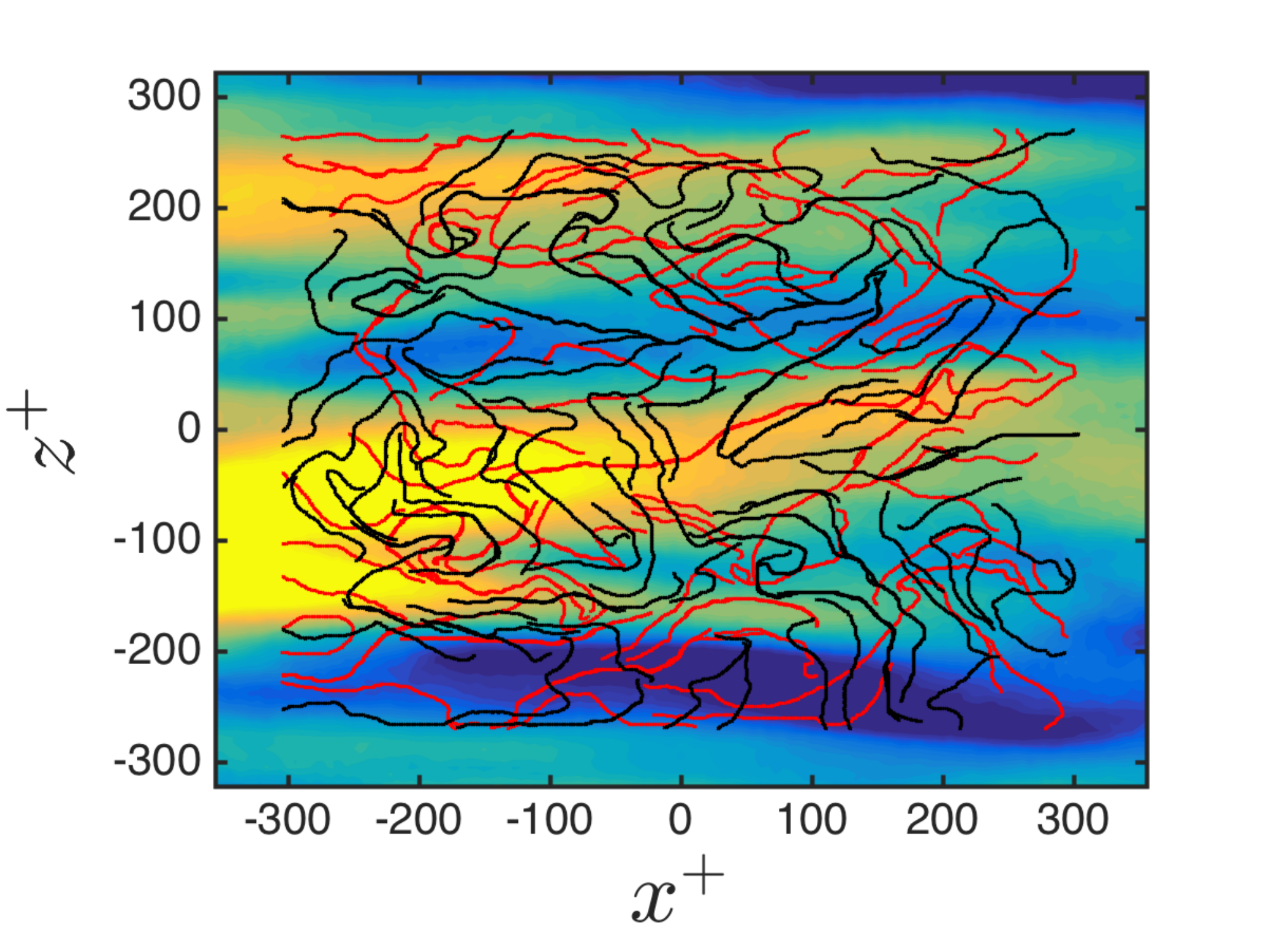} } 
\subfigure[$P_{U_1}+LCS_{t^{+}=37.5}$]{ \label{c} \includegraphics[
width=0.48\textwidth]{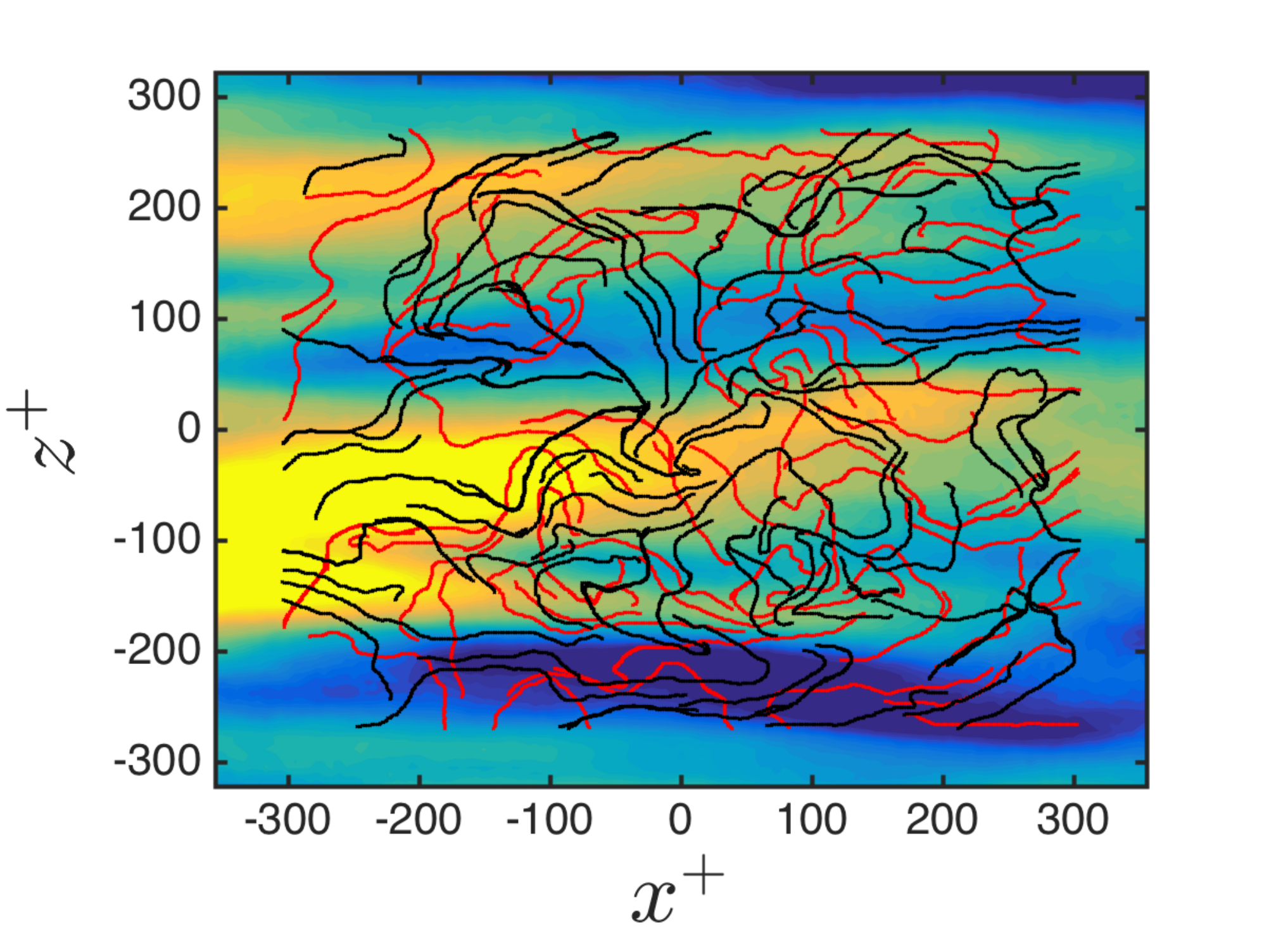} } 
\subfigure[$P_{U_1}+LCS_{t^{+}=49}$]{ \label{d} \includegraphics[
width=0.48\textwidth]{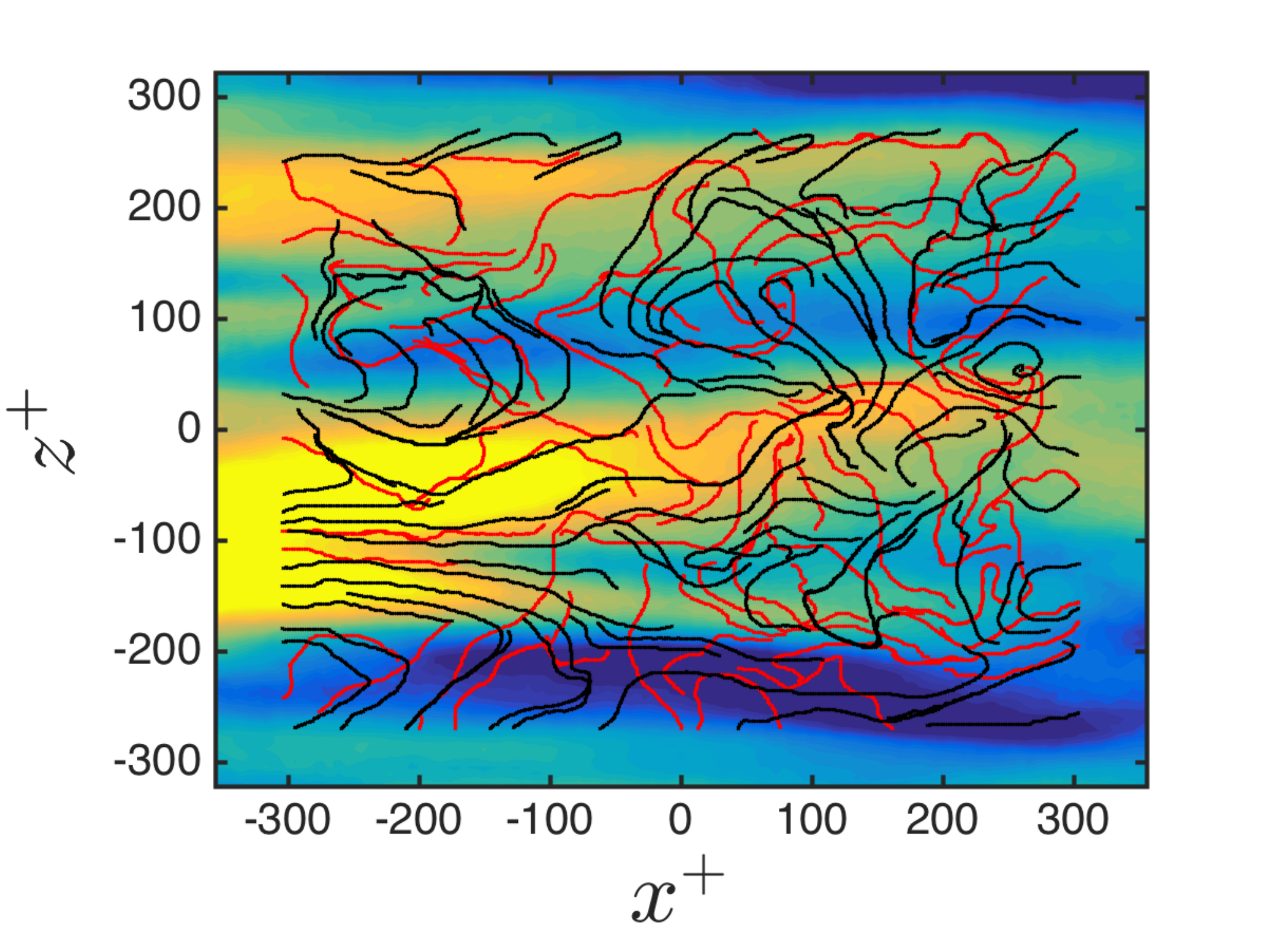} }
\end{center}
\caption{\label{} The first POD mode with LCSs at four different times: (a) $t^{+}=1.29$; (b) $t^{+}=18$; (c) $t^{+}=37.5$; (d) $t^{+}=49$.}
\end{figure}

\begin{figure}
\begin{center}
\subfigure[$LCS_{P_{U_1}}$]{ \label{a} \includegraphics[
width=0.478\textwidth]{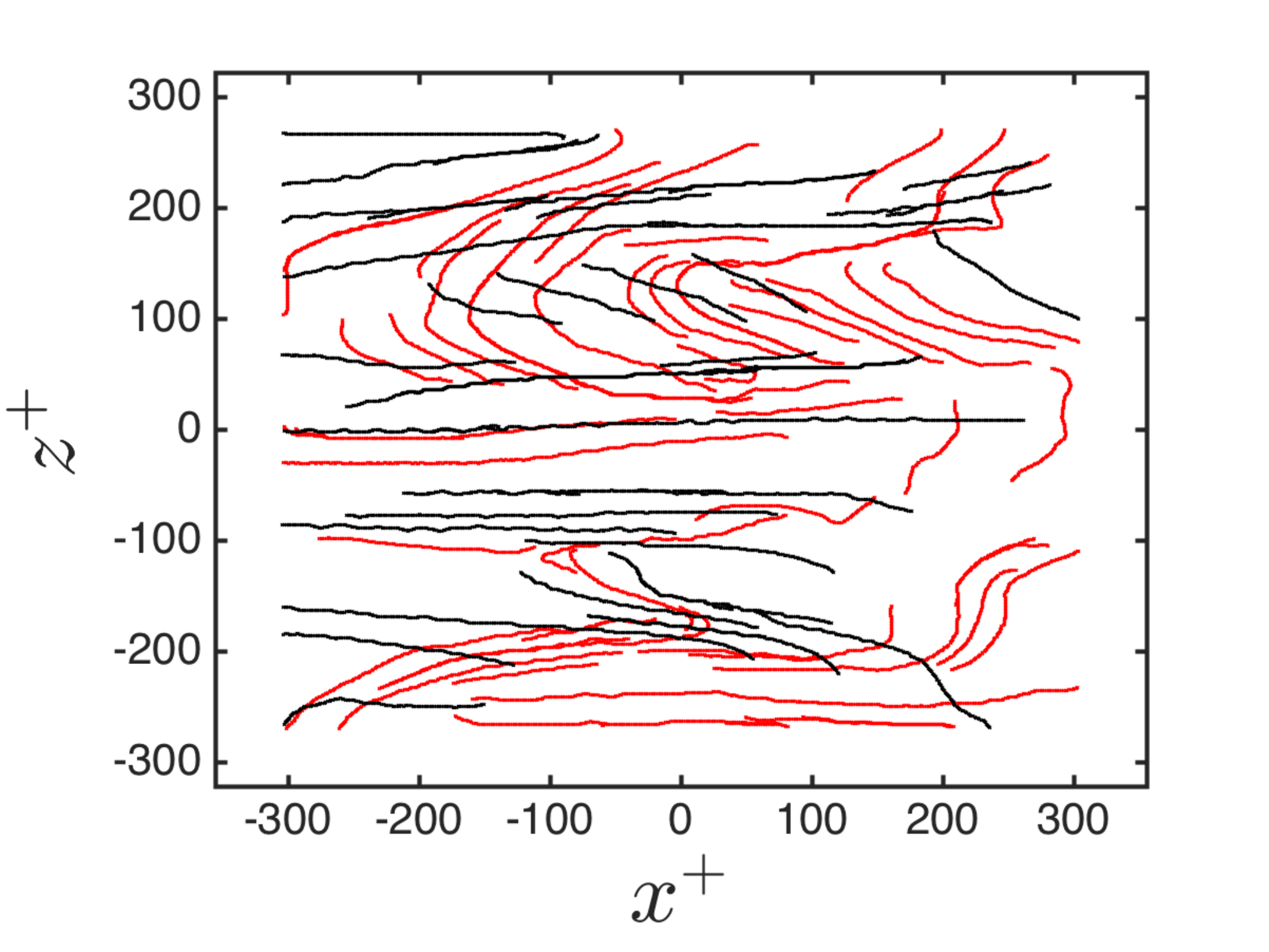} }
\subfigure[$LCS_{P_{U_2}}$]{ \label{b} \includegraphics[
width=0.478\textwidth]{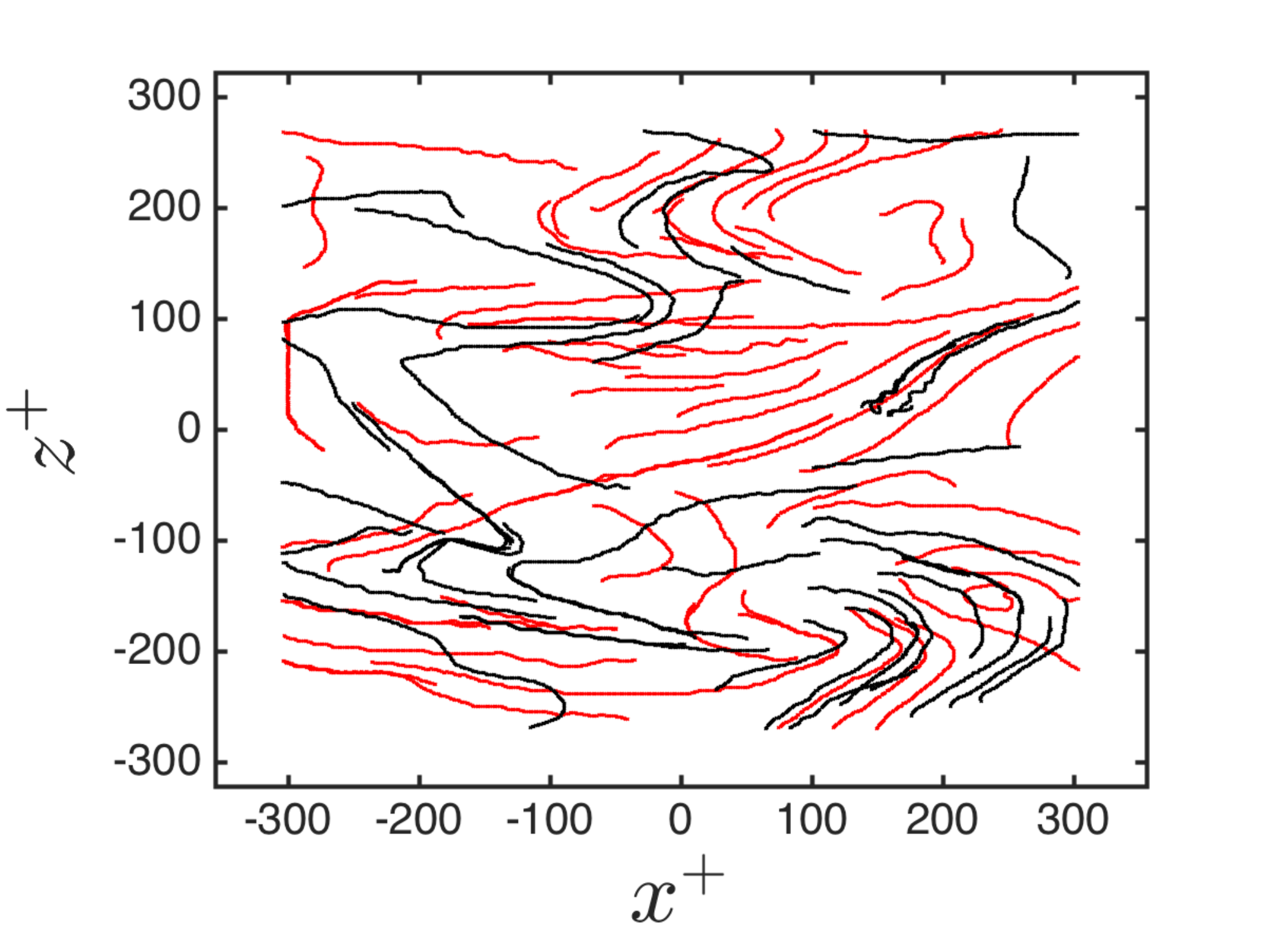} }
\subfigure[$LCS_{P_{U_3}}$]{ \label{c} \includegraphics[
width=0.478\textwidth]{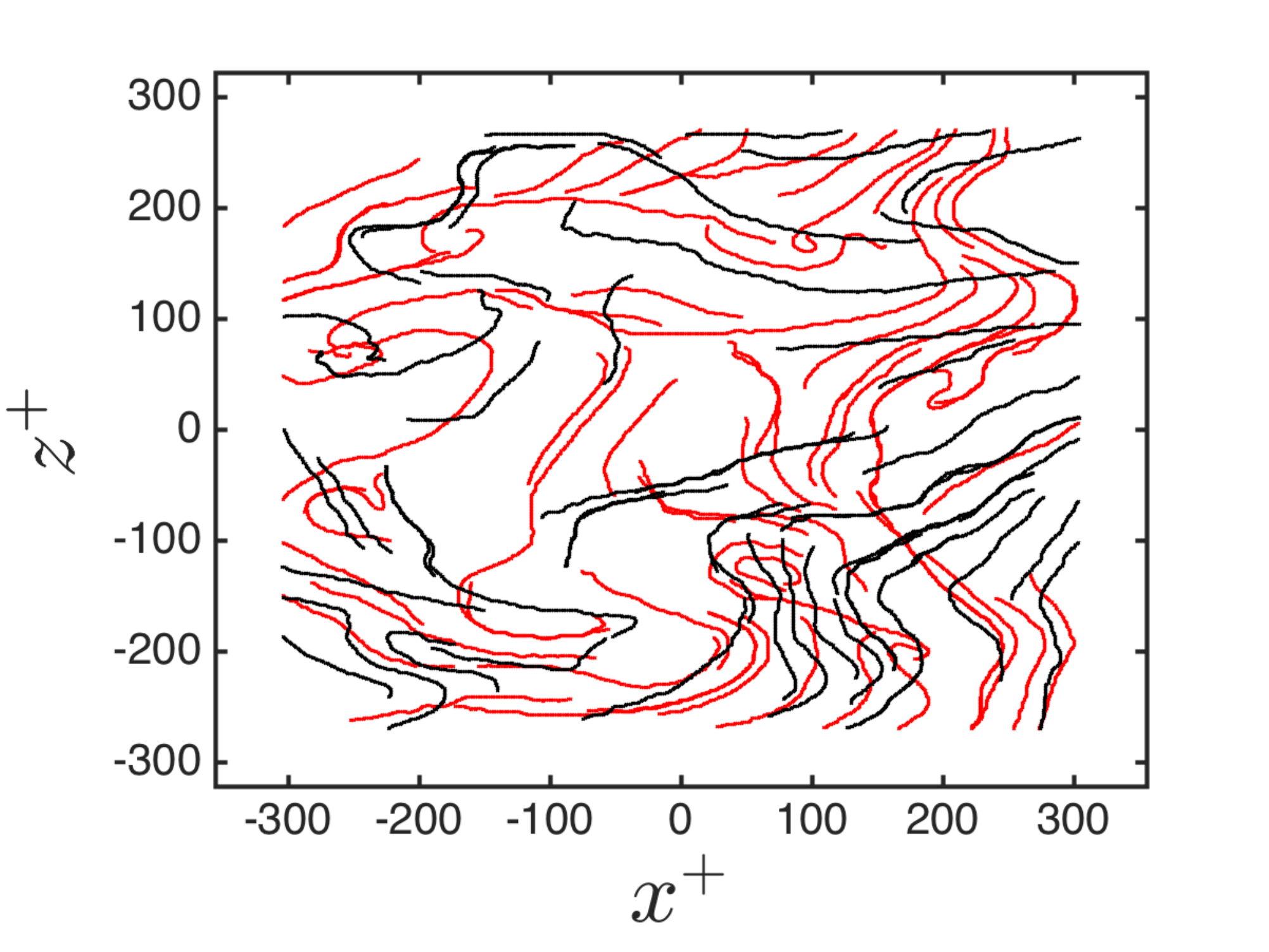} } 
\subfigure[$LCS_{P_{U_4}}$]{ \label{d} \includegraphics[
width=0.478\textwidth]{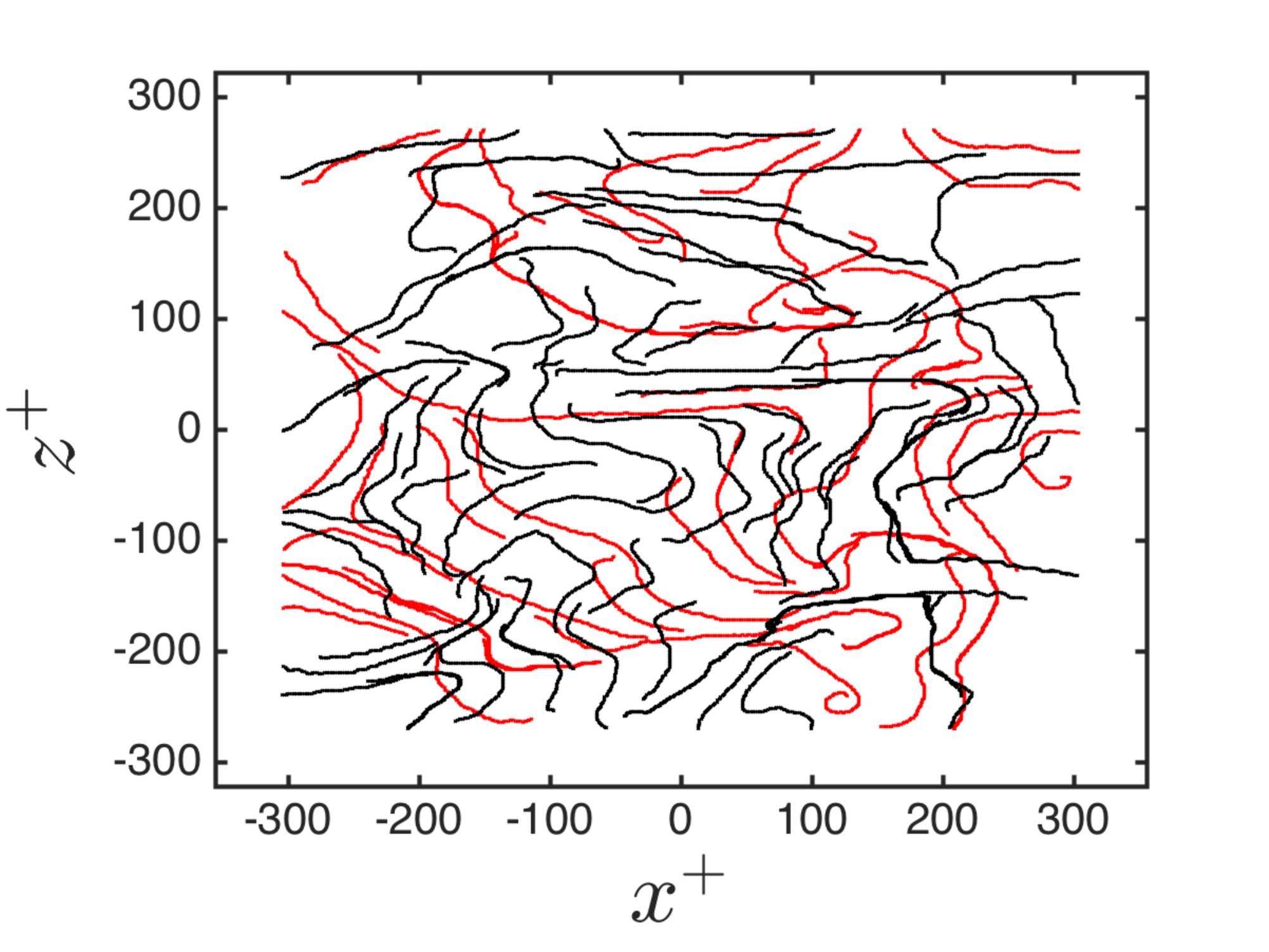} }
\subfigure[$LCS_{P_{U_5}}$]{ \label{e} \includegraphics[
width=0.478\textwidth]{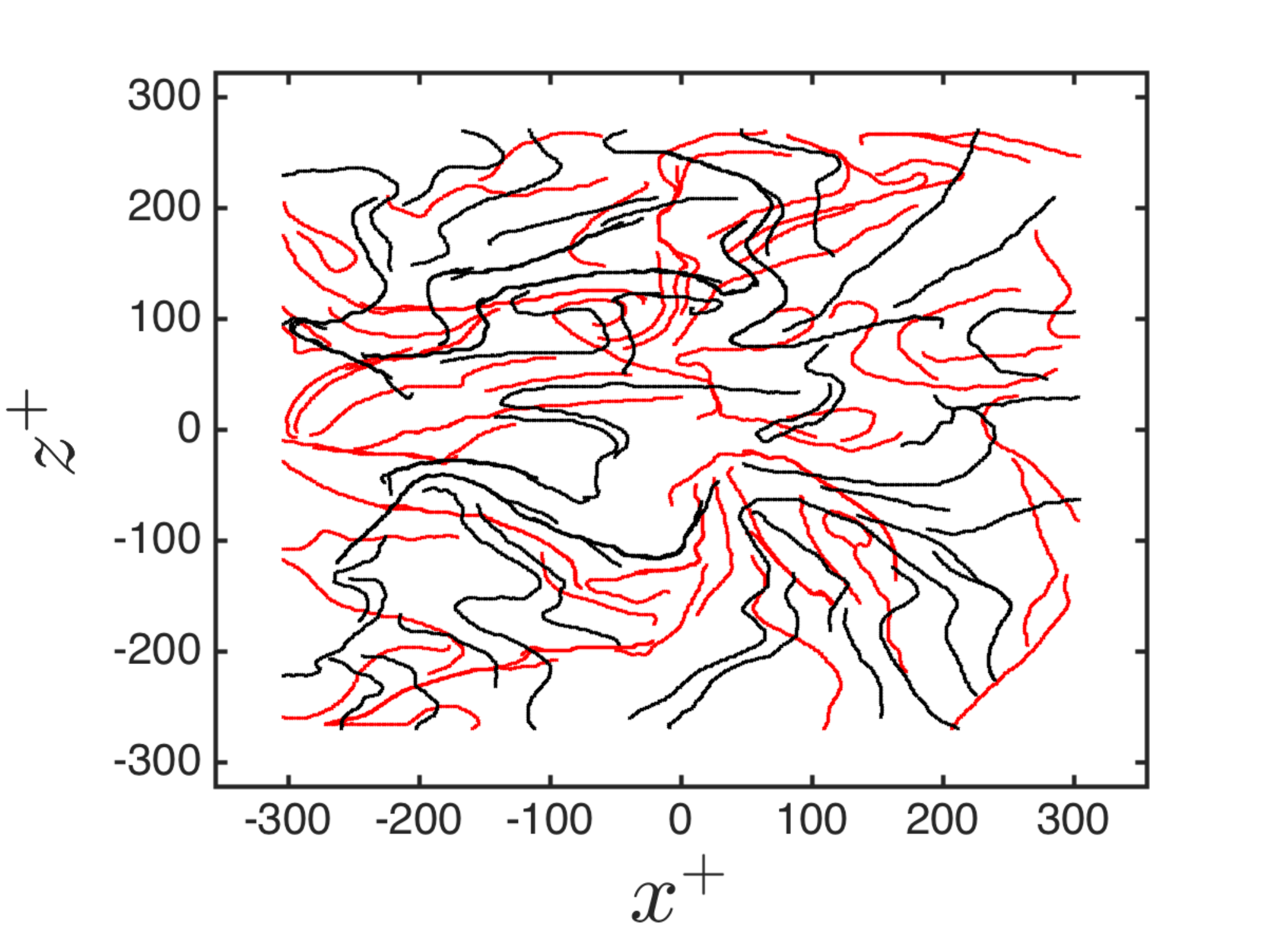} } 
\subfigure[$LCS_{P_{U_{20}}}$]{ \label{f} \includegraphics[
width=0.478\textwidth]{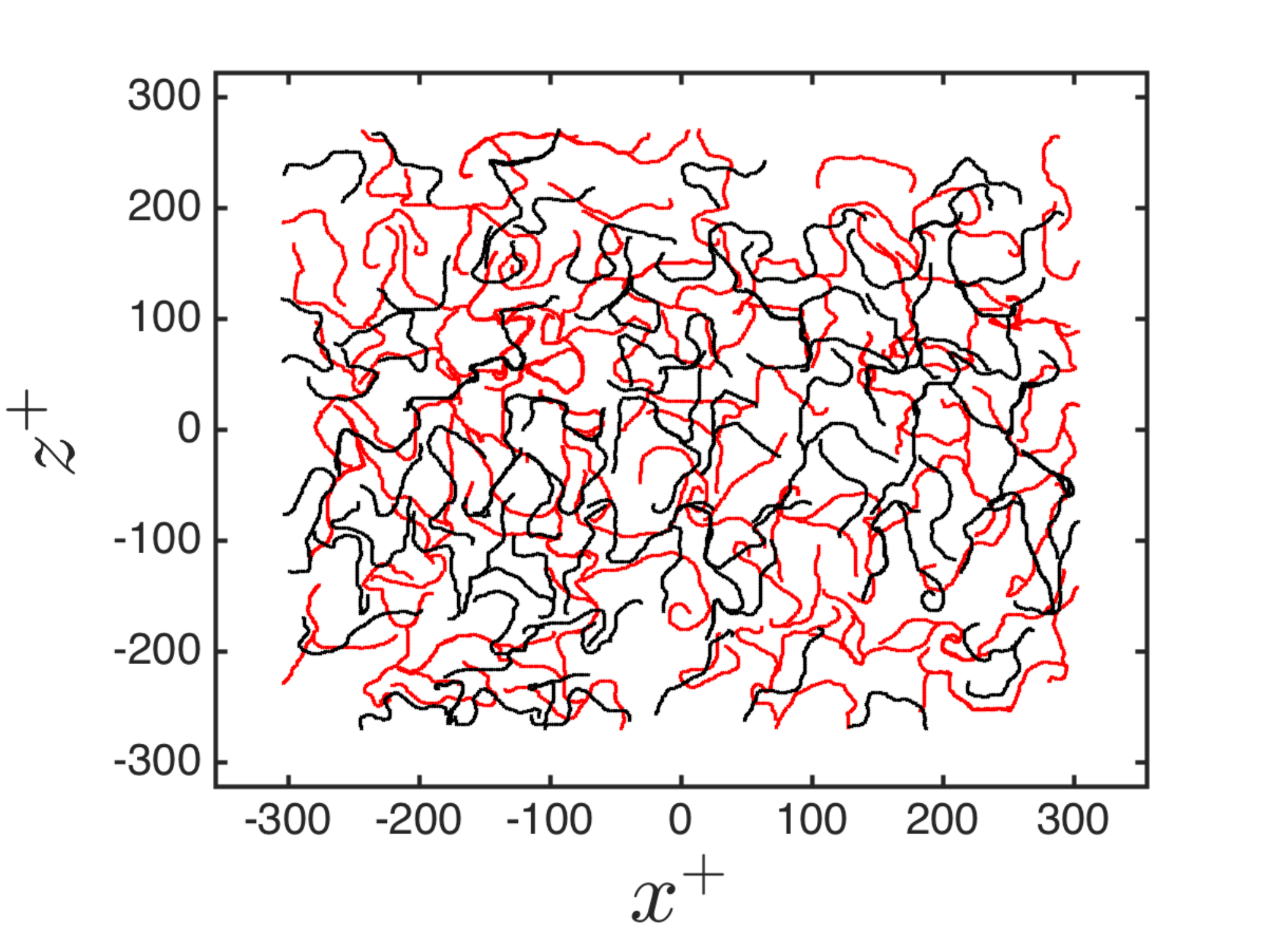} }
\end{center}
\caption{\label{}Reconstructed LCSs at $t^+ = 1.29$ from: (a) the first POD mode; (b) the second POD mode; (c) the third POD mode; (d) the fourth POD mode; (e) the fifth POD mode; (f) the twentieth POD mode. Red lines (attracting) and black lines (repelling)}
\end{figure}

Figure 10 shows the LCSs computed from a reconstructed velocity field using individual POD modes, namely one through five and twentieth at $t^+ = 1.29$. The features of the LCS are entirely different from the LCSs computed from the full field data. Figure 10(a) shows the LCSs from the first mode being long and smooth especially for the repelling trajectories. This behavior is expected since the first mode contains the most energy and therefore coherence. The attracting trajectories exhibit similar behavior especially in the region between $z^+$= 0 and -300. Above this region, the attracting trajectories show subsequent bending. Figure 10(b) represents the reconstructed LCSs from the second mode, where repelling LCSs show a moderate decrease in length while increasing in their bending and tangling. Attracting LCSs also decrease in length but less than repelling LCSs. The most bent trajectories are located at $x^{+} \approx 200$ and $z^{+} \approx \pm 200$. Figures 10(c-e) show reconstructed LCSs from the third through fifth modes, where the bending in the repelling LCSs is progressively pronounced as well as shortening in length as a function of increasing mode number. The structures are far more complex; a larger degree of incoherence than those observed in the first two modes is apparent. The attracting LCSs decrease in length and further bending occurs especially at $x^{+} < 0$. In figure 10(f), the reconstructed LCSs from the twentieth mode is characterized by a completely different behavior from previous modes 1 through 5, where the LCSs do not show a preference in direction and the size of the structures and can be thought largely homogeneous and isotropic.   

Figure 11 shows reconstructed LCSs from the first POD mode at four different times $t^{+}= 1.29, 18, 37.5$ and 49. Variation of attracting and repelling lines is minimal though the attracting structures are prone to further curvature, whereas repelling structures remain relatively constant in size and shape. The evolution of these structures can mostly be characterized as translation in space due to the convection. With time increasing, the shrinkage in LCS length is less pronounced that the full data LCSs, where LCSs maintain their length especially for the repelling trajectories. One can note the evolution of the LCSs in time for this particular mode is then captured while retaining the coherence within the domain.

\begin{figure}
\begin{center}
\subfigure[$LCS_{t^{+}= 1.29}$]{ \label{a} \includegraphics[
width=0.48\textwidth]{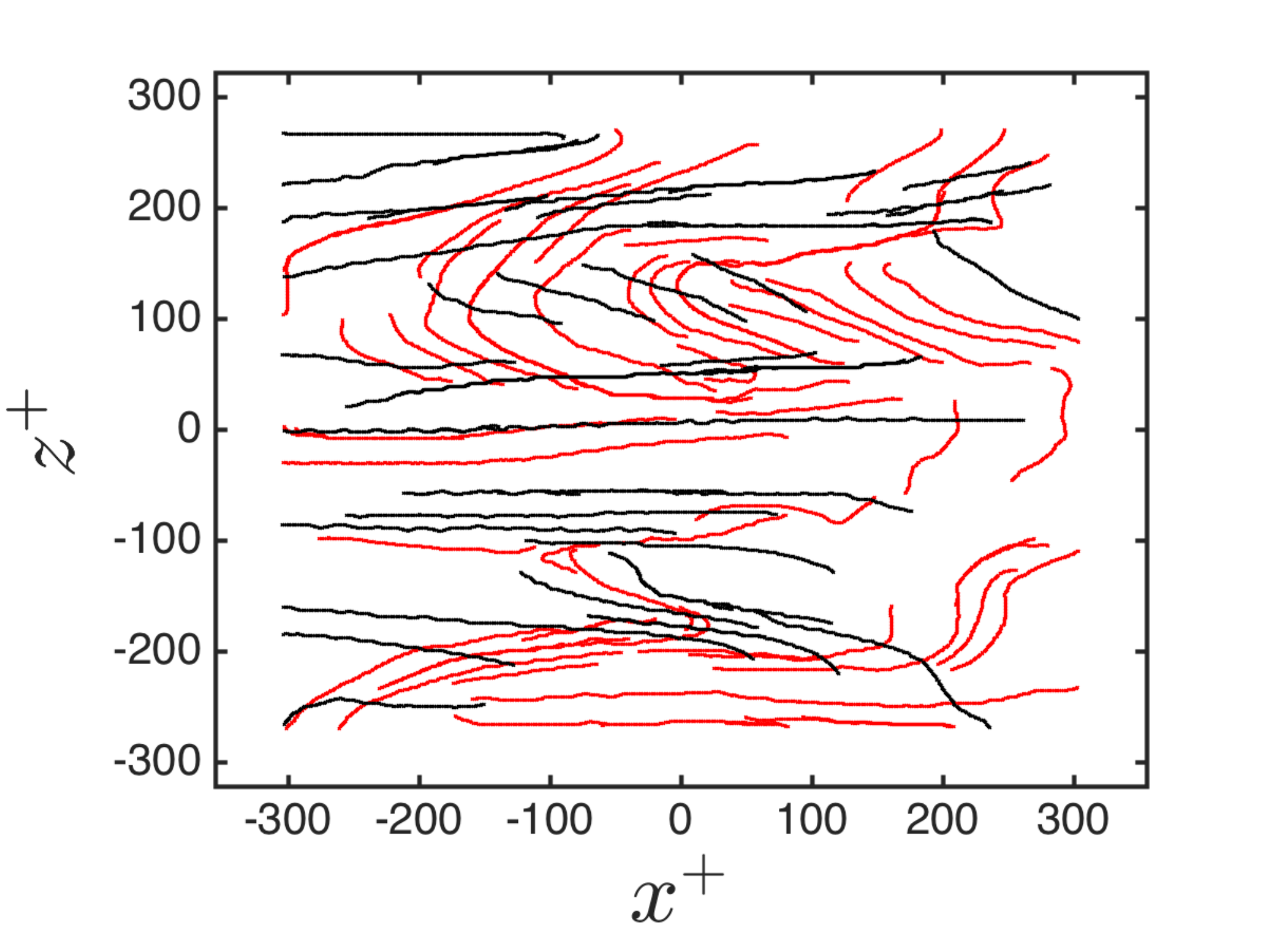} }
\subfigure[$LCS_{t^{+}= 18}$]{ \label{b} \includegraphics[
width=0.48\textwidth]{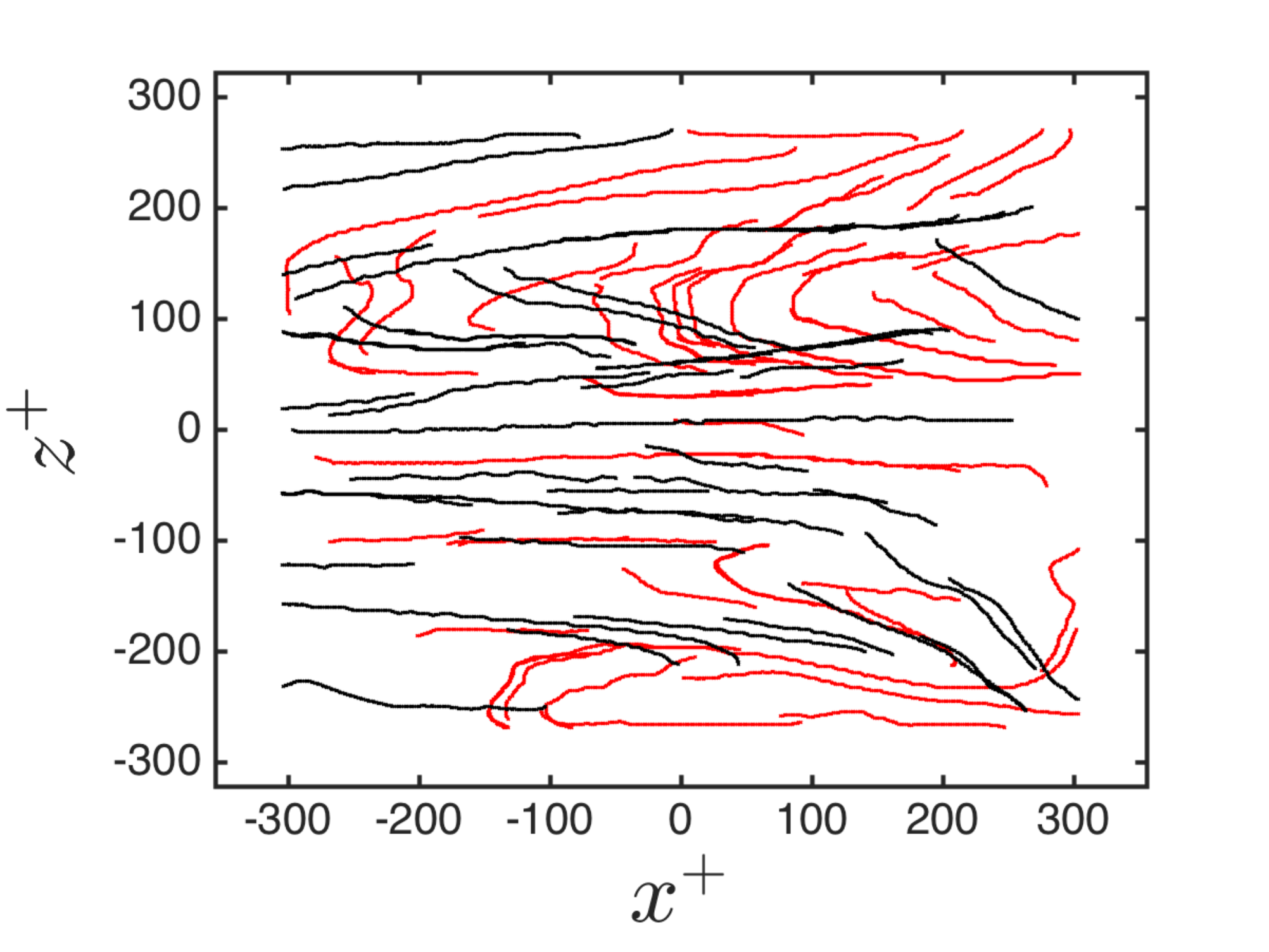} }
\subfigure[$LCS_{t^{+}= 37.5}$]{ \label{c} \includegraphics[
width=0.48\textwidth]{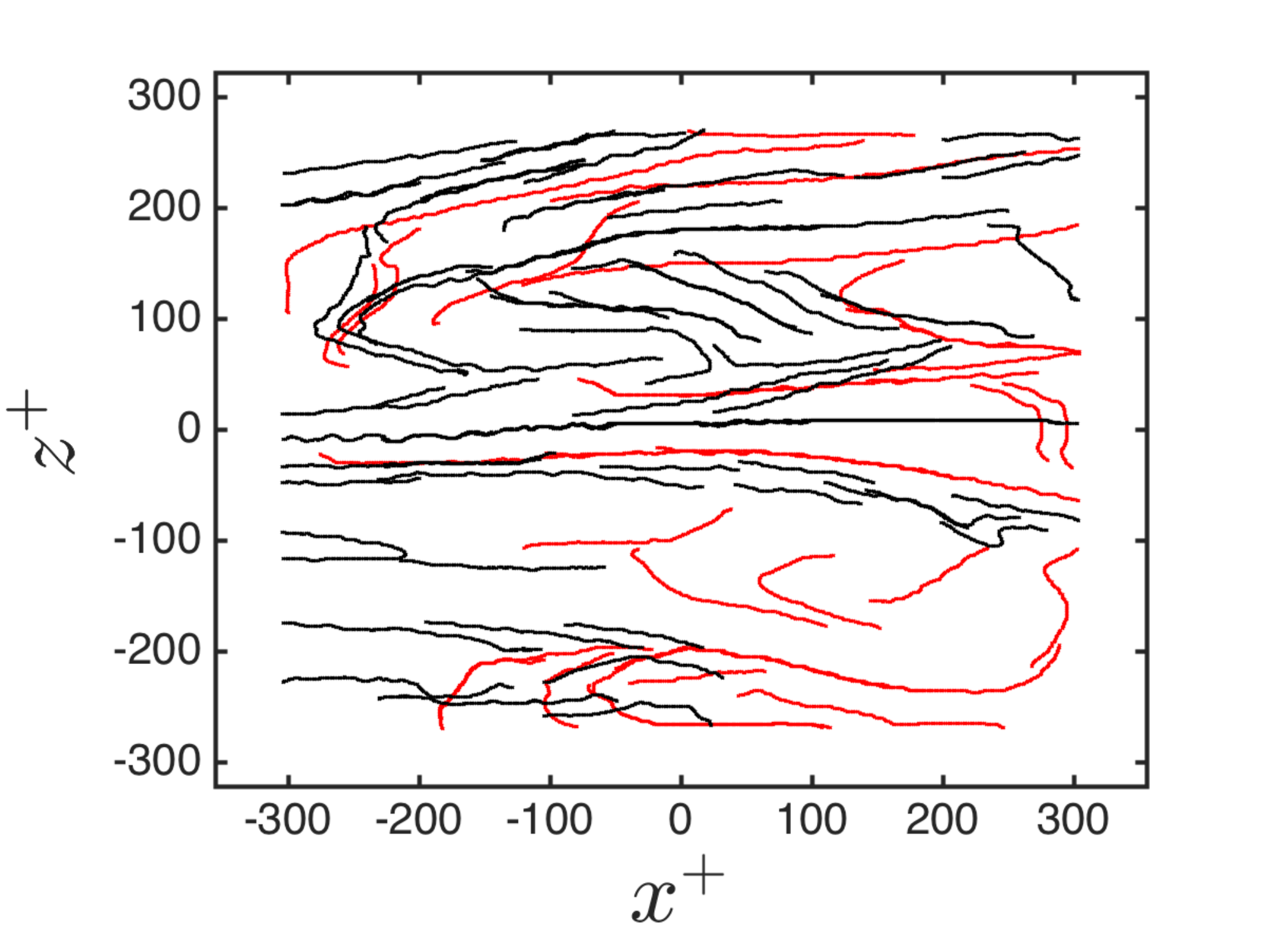} } 
\subfigure[$LCS_{t^{+}= 49}$]{ \label{d} \includegraphics[
width=0.48\textwidth]{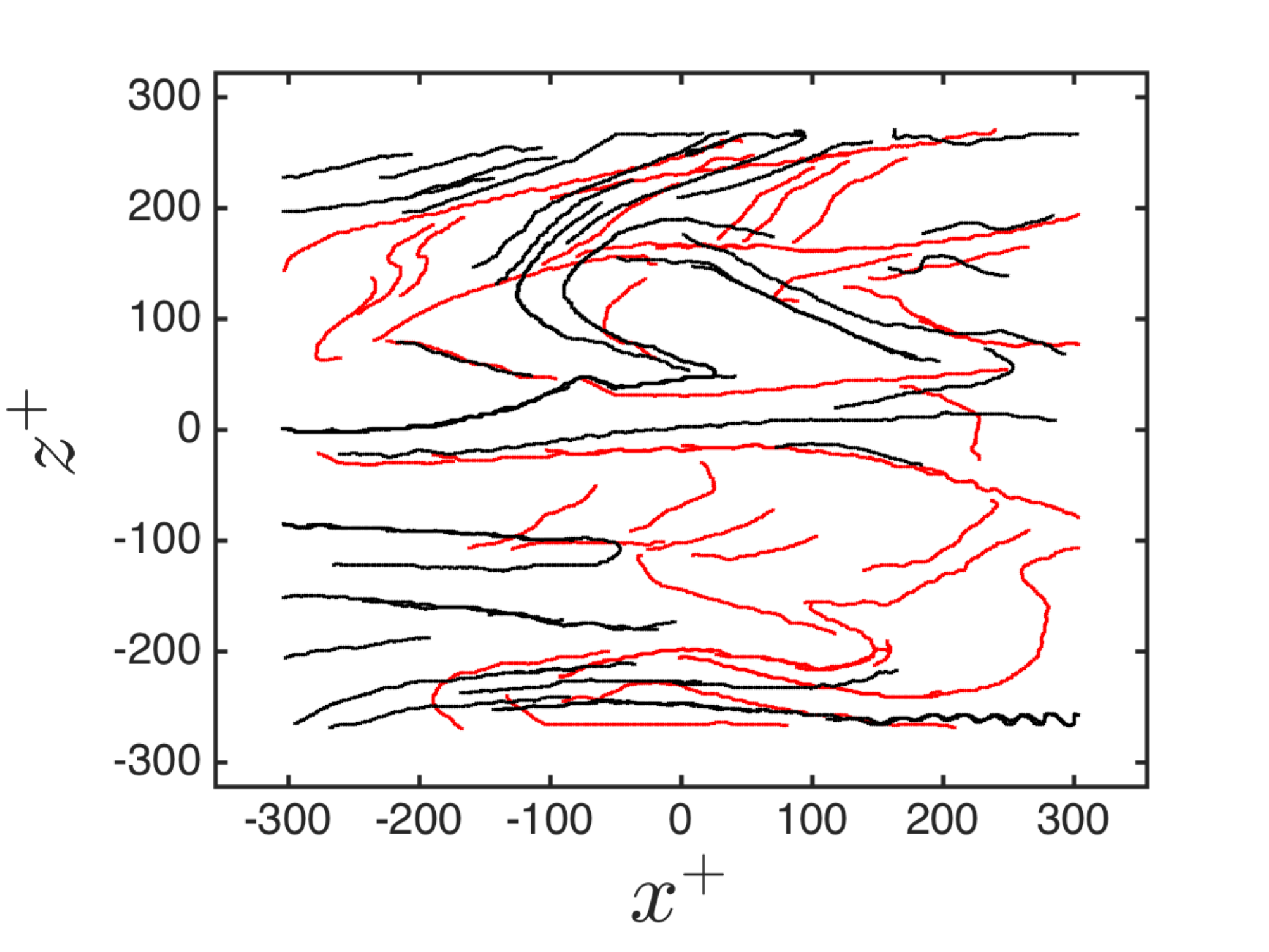} }
\end{center}
\caption{\label{}Reconstructed LCSs from the first POD mode at four different times: (a) $t^{+}=1.29$; (b) $t^{+}=18$; (c) $t^{+}=37.5$; (d) $t^{+}=49$. Red lines (attracting) and black lines (repelling).}
\end{figure}

The resemblance between the reconstructed LCSs obtained from the truncated fields and the particular POD modes are explored in figures 12. Particular POD modes, from mode number one through five and mode 20, with the LCSs overlayed capture the most prominent features attributed to the particular mode. Clearly, LCSs based on reconstructed fields still delineate the flow structures. The LCSs from the first POD mode agrees well with the structure of the first POD mode, see figures 12(a). The LCSs of the velocity field based on reconstruction using only the second POD mode capture the structure and mostly take the shape of second POD mode as shown in figure 12(b). In the same manner the LCSs from the third POD mode based reconstructed field delineate the dominant features of third POD modes as shown in figure 12(c). The reconstructed LCSs from the fourth POD mode manifest well matching with the POD mode where the trajectories coincide with the coherent structures as shown in figure 12(d). Figure 12(e) presents the reconstructed LCSs from the fifth POD mode and show also a satisfactory matching with the $P_{U_5}$ mode. Finally, the reconstructed LCSs from the twentieth POD mode are relatively irrelevant as LCSs distributed over all the domain are incredibly small with a large number of tangles. Thus, the trajectories do not mate the structures of the twentieth POD mode as shown in figures 12(f).  

\begin{figure}
\begin{center}
\subfigure[$P_{U_1}+LCS _{P_{U_1}}$]{ \label{a} \includegraphics[
width=0.48\textwidth]{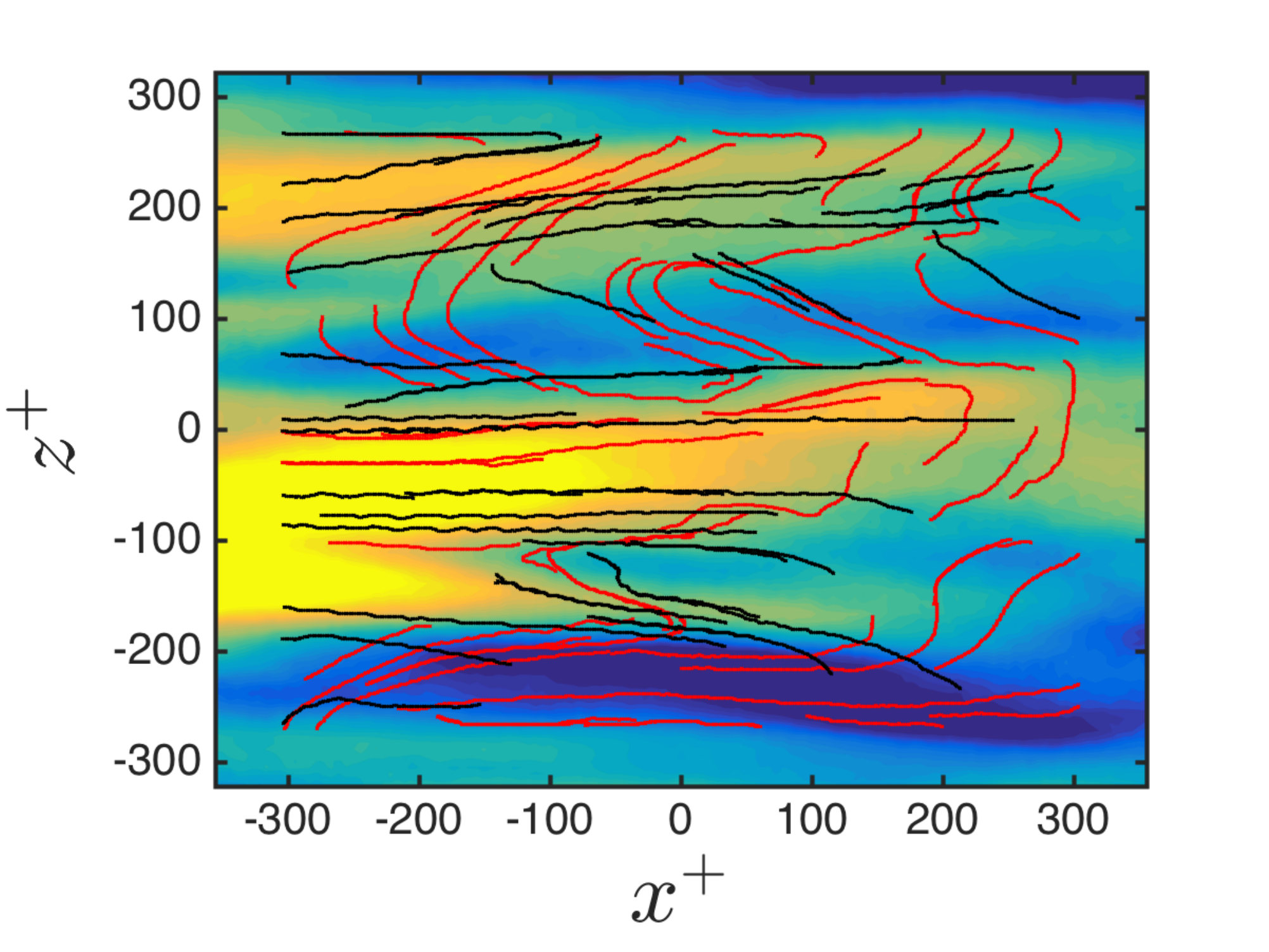} }
\subfigure[$P_{U_2}+LCS_{P_{U_2}}$]{ \label{b} \includegraphics[
width=0.48\textwidth]{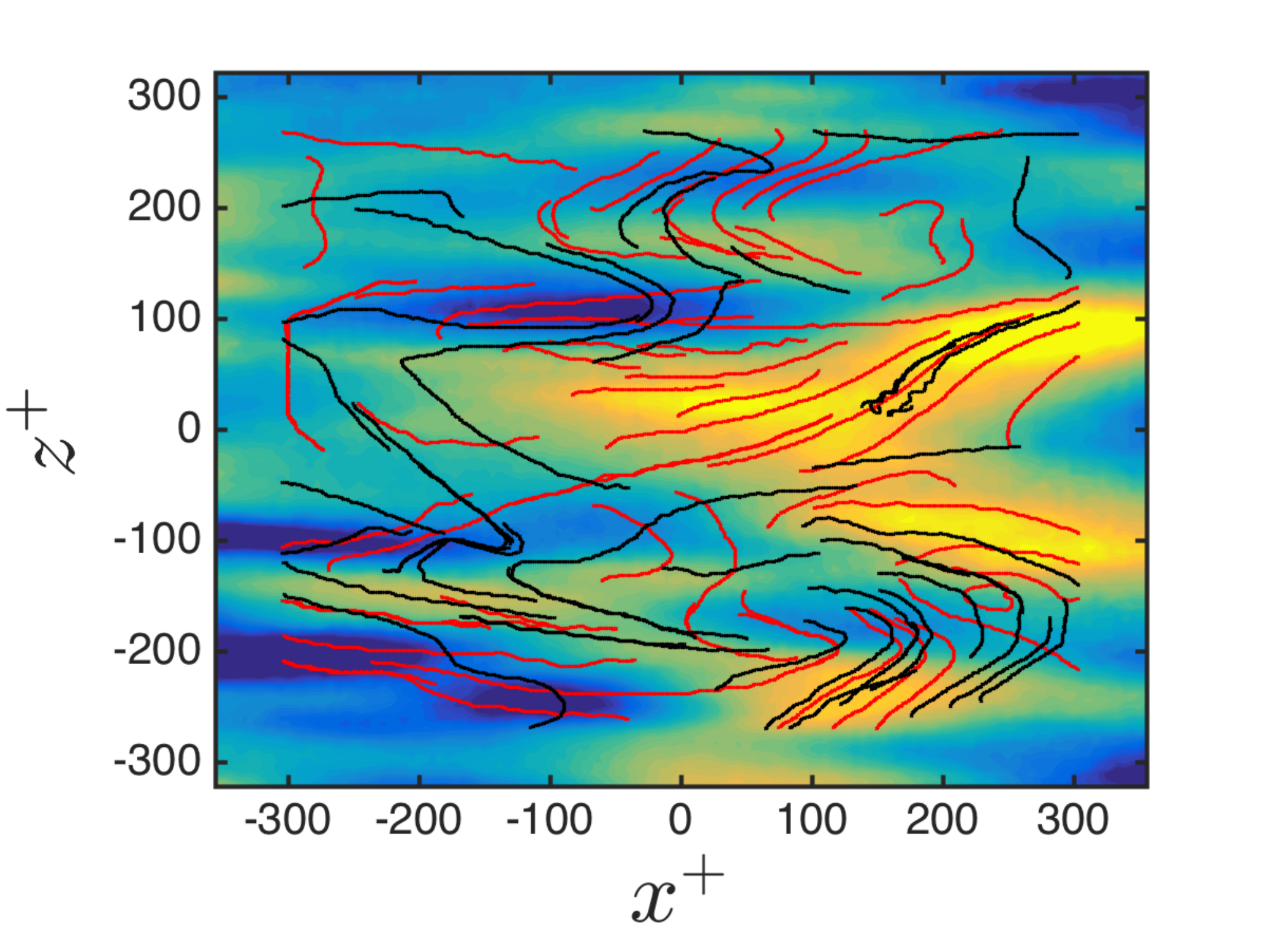} }
\subfigure[$P_{U_3}+LCS_{P_{U_3}}$]{ \label{c} \includegraphics[
width=0.48\textwidth]{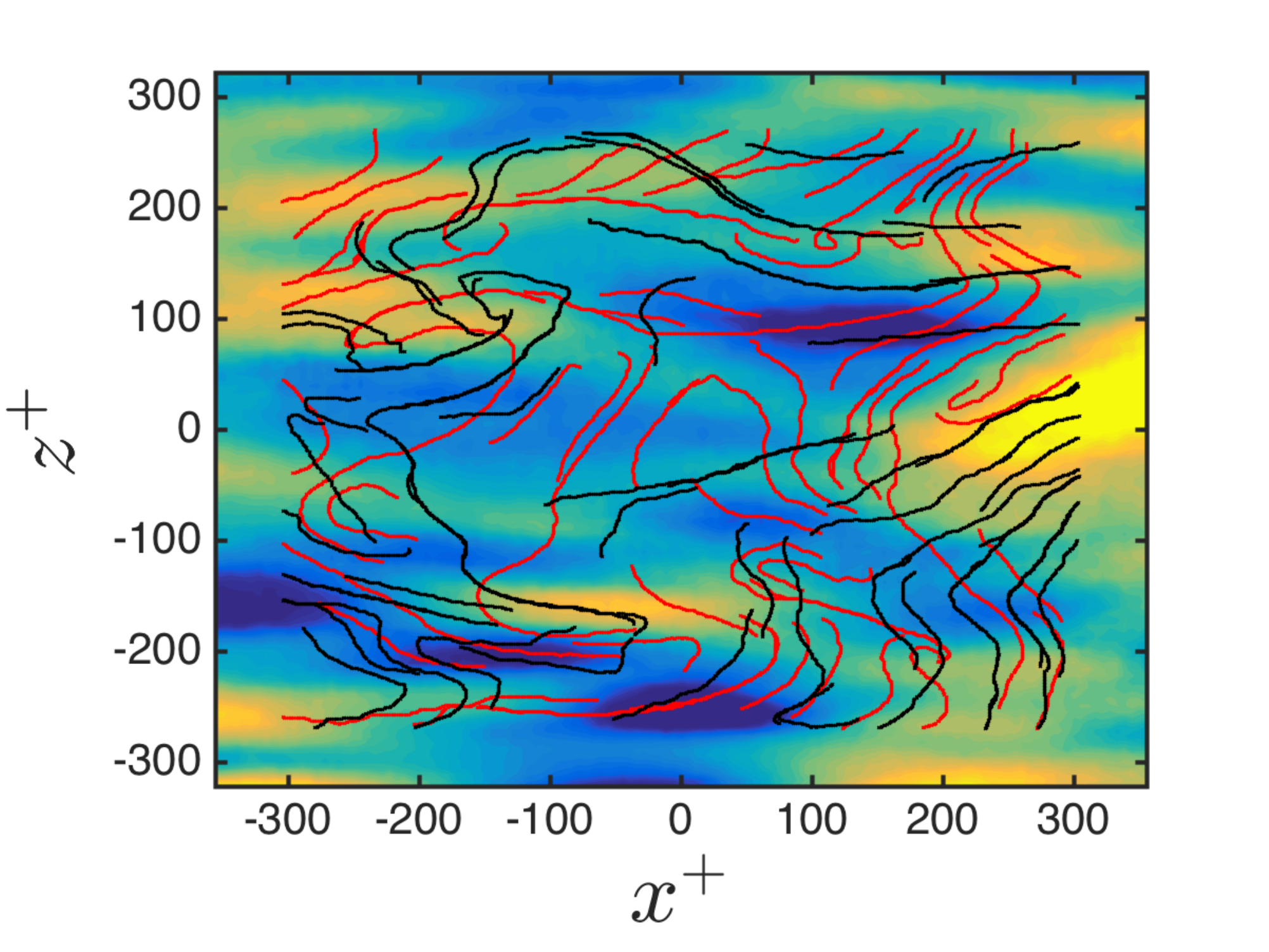} } 
\subfigure[$P_{U_4}+LCS_{P_{U_4}}$]{ \label{d} \includegraphics[
width=0.48\textwidth]{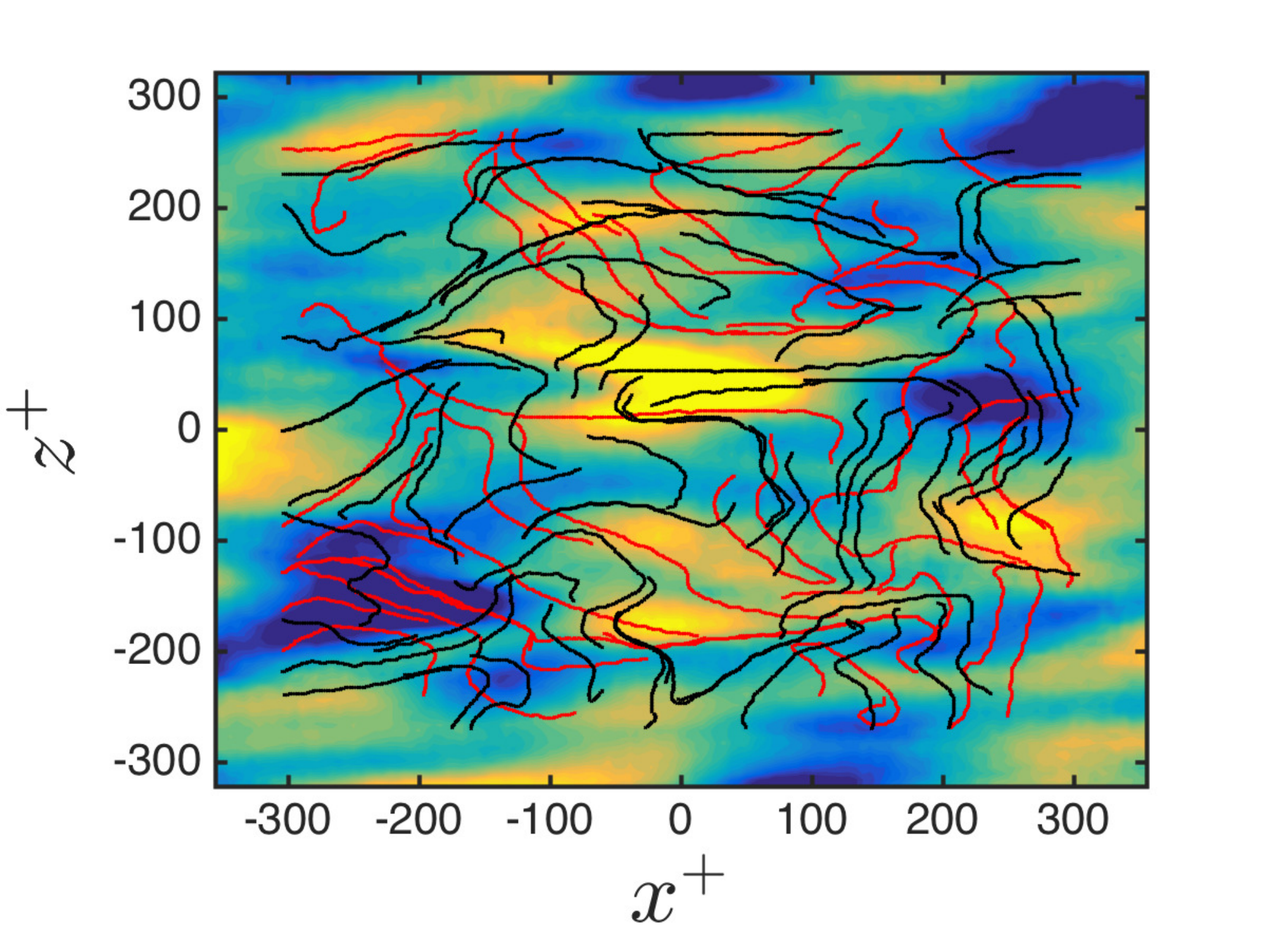} }
\subfigure[$P_{U_5}+LCS_{P_{U_5}}$]{ \label{e} \includegraphics[
width=0.48\textwidth]{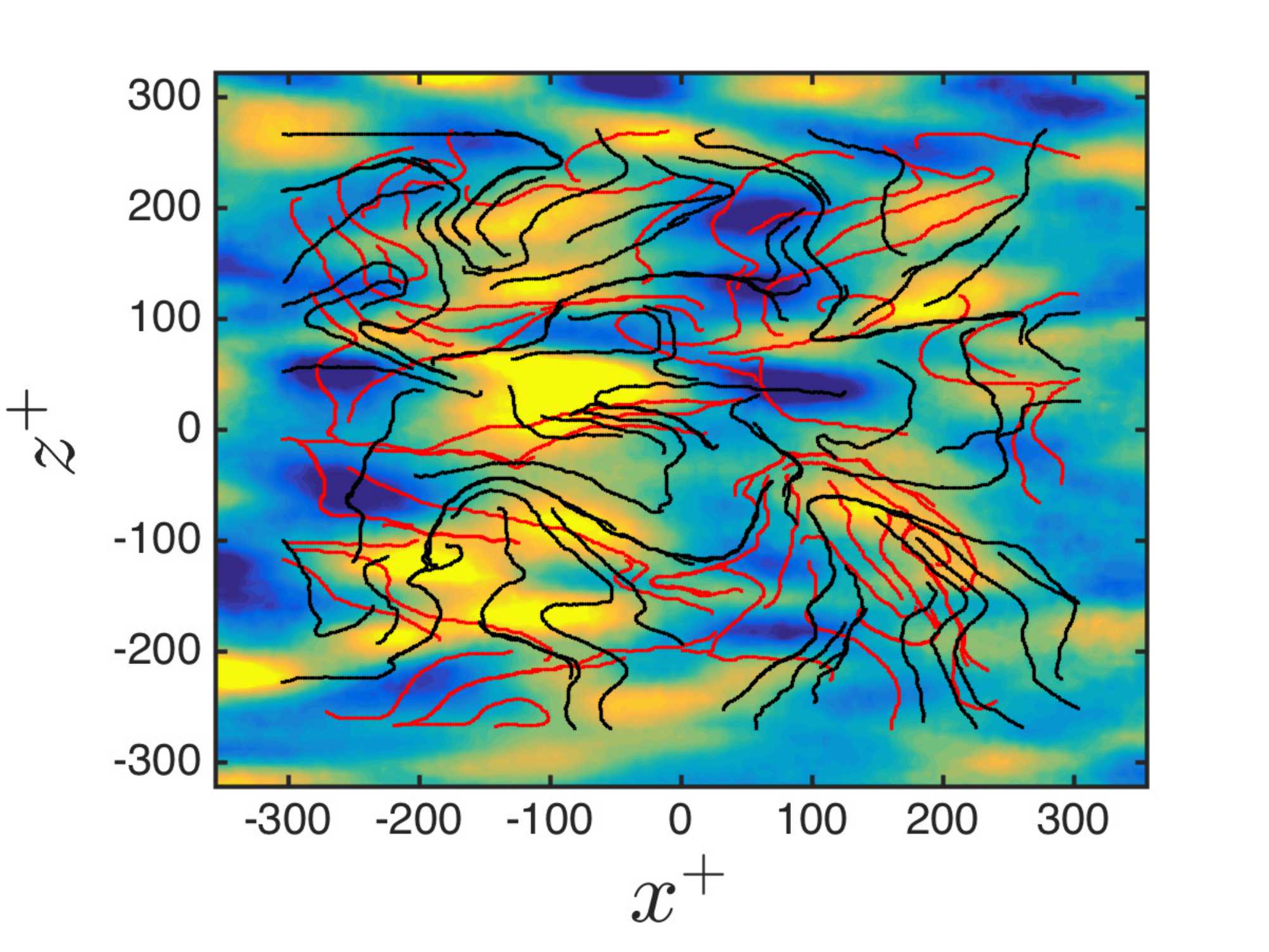} } 
\subfigure[$P_{U_{20}}+LCS_{P_{U_{20}}}$]{ \label{f} \includegraphics[
width=0.48\textwidth]{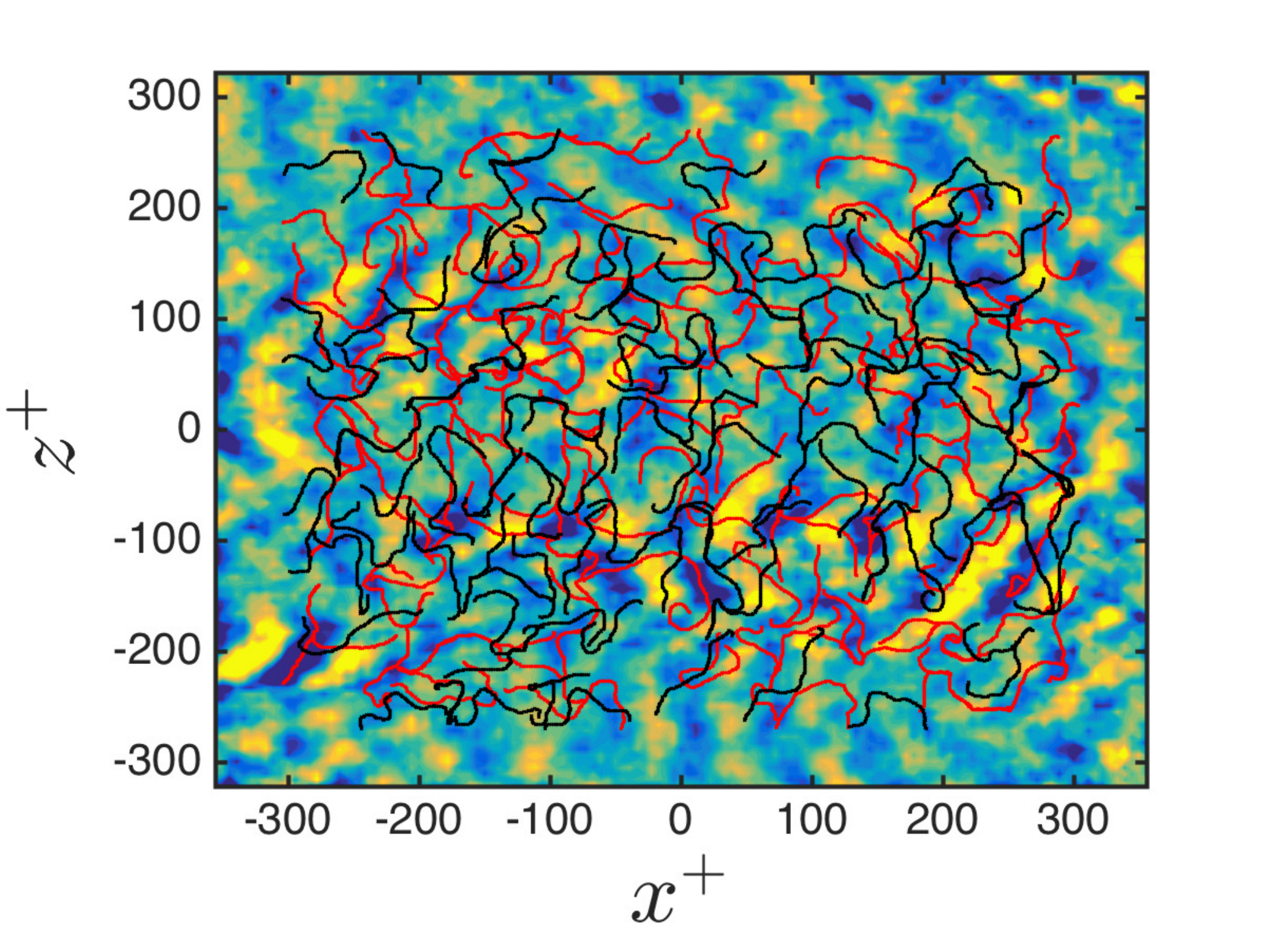} }
\end{center}
\caption{\label{} POD modes with Reconstructed LCSs from the specific modes: (a) the first POD mode ; (b) the second POD mode; (c) the third POD mode; (d) the fourth POD mode; (e) the fifth POD mode; (f) the twentieth POD mode.}
\end{figure}

DMD algorithm is also utilized to reconstruct the flow field and applied the LCS approach to rebuild new LCSs based frequency level. The first five of the DMD modes which hold the lowest frequencies as shown in figure 4 are used to reconstruct the new LCSs. These modes correspond to frequencies: 0, 230, 475, 700 and 910. Mode twenty is used as well to reconstruct the LCSs of higher frequency level. Figure 13 illustrates the overlying of the new LCSs with the particular DMD modes that are used to reconstruct these new trajectories. The same observation shown in the POD case is hold in the DMD, where the new LCSs depicts the structures of the modes. Thus, the length and the degree of the entanglement of the new LCSs are highly correlated with the coherence and the perturbation of the incoherence structures, in other words, the frequency of the structures. The longest LCSs with the least entanglements correspond the lowest frequency, and vice versa.  

\begin{figure}
\begin{center}
\subfigure[$D_{U_1}+LCS _{D_{U_1}}$]{ \label{a} \includegraphics[
width=0.48\textwidth]{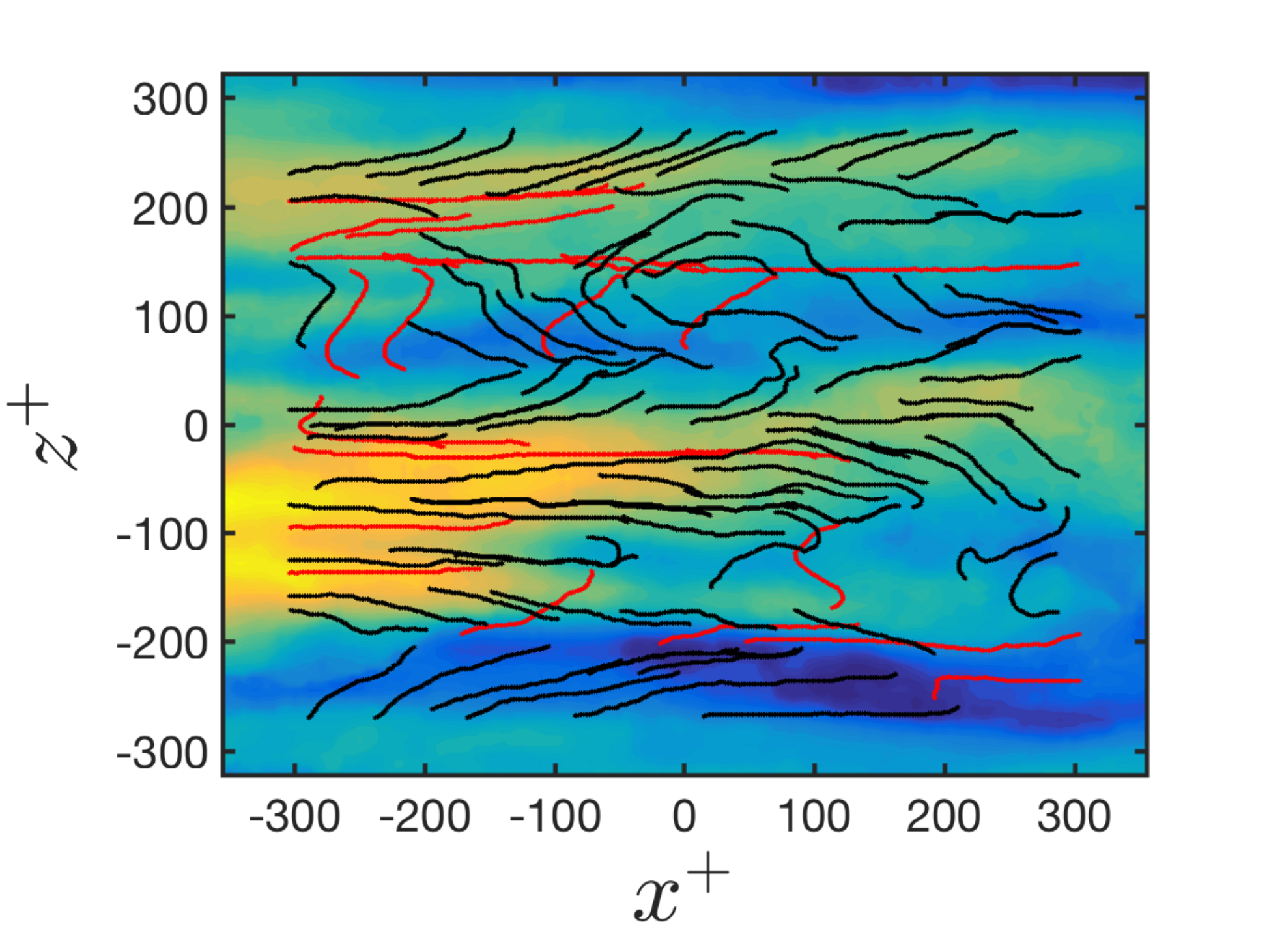} }
\subfigure[$D_{U_2}+LCS _{D_{U_2}}$]{ \label{b} \includegraphics[
width=0.48\textwidth]{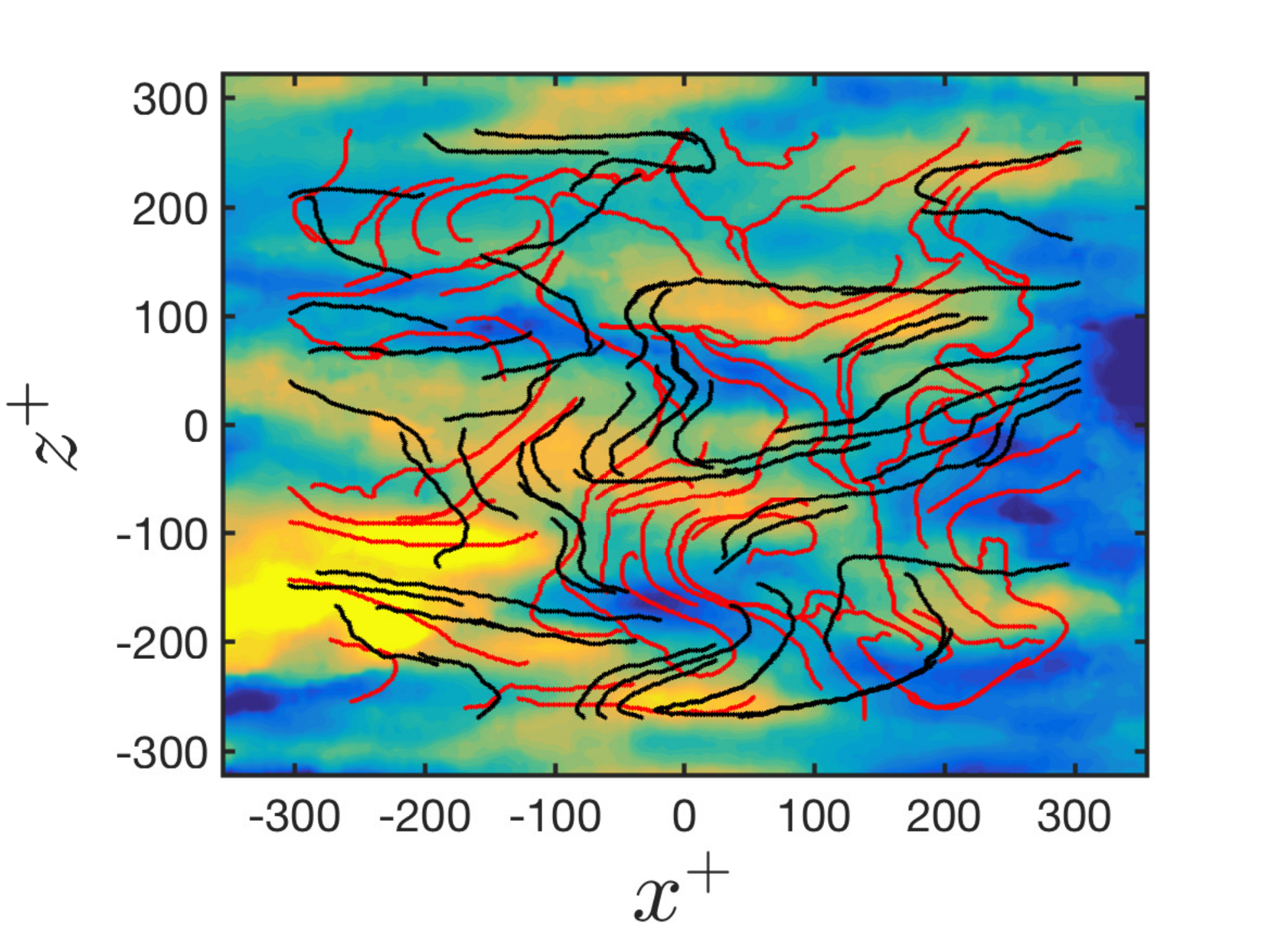} }
\subfigure[$D_{U_3}+LCS _{D_{U_3}}$]{ \label{c} \includegraphics[
width=0.48\textwidth]{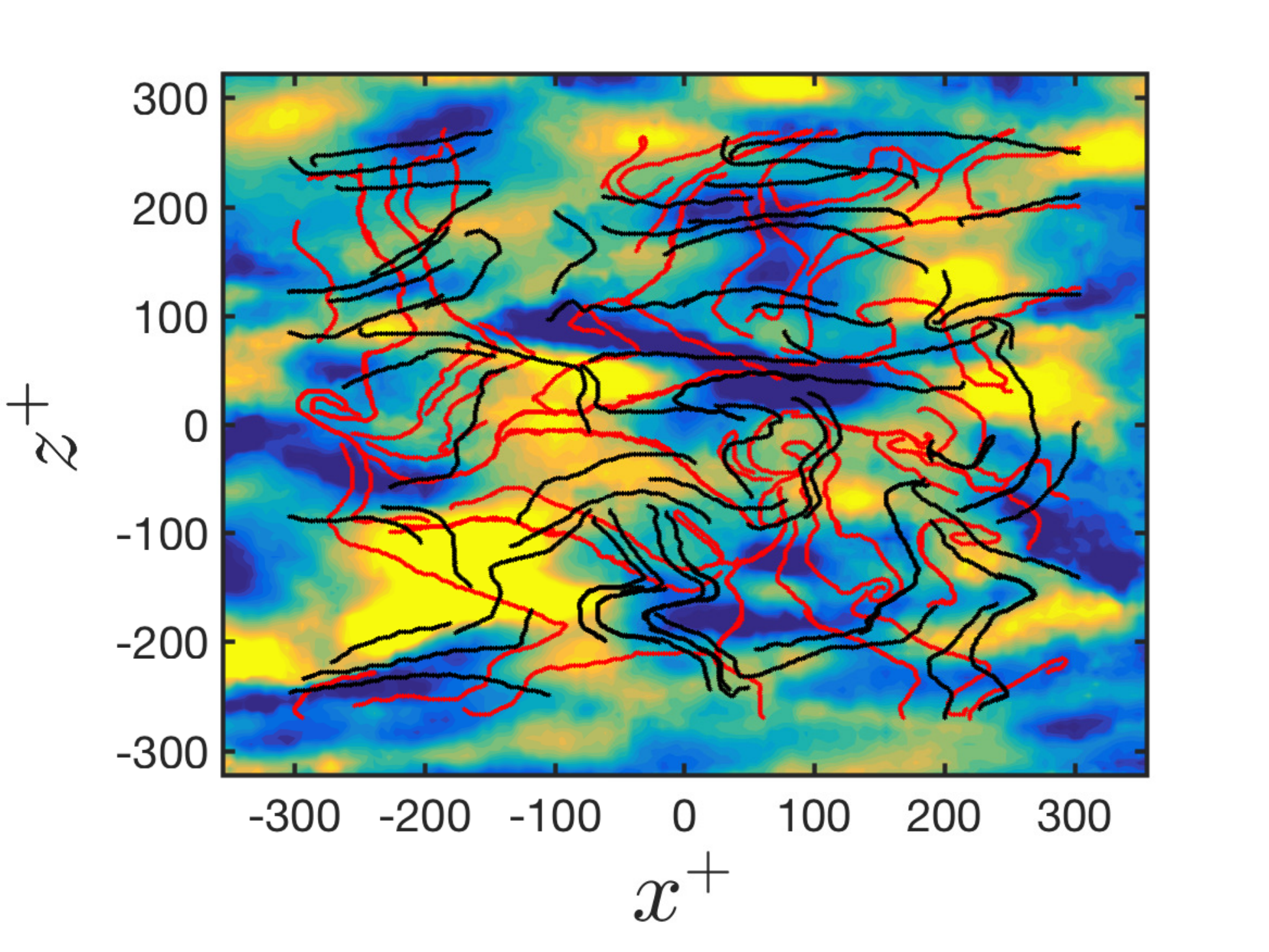} } 
\subfigure[$D_{U_4}+LCS _{D_{U_4}}$]{ \label{d} \includegraphics[
width=0.48\textwidth]{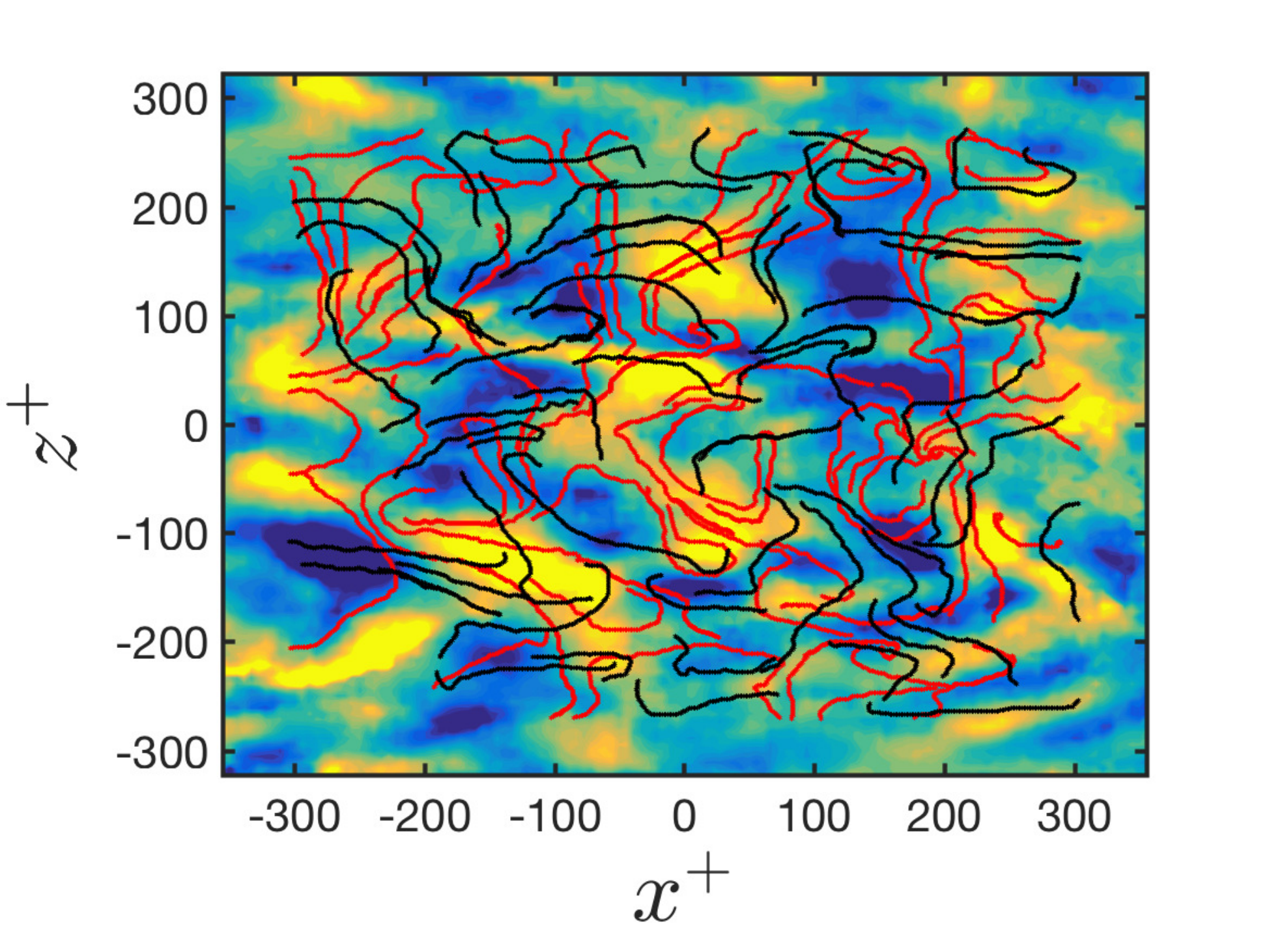} }
\subfigure[$D_{U_5}+LCS _{D_{U_5}}$]{ \label{e} \includegraphics[
width=0.48\textwidth]{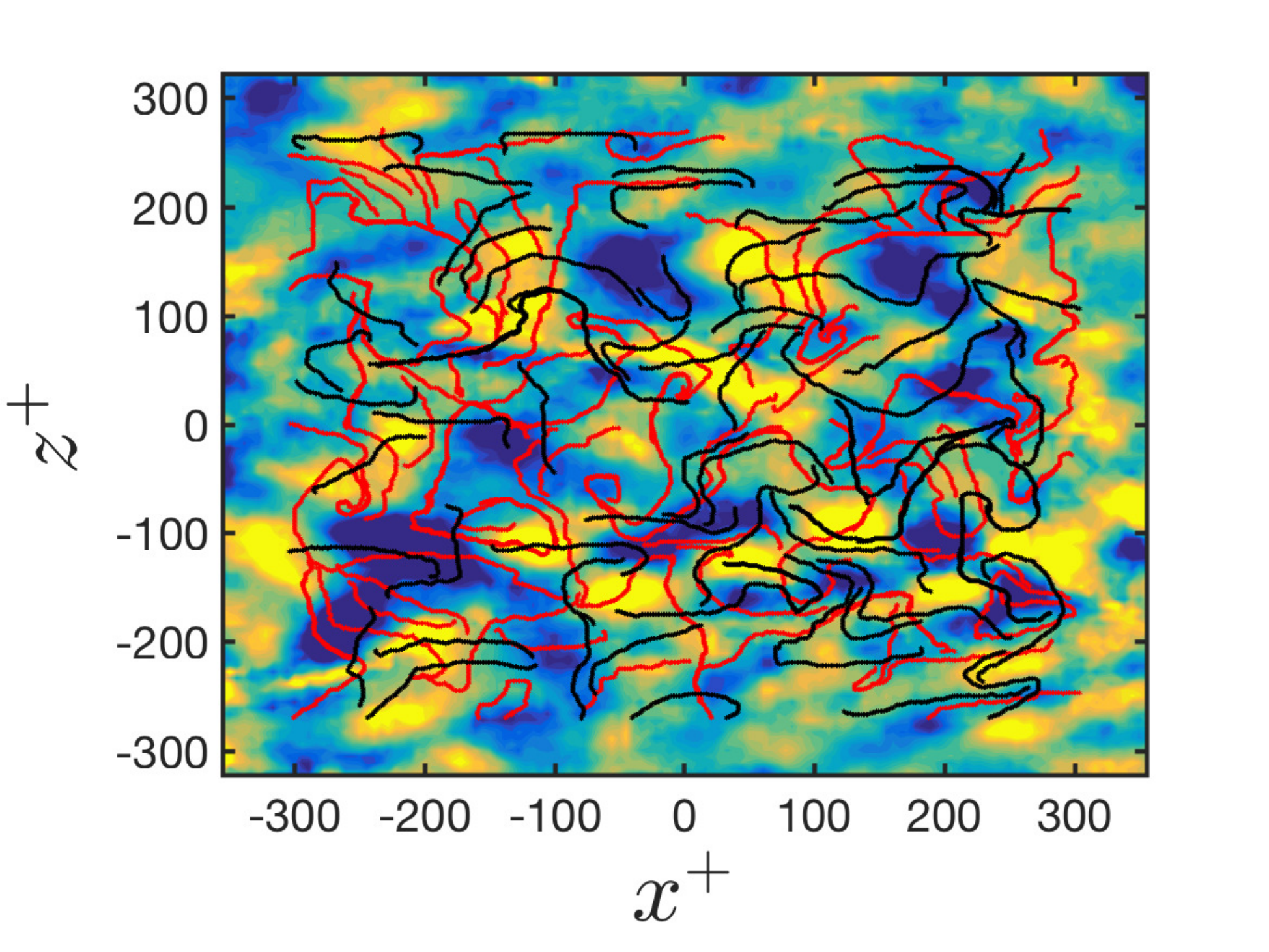} } 
\subfigure[$D_{U_6}+LCS _{D_{U_{20}}}$]{ \label{f} \includegraphics[
width=0.48\textwidth]{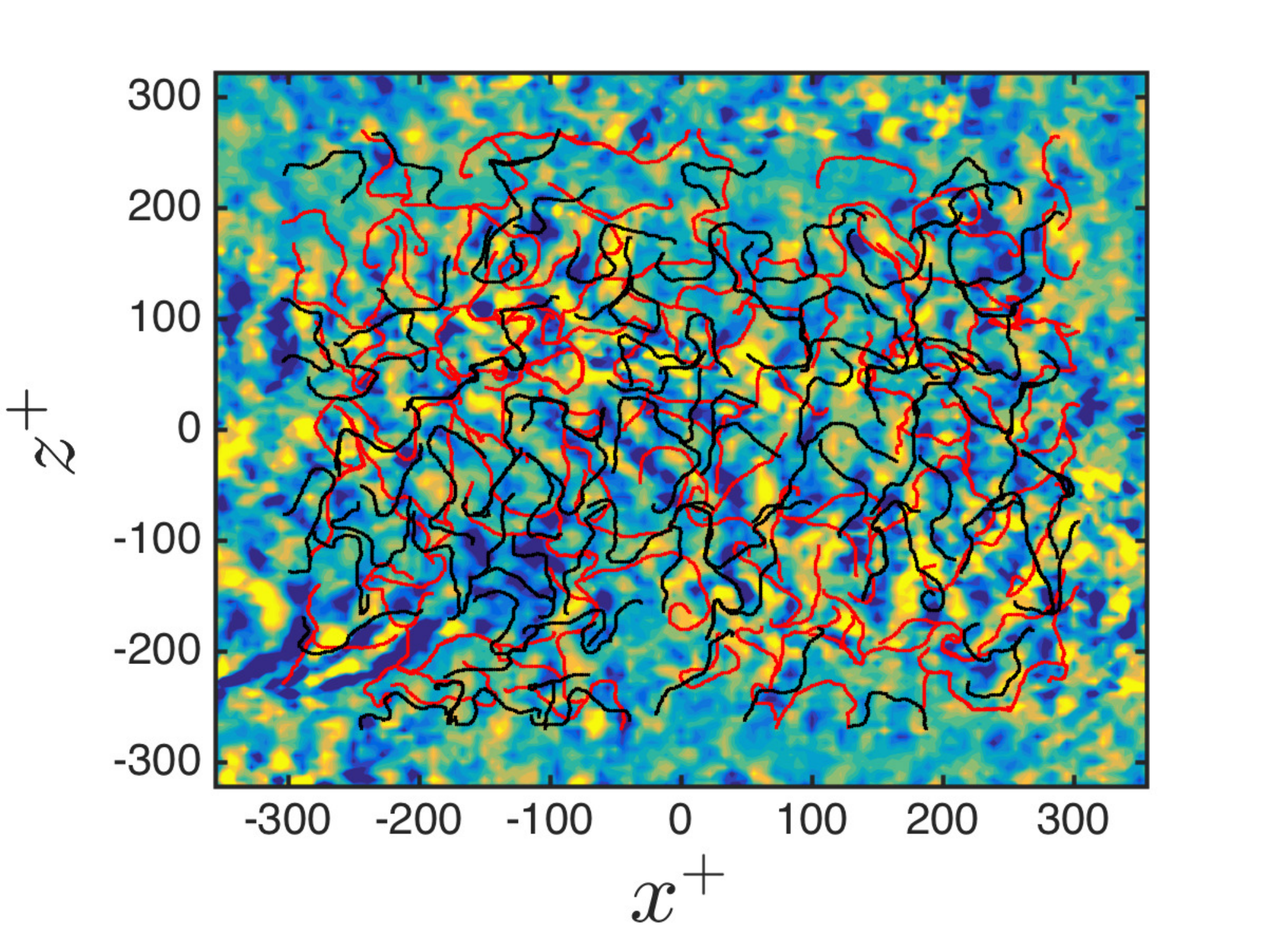} }
\end{center}
\caption{\label{} DMD modes with Reconstructed LCSs from the specific modes: (a) the first DMD mode; (b) the second DMD mode; (c) the third DMD mode; (d) the fourth DMD mode; (e) the fifth DMD mode; (f) the twentieth DMD mode.}
\end{figure}

To show the effect of the  reconstruction as combination of the POD modes on the resulting LCSs, the fluctuation velocity fields are reconstructed using the first two, first three, first six. In addition, the fields reconstructed using the POD modes are observed which are characterized as not dominant, namely from the seventh POD mode up to fortieth POD modes. Figure 14(a) presents the LCSs of the velocity field which was reconstructed using the first two POD modes at $t^{+}=1.29$. The repelling and attracting LCSs remain coherent, with a length of approximately extending from $x^+ = -300$ to $300$ in the regions of $z^{+} \approx \pm 0$ and these encompass areas of high energy per the POD modes themselves. Figure 14(b) represents the LCSs from the first three POD modes. Although there are small differences between the POD modes represented in the contour of the figure 16(a) and figure 14(b), the reconstructed LCSs show large difference between two cases which confirm that LCSs are sensitive even for small changes in the turbulence kinetic energy. Figure 14(c) represents the LCSs from the first six POD modes. Here, attracting and repelling LCSs are diverging in shape when compared with the reconstruction using a less number of modes as presented in figures 14 (a) and (b). However, LCSs of the six modes are longer than the case of total field data (see figure 6(a)). Furthermore,  the trajectories still orient themselves with the mode structure. The LCS trajectories are sensitive to the critical points, where the velocity is null, and their near fields. This is identified as green and blue in the figure. The near zero velocity regions bend the trajectories.  Increasing the number of POD modes in reconstruction leads to more complex flow, hence more critical points and therefore more bending in LCS trajectories. To show the effect of the incoherence on the attracting and repelling LCSs, the velocity field is reconstructed using modes from 7 to 40 which represent merely 25\% of the total energy. Figure 14(d) shows the sum of these modes with the reconstructed LCSs. It is clear that the length of the trajectories depends on the number of the modes that are used to reconstruction of the velocity field. The LCSs appear to be homogeneous in their size and distribution throughout the domain and retain their shape with advected time in this case.

\begin{figure}
\begin{center}
\subfigure[${\sum_{1}^{2} P_{U_n}}+LCS _{\sum_{1}^{2} P_{U_n}}$]{ \label{a} \includegraphics[
width=0.48\textwidth]{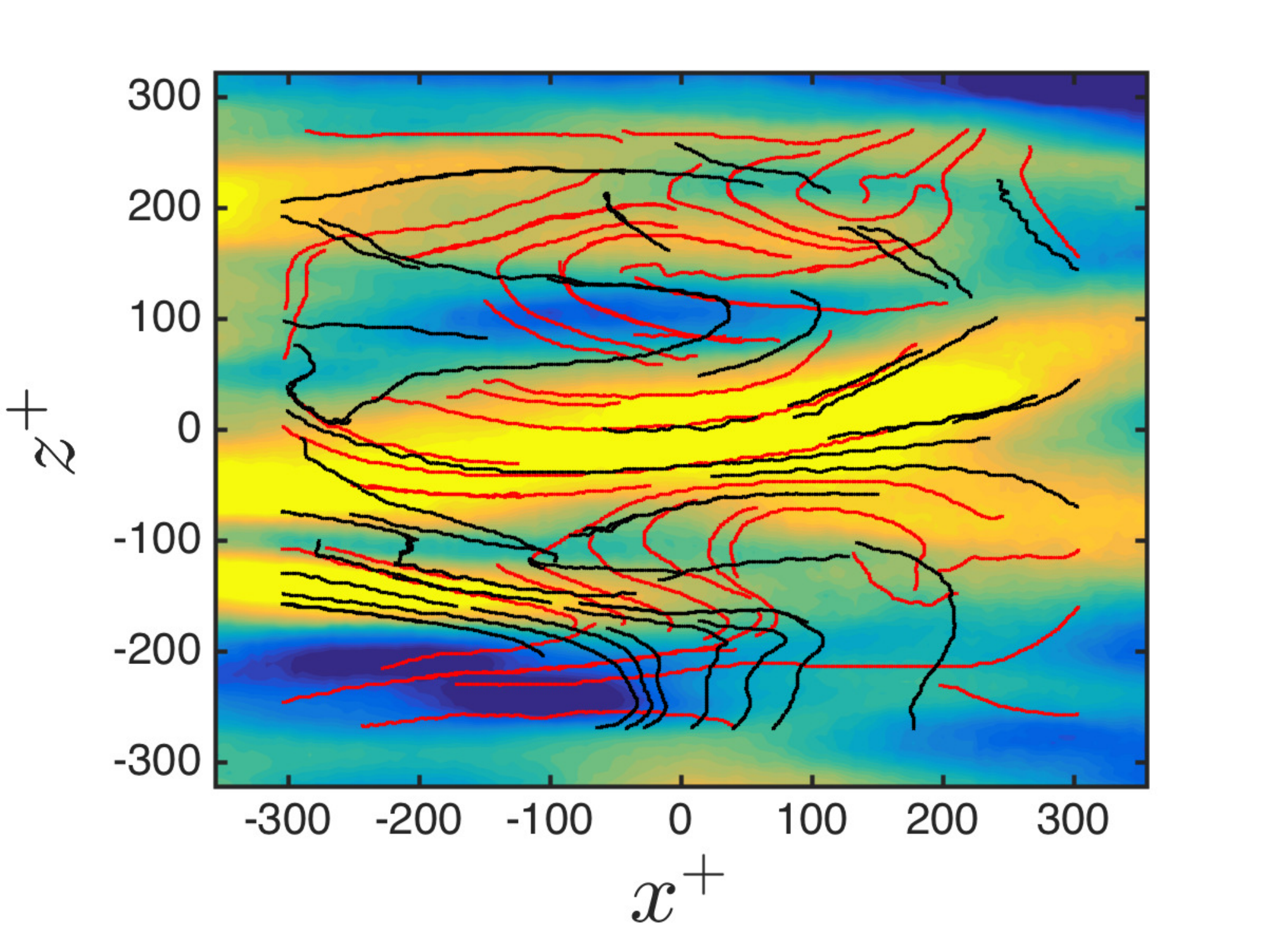} }
\subfigure[${\sum_{1}^{3} P_{U_n}}+LCS _{\sum_{1}^{3} P_{U_n}}$]{ \label{b} \includegraphics[
width=0.48\textwidth]{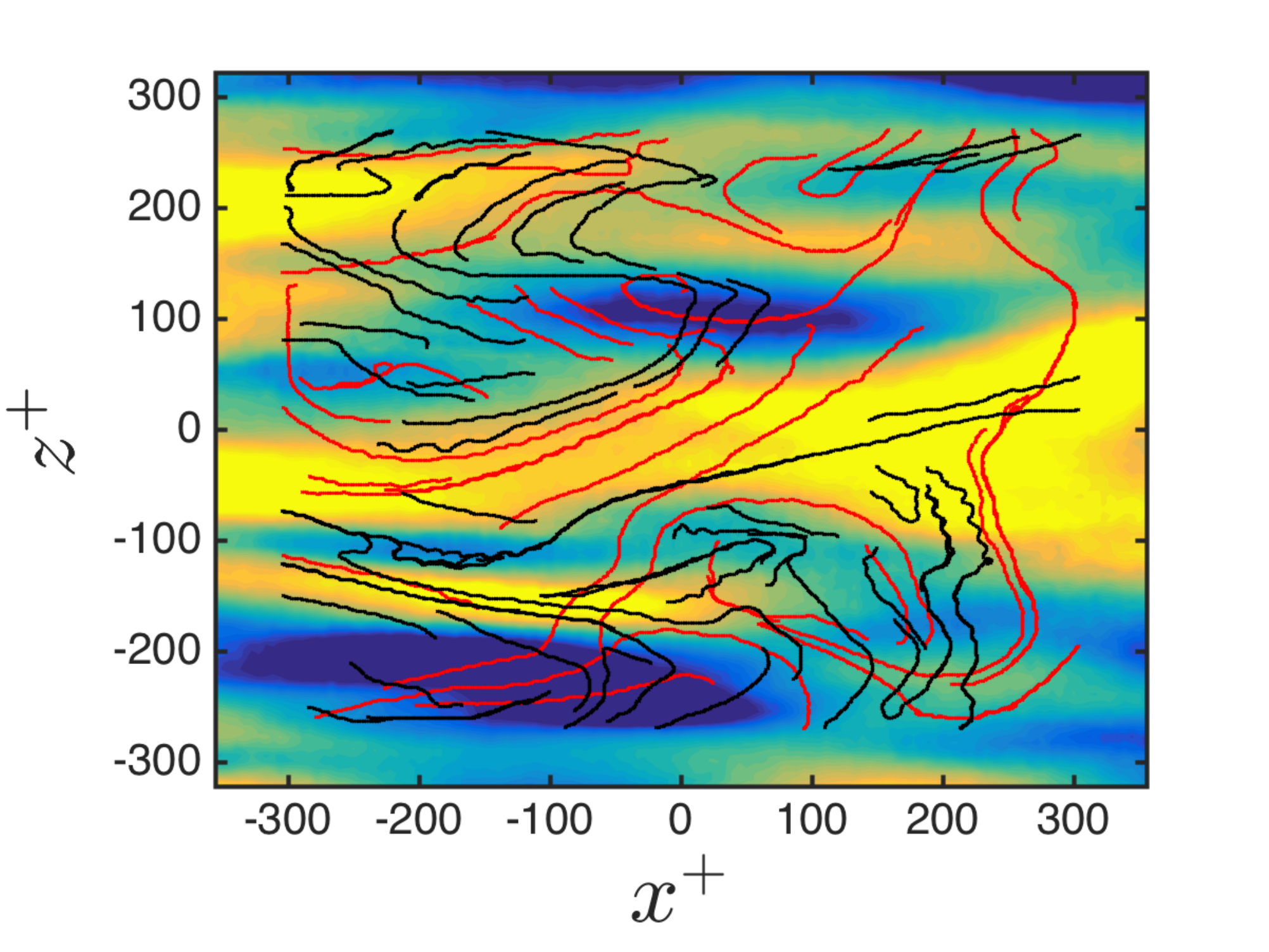} }
\subfigure[${\sum_{1}^{6} P_{U_n}}+LCS _{\sum_{1}^{6} P_{U_n}}$]{ \label{c} \includegraphics[
width=0.48\textwidth]{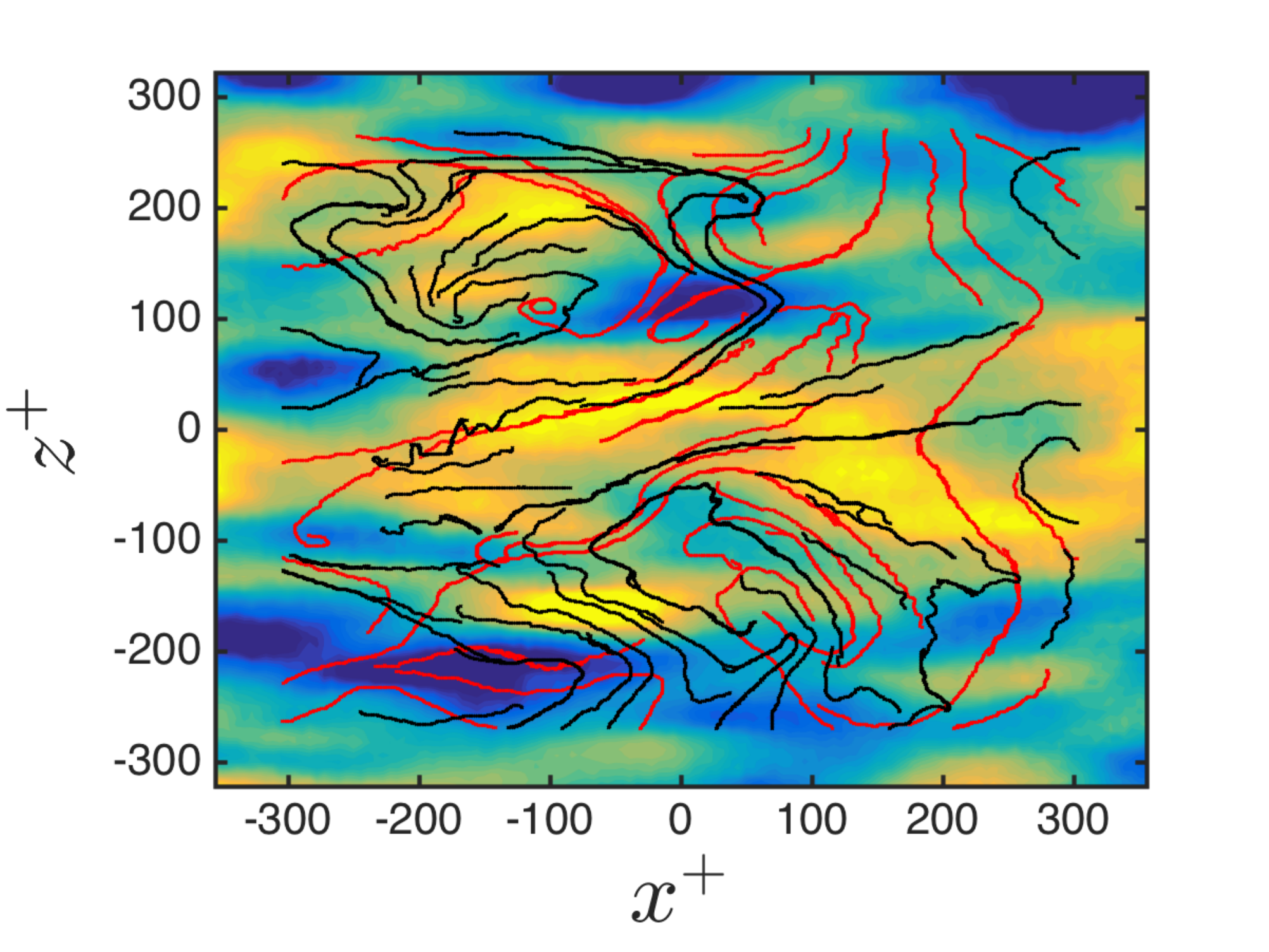} } 
\subfigure[${\sum_{7}^{40} P_{U_n}}+LCS _{\sum_{7}^{40} P_{U_n}}$]{ \label{d} \includegraphics[
width=0.48\textwidth]{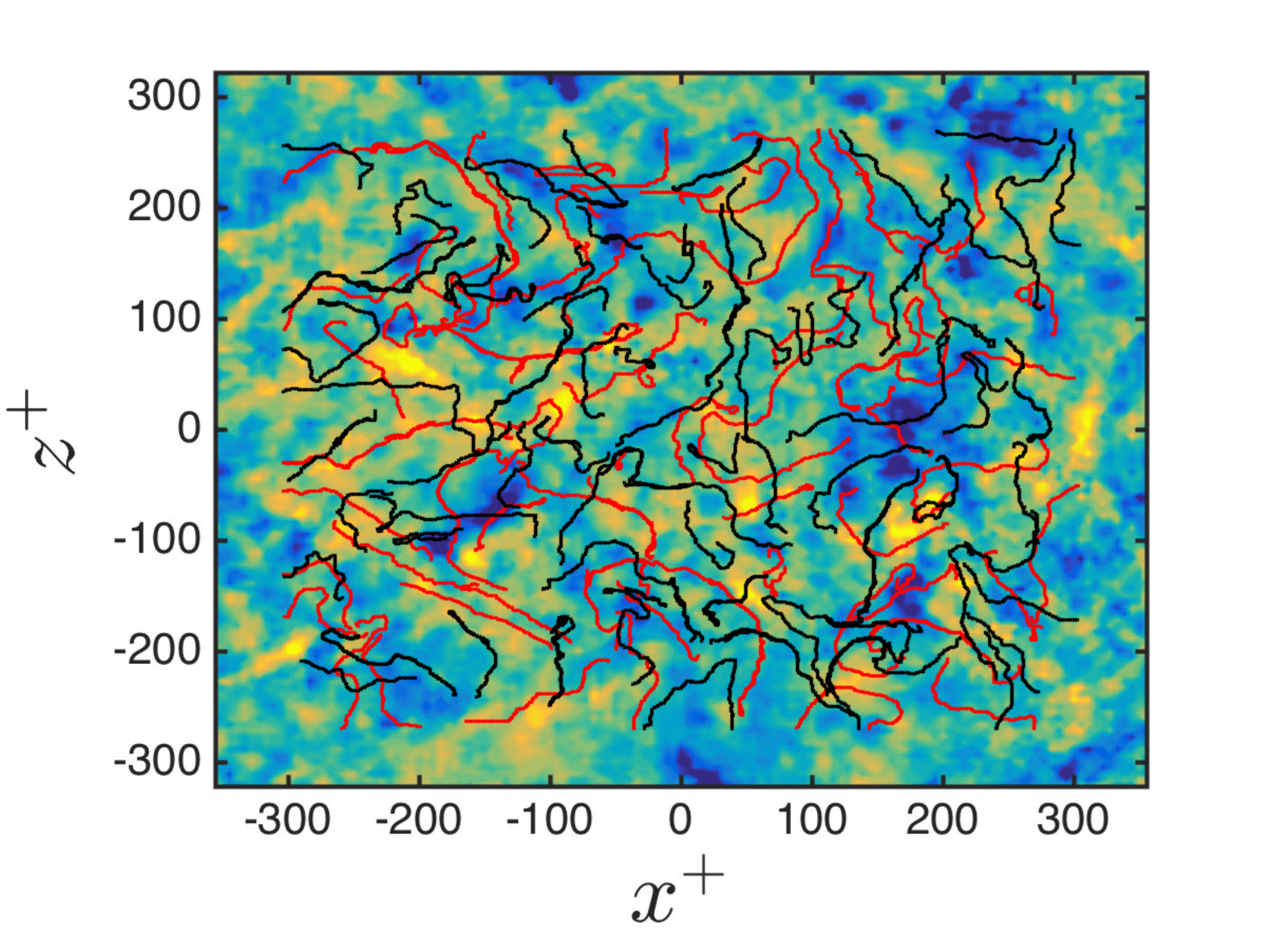} }
\end{center}
\caption{\label{}Reconstructed LCSs with POD modes: (a) the first two POD modes with reconstructed LCSs from the first two POD modes; (b) the first three POD modes with reconstructed LCSs from the first three POD modes; (c) the first six POD modes with reconstructed LCSs from the first six POD modes; (d) the last 34 POD modes with reconstructed LCSs from the last 34 POD modes. Red lines (attracting) and black lines (repelling).}
\end{figure}

\subsection{\label{Results5} Intersection Points}

The location of the intersection points between the attracting and repelling LCSs are investigated as these pertain to regions of low momentum within the domain.  Attracting and repelling LCSs  are matched with the normalized fluctuation velocity, ($u_f$=$u$/$u_{\tau}$), at four different times, $t^{+}=1.92, 18, 37.5$ and 49 as shown in figure 15. The trajectories portray the velocity contour and also tend to adhere to the boundaries highlighting particular features with a preference of high momentum. Most bending in trajectories occurs when these trajectories pass regions of null velocity and bound by positive and negative velocities. The intersection points are located at the low velocity (near to zero) region at the beginning of the advection time and LCSs remain concave or convex in shape. When advection time increases, few intersecting points between repelling and attracting LCSs appear in high velocity regions although most points remain in the low velocity regions. Previous studies have visualized vortices in the turbulent boundary layer and connected them with the low momentum regions where these tend to coincide \citep{dennis201}. Hence, the connection between intersection points and low speed region of the fluctuation velocity contour can also indicate the location of the vortices. 

\begin{figure}
\begin{center}
\subfigure[$(u_{f}+LCS)_{t^{+}=1.29}$]{ \label{a} \includegraphics[
width=0.48\textwidth]{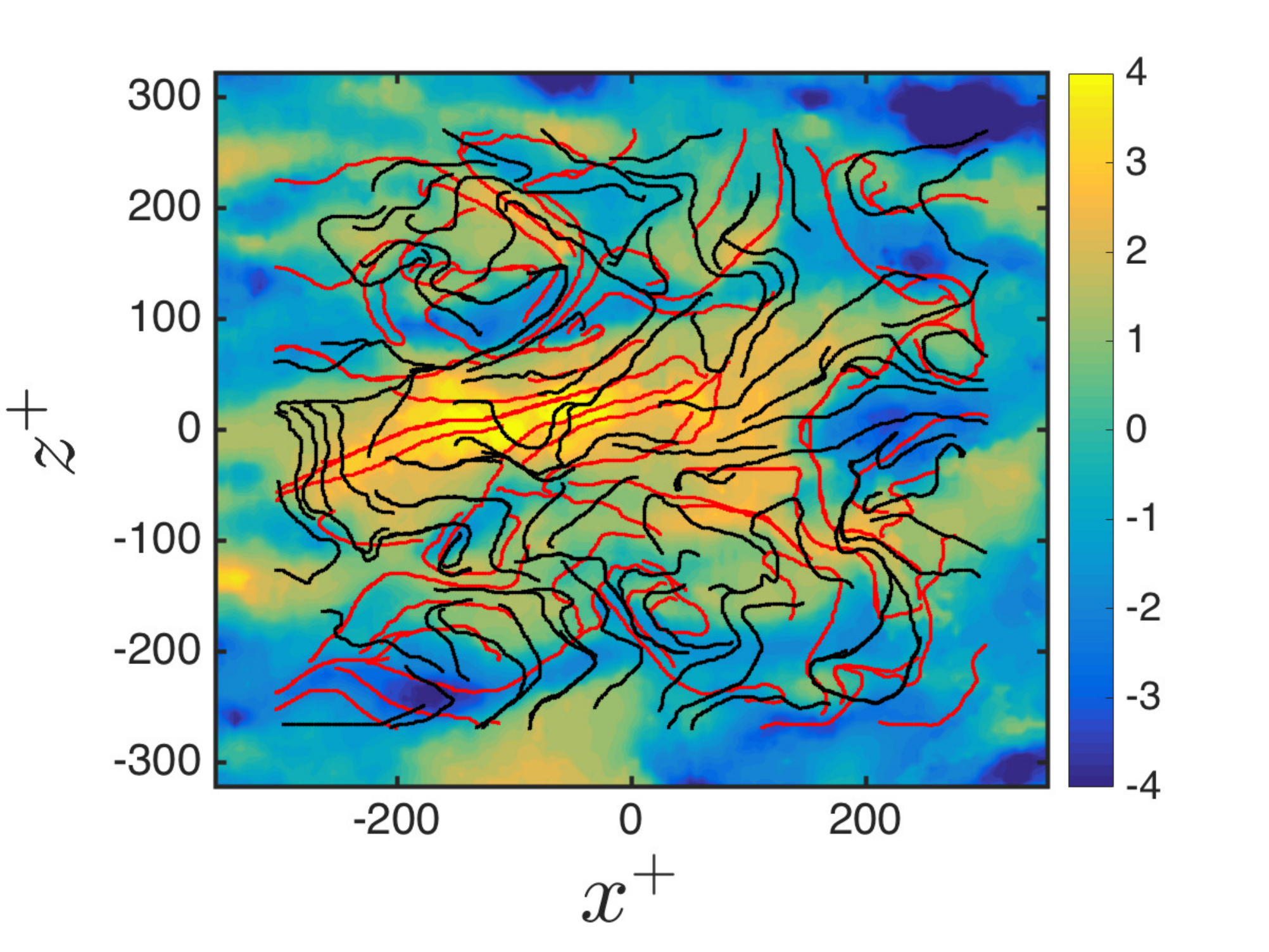} }
\subfigure[$(u_{f}+LCS)_{t^{+}=18}$]{ \label{b} \includegraphics[
width=0.48\textwidth]{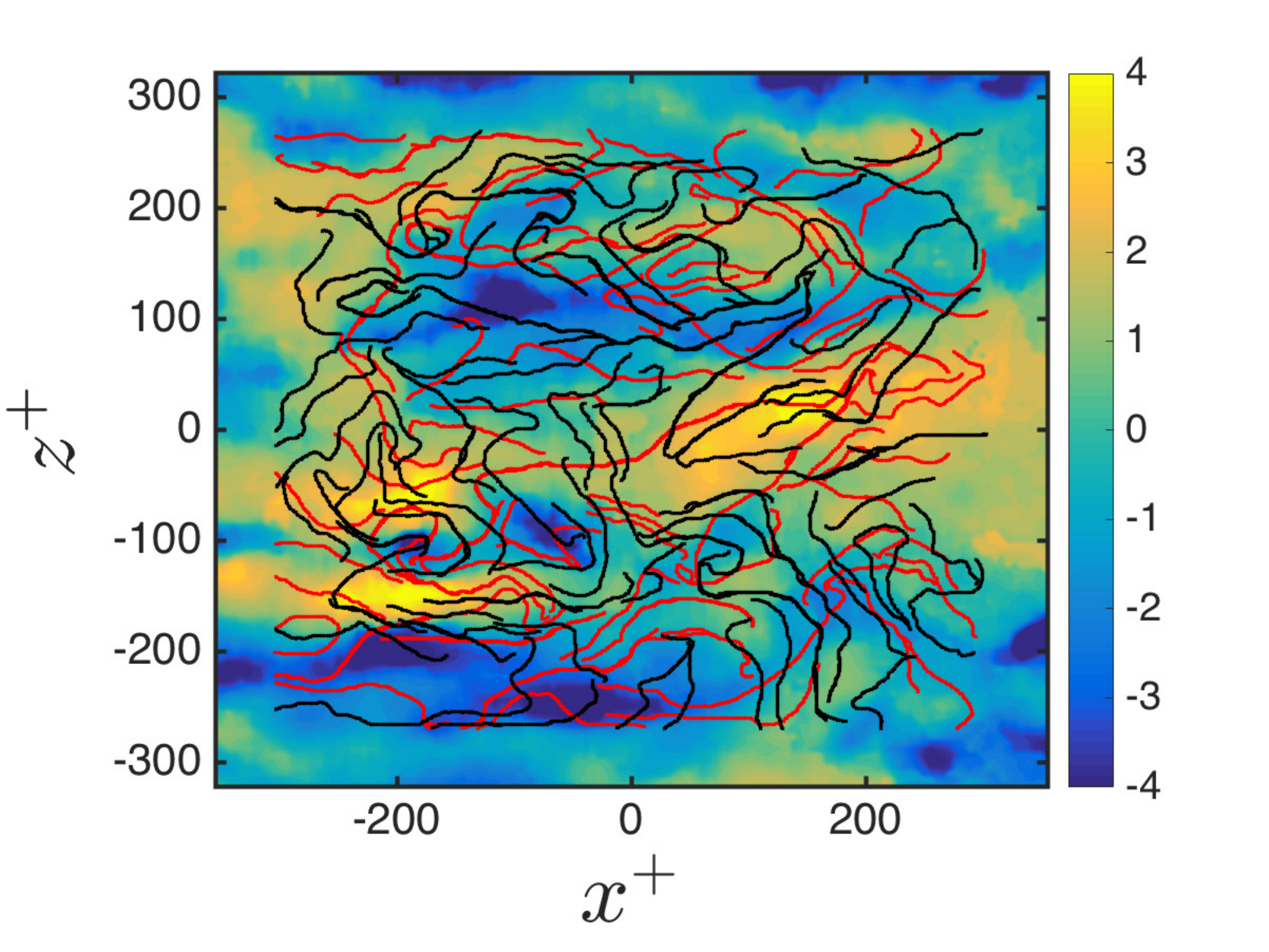} } 
\subfigure[$(u_{f}+LCS)_{t^{+}=37.5}$]{ \label{c} \includegraphics[
width=0.48\textwidth]{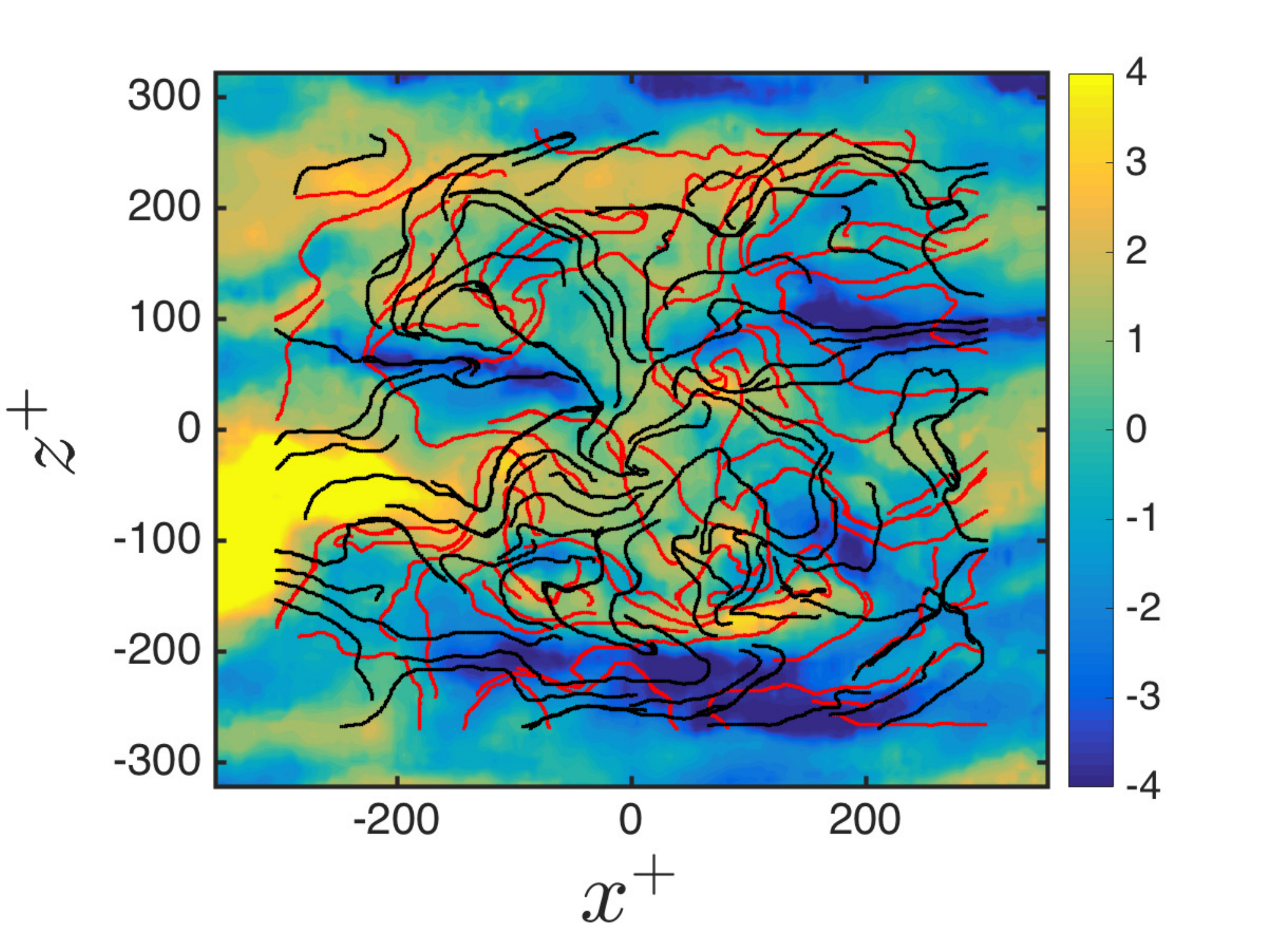} } 
\subfigure[$(u_{f}+LCS)_{t^{+}=49}$]{ \label{d} \includegraphics[
width=0.48\textwidth]{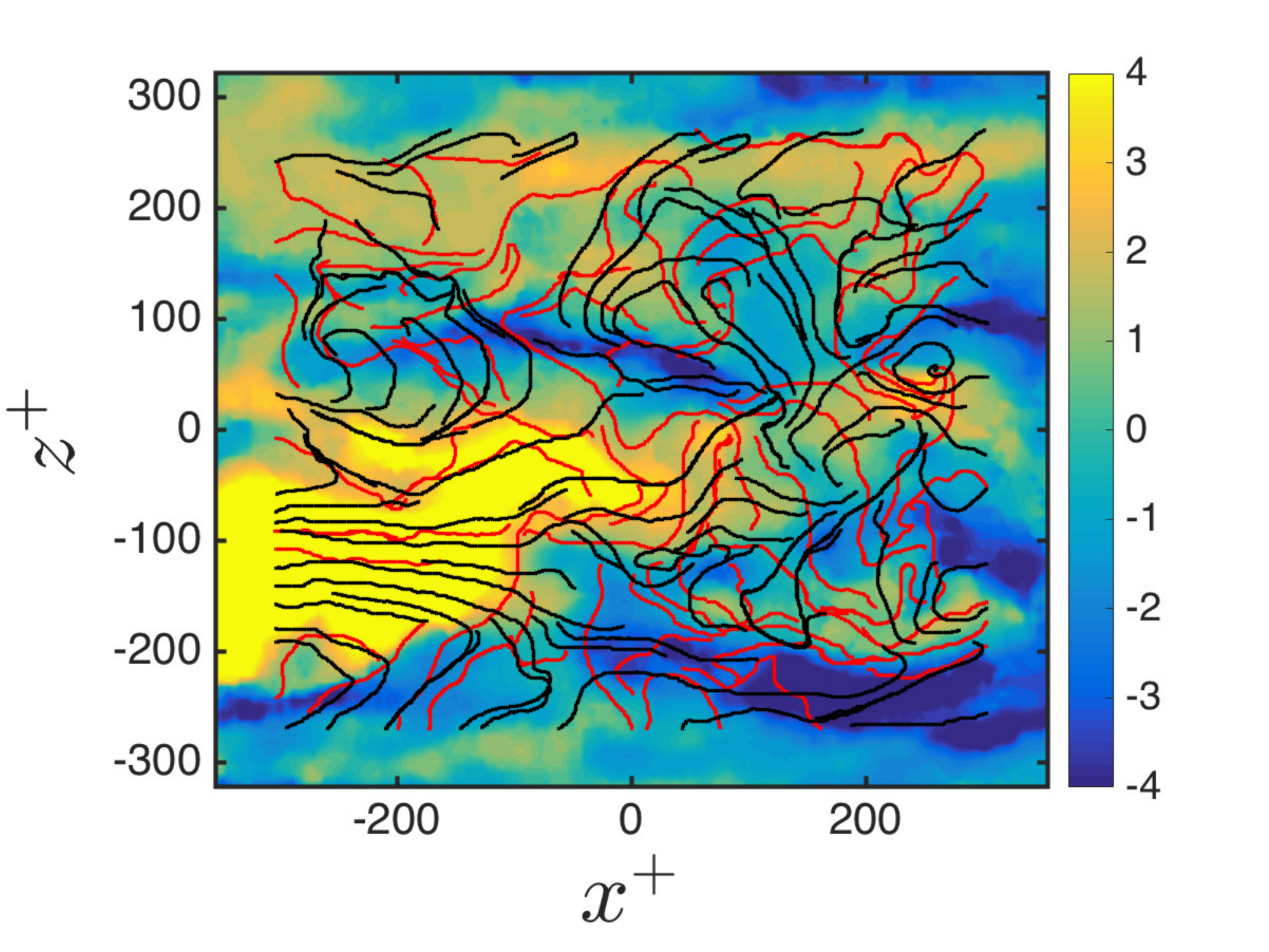} }
%\\
\end{center}
\caption{\label{}Normalized fluctuation velocity with LCS at four different times: (a) $t^{+}=1.29$; (b) $t^{+}=18$; (c) $t^{+}=37.5$; (d) $t^{+}=49$. Red lines (attracting) and black lines (repelling).}
\end{figure}

To ensure the location of the intersection points based on particular energy-containing modes, LCSs are overlayed on the fluctuation velocity as shown in figure 16 to look for the regions of the intersections. Figure 16(a) shows the  LCSs from the first POD mode and the normalized fluctuation velocity (original). Few intersection points are present, but these are nevertheless located at near-zero velocity regions and most bending is present on the repelling LCSs located at low speed region. Figure 16(b) presents overlaying between the fluctuation velocity and the LCSs based on the second POD mode.

The number of the intersection points is increased, but their location remains in near-zero regions. Remarkably, bending of the structures remains in areas of low momentum. Figure 16(c) is similar to figures 16(a) and (b), but now containing the reconstructed LCS employing the third POD mode as to show the dependence between the intersection points and a mode-to-mode increase in the LCS reconstruction. The intersecting points between the repelling and attracting structures continue to reside in areas of low momentum regions and increase in number. For example, large number of these crossings are found in the domain at $x^+ = 100$ and $z^+ \approx -200$. Figure 16(d) displays the reconstructed LCSs from $P_{U_4}$ with the fluctuation velocity and continuous to manifest the same trend as with modes 1 through 3. The scenario holds for the case of the reconstruction LCSs from the fifth POD mode as shown in figure 16(e), but clearly the complexity increases,  therefore making it more difficult to assess. Finally, Figure 16(f) presents the matching between the fluctuation velocity and the LCSs based on the twentieth POD modes. The intersection points have amassed in increasing number and are distributed over the entire domain. The preference between high/low-momentum region is lost, thus pointing towards the homogeneity and isotropic nature of this particular mode. in high, low and zero momentum regions.

\begin{figure}
\begin{center}
\subfigure[$u_{f}+LCS_{P_{U_1}}$]{ \label{a} \includegraphics[
width=0.48\textwidth]{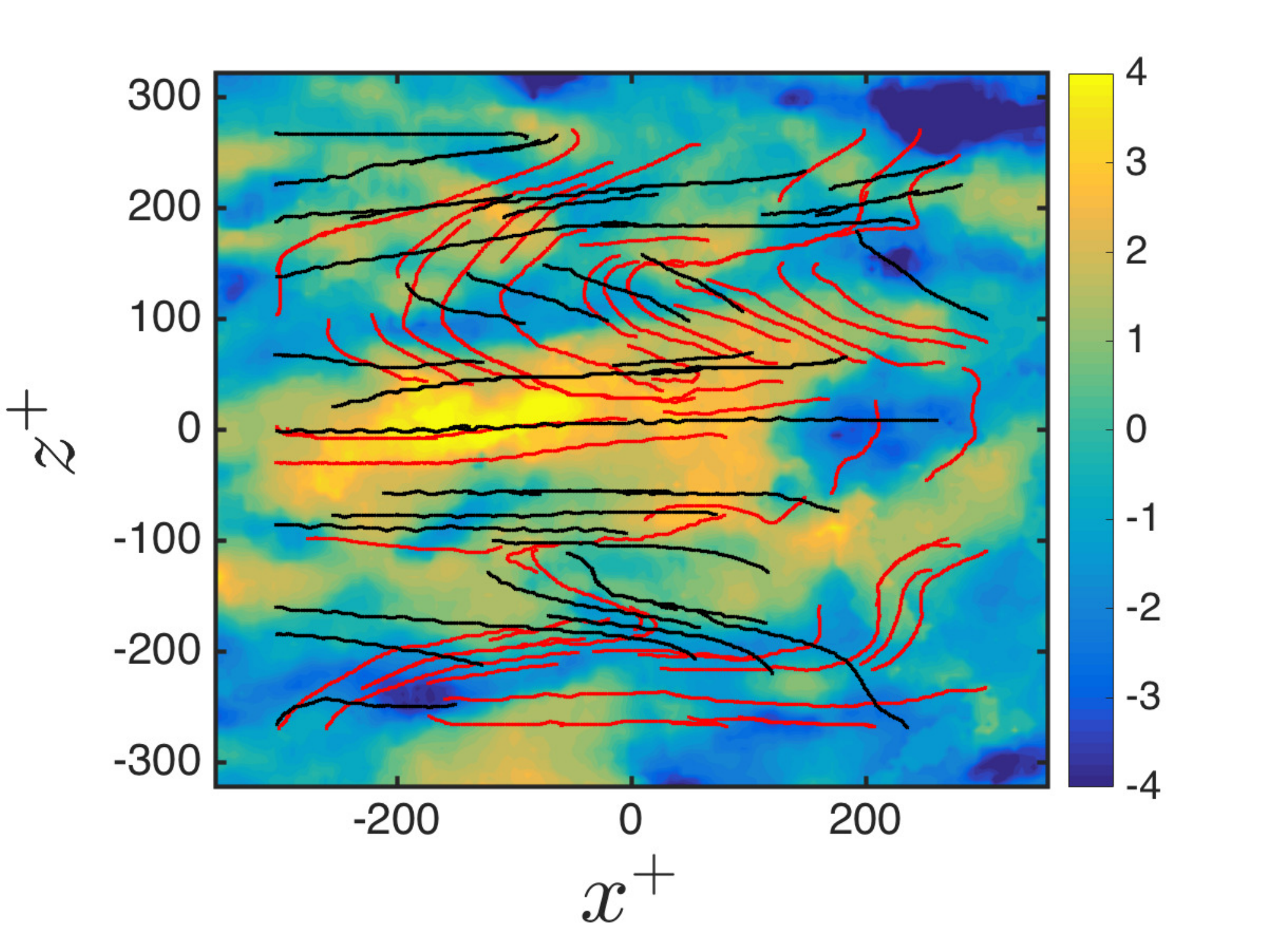} }
\subfigure[$u_{f}+LCS_{P_{U_2}}$]{ \label{b} \includegraphics[
width=0.48\textwidth]{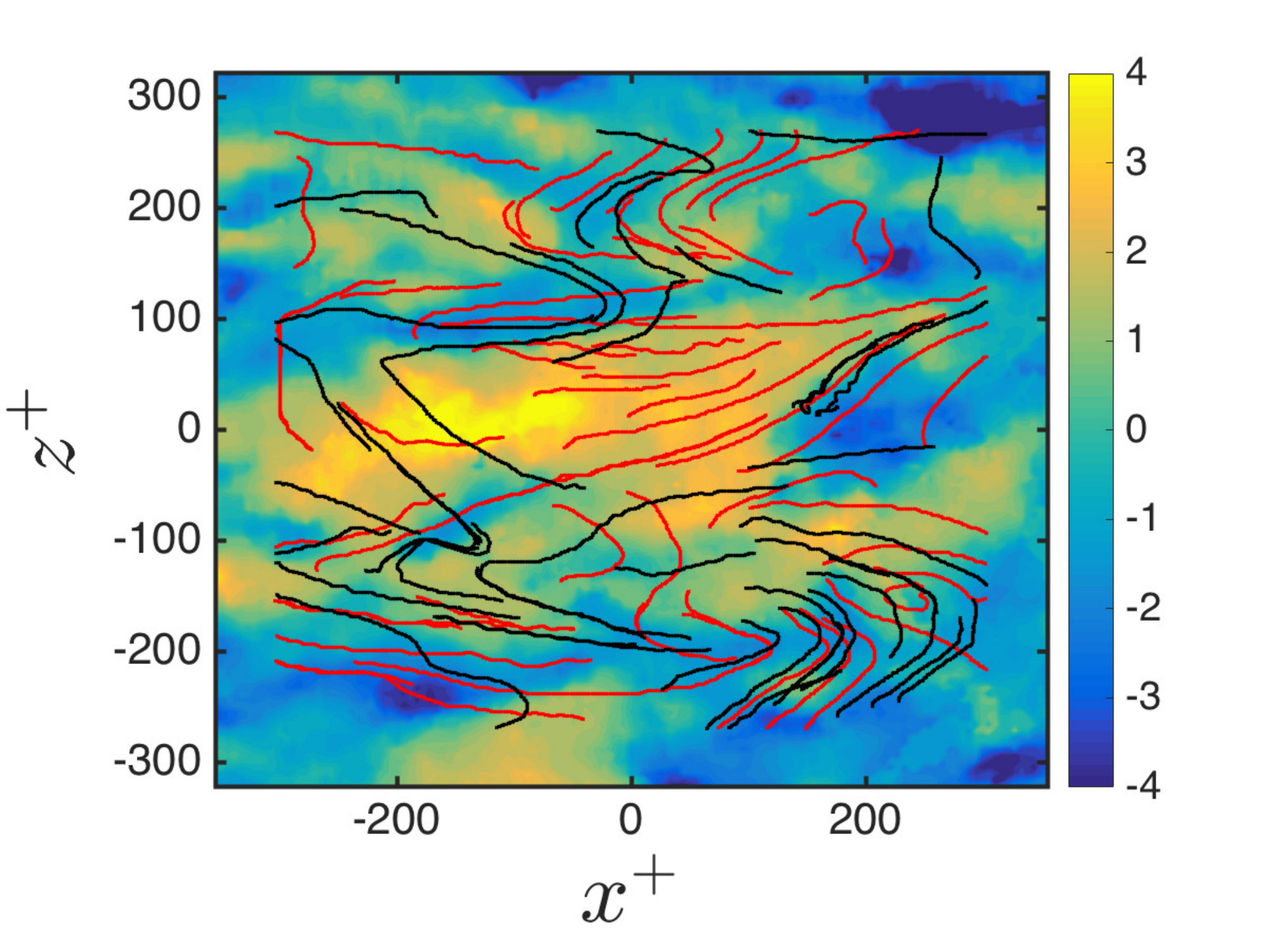} } 
\subfigure[$u_{f}+LCS_{P_{U_3}}$]{ \label{c} \includegraphics[
width=0.48\textwidth]{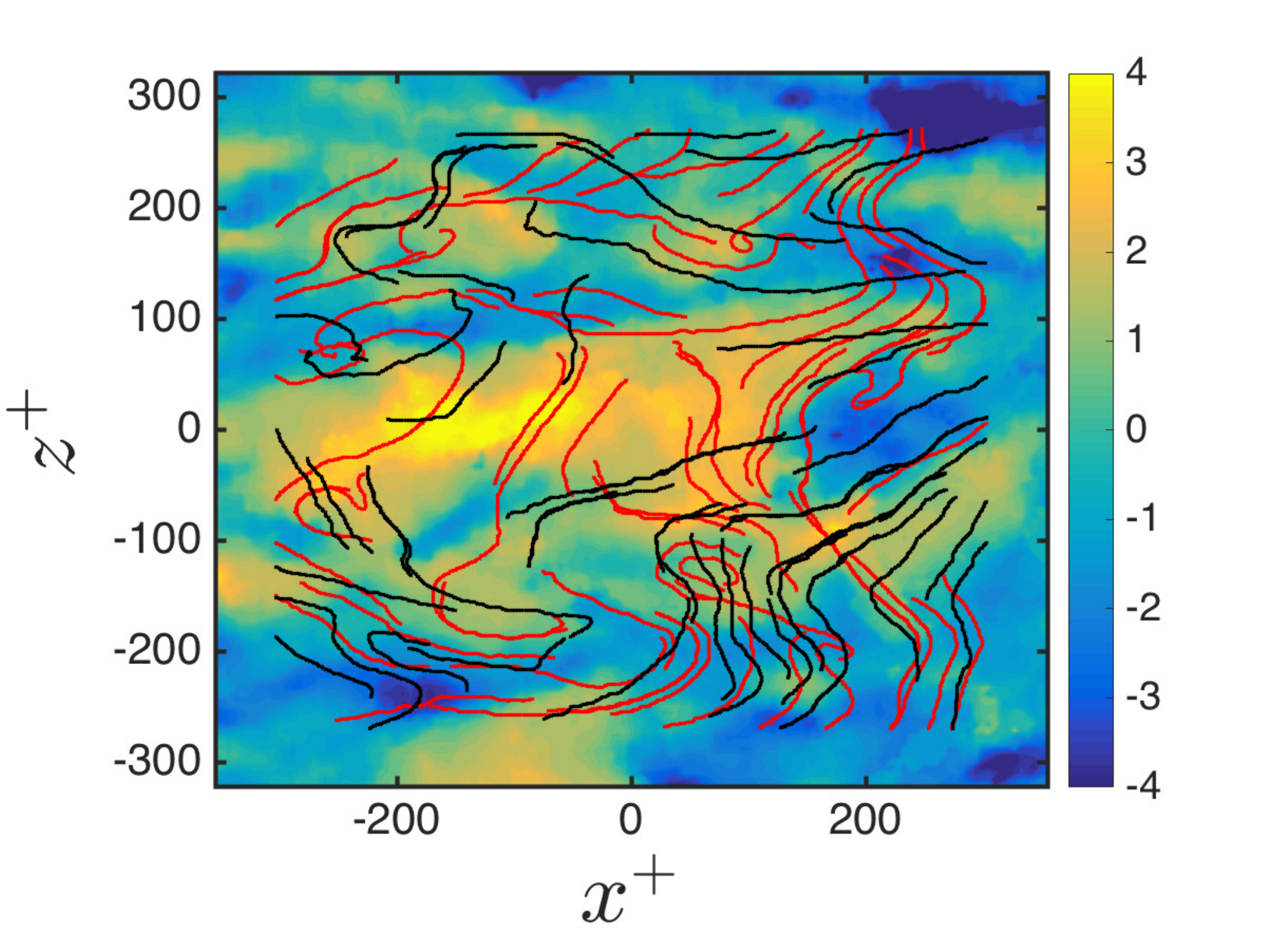} } 
\subfigure[$u_{f}+LCS_{P_{U_4}}$]{ \label{d} \includegraphics[
width=0.48\textwidth]{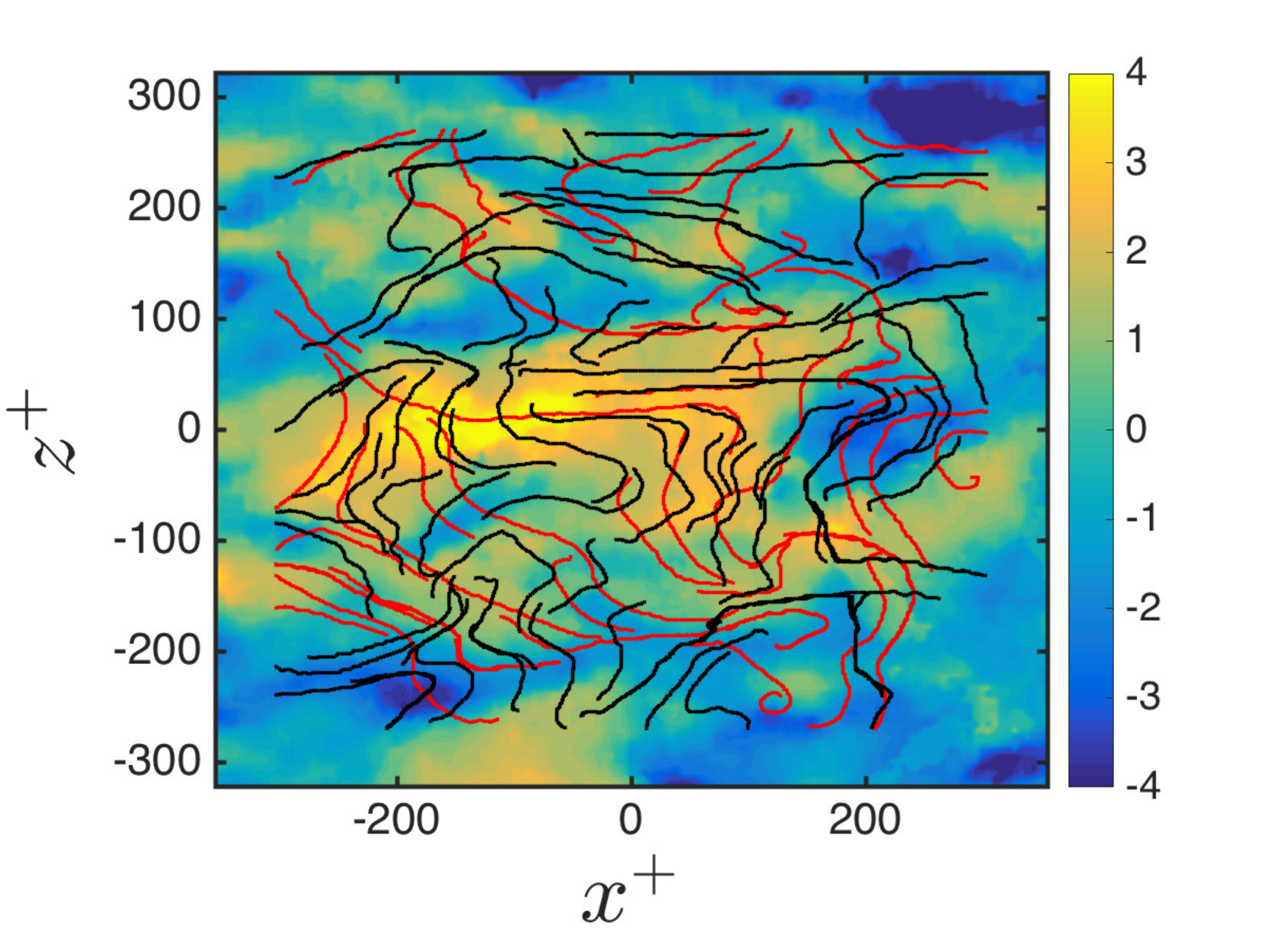} }
\subfigure[$u_{f}+LCS_{P_{U_5}}$]{ \label{e} \includegraphics[
width=0.48\textwidth]{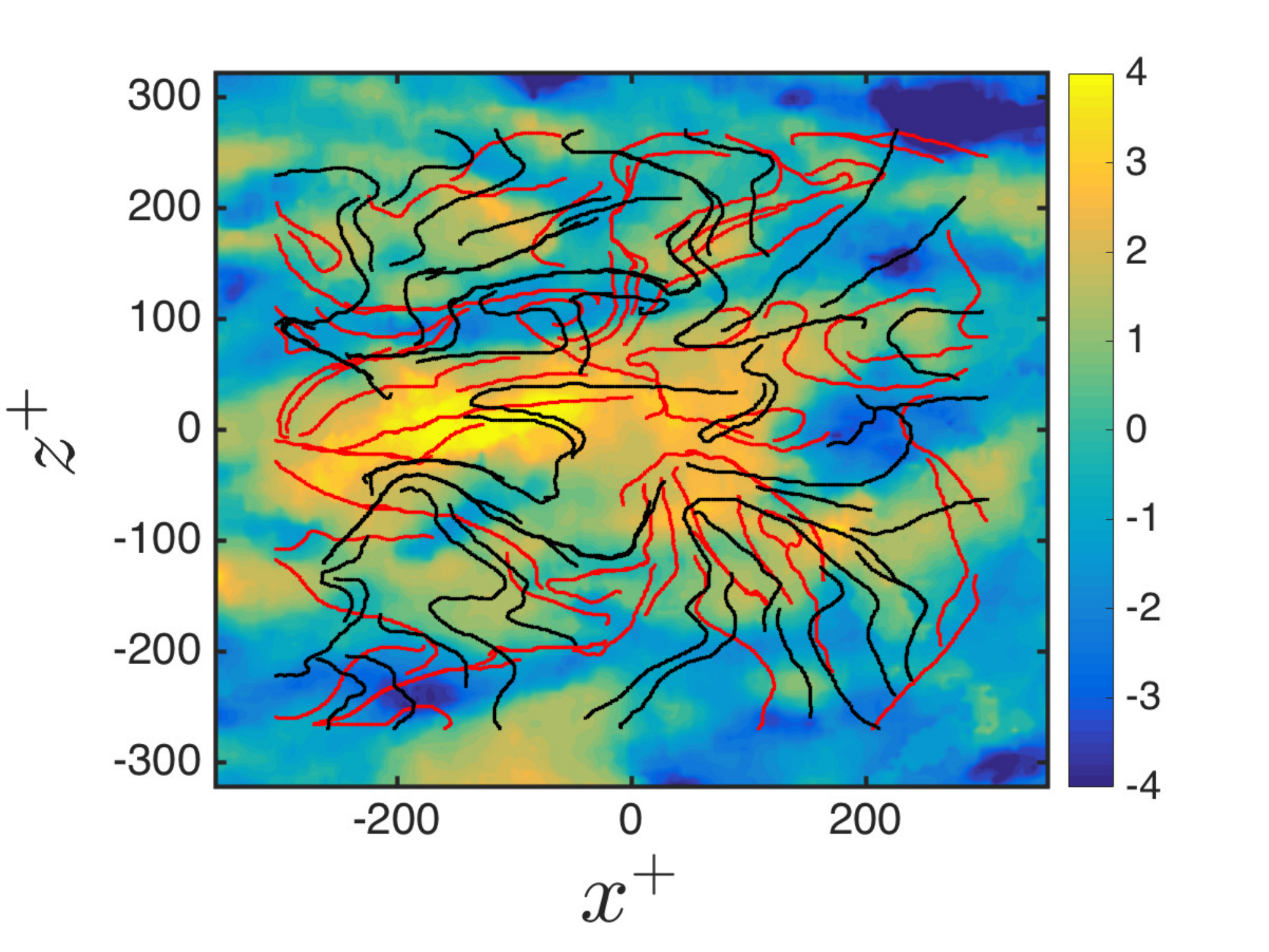} } 
\subfigure[$u_{f}+LCS_{P_{U_{20}}}$]{ \label{f} \includegraphics[
width=0.48\textwidth]{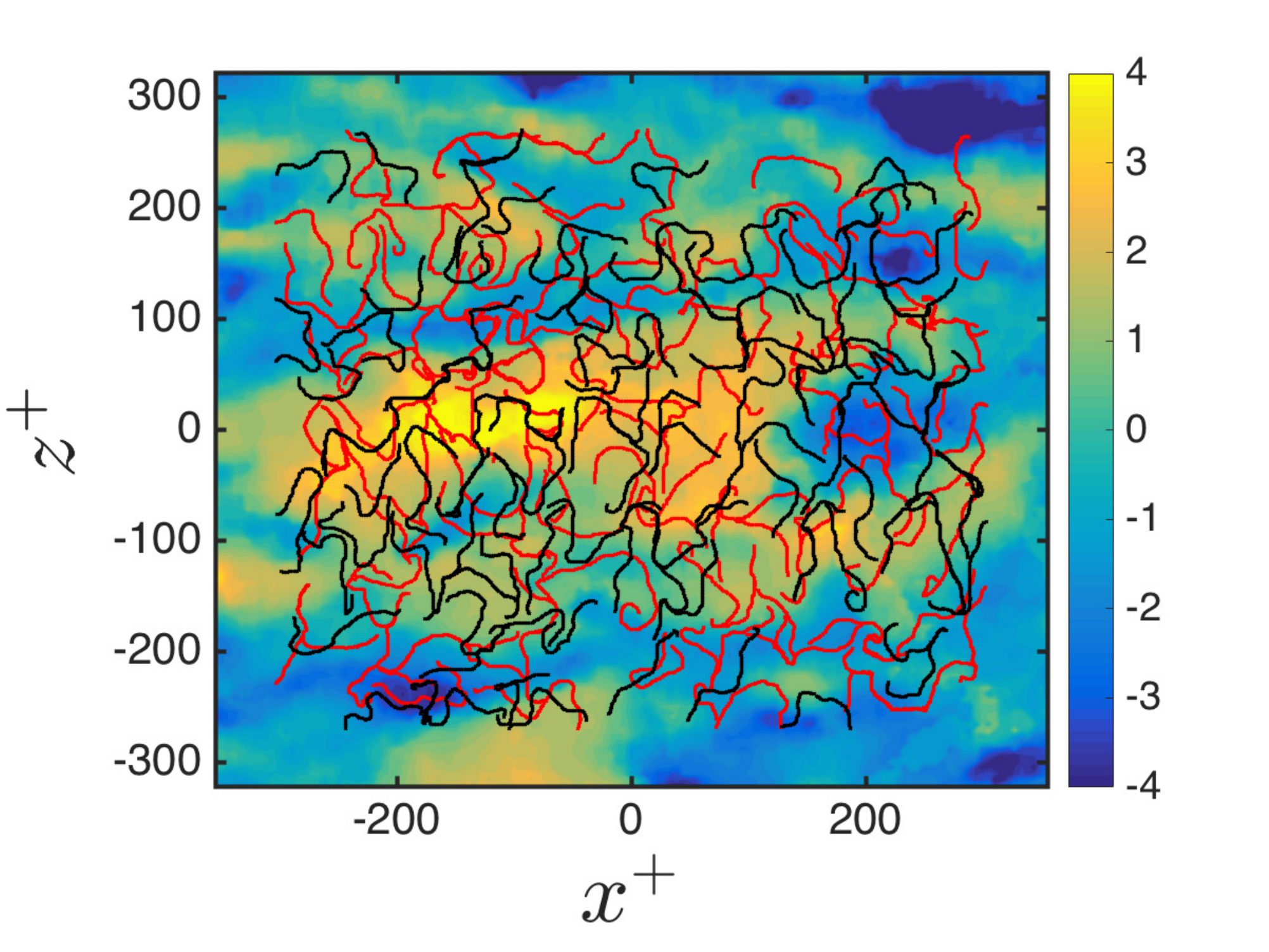} }
\end{center}
\caption{\label{}Normalized fluctuation velocity at $t^{+}=1.29$ with reconstruction LCSs from: (a) the first POD mode; (b) the second POD mode; (c) the third POD mode; (d) the fourth POD mode; (e) the fifth POD mode; (f) the twentieth POD mode.}
\end{figure}

\section{Conclusions}\label{con}

 Proper orthogonal decomposition, dynamic mode decomposition and Lagrangian coherent structures are studied to gain insight in structures of turbulent boundary layer flow. Low order descriptors are used to obtain energy content and frequency information of the flow. Decomposed flow {\it{via}} POD show that the first six modes can reconstruct 75\% of the total kinetic energy furthermore, the first six modes obtained {\it{via}} DMD possess the lowest frequencies in comparison with the remaining modes. POD and DMD modes can effectively represent the coherent and incoherent structures of the flow and manifest the low speed streak events that occur in the turbulent boundary layer flow. 
 
Repelling and attracting LCSs are used to extract the coherent structure of the boundary layer. Shapes of the attracting and repelling lines vary with advection time as a result of the temporal coherence of the flow structure. With increasing time, the length of the trajectories tends to shrink and more tangling between the attracting and repelling LCSs occur. The attracting and repelling LCSs are matched with POD and DMD modes in order to understand the relationship between the frameworks and the respective representations. LCSs extracted from the full data also cover all the POD and DMD modes. However, it is difficult to extract the patterns and trends from the LCS as these are rather complex in shape. The computational sensitivity of the LCS is an important factor that cause over estimated LCSs. Although there are many treatments that reduce the errors in the computation, there are limitations for the experimental data, in particular for the turbulent boundary layer.

Small scale structures work as noise causing error in the LCSs calculation. The main aim of this study is to explore new techniques to reduce the error that may be due to the small scales. This eventually may help us to uncover the structure of the flow. POD and DMD are useful to filter according to associated kinetic energy and frequency, and thereafter reconstruction the velocity depending on the modes that have the largest coherent structure and lowest frequency solely. The reconstruction velocity is used to generate new LCSs. The energy and frequency content in the mode has a direct effect on the length of the LCSs, where the longest attracting and repelling LCSs extracted from the first mode and the shortest one come from the mode twenty. The reconstructed LCSs are well matched with the POD and DMD modes although they are reconstructed from individual modes. The linearity of the POD is used to reconstruct new LCSs using the first two mode, using the first three mode, using the first six mode and using the remaining. With increasing number of POD modes involved in the reconstruction, the complexity of the flow increases due to the increasing the strength of the energy. With increasing flow complexity, the shape and the length of the LCSs decreases with increasing the number of the modes included to generate LCSs. The intersection points between the attracting and repelling are also investigated. For full data LCSs, intersection point are distributed over all high and low momentum regions. For the reconstructed LCSs, most locations are found close to the low momentum regions. Finally, most bending occurring in the trajectories is located when they pass between high and low velocity regions. In conclusion, tracking the low order decomposition and the Lagrangian coherent structures can accurately distinguish the structure of the turbulent boundary layer.

To further explore, the quantification of the LCS length is required to fully understand the nature of the LCS and further investigation for the location of the intersection point with the critical point location may be given a direct indication of the physical meaning to intersection location.

\section*{Acknowledgments} 

This work is funded by the National Science Foundation (NSF-CBET-1034581) for which the authors are thankful. 

\bibliography{NaseemTutkunCal}

\begin{thebibliography}{57}
\expandafter\ifx\csname natexlab\endcsname\relax\def\natexlab#1{#1}\fi
\expandafter\ifx\csname bibnamefont\endcsname\relax
  \def\bibnamefont#1{#1}\fi
\expandafter\ifx\csname bibfnamefont\endcsname\relax
  \def\bibfnamefont#1{#1}\fi
\expandafter\ifx\csname citenamefont\endcsname\relax
  \def\citenamefont#1{#1}\fi
\expandafter\ifx\csname url\endcsname\relax
  \def\url#1{\texttt{#1}}\fi
\expandafter\ifx\csname urlprefix\endcsname\relax\def\urlprefix{URL }\fi
\providecommand{\bibinfo}[2]{#2}
\providecommand{\eprint}[2][]{\url{#2}}

\bibitem[{\citenamefont{Theodorsen}(1952)}]{T}
\bibinfo{author}{\bibfnamefont{T.}~\bibnamefont{Theodorsen}}, in
  \emph{\bibinfo{booktitle}{Proceedings of the 2nd Mid-western Conference on
  Fluid Mechanics}}, edited by \bibinfo{editor}{\bibfnamefont{O.~S.}
  \bibnamefont{University}} (\bibinfo{year}{1952}), pp. \bibinfo{pages}{1--19}.

\bibitem[{\citenamefont{Robinson}(1991)}]{R1991}
\bibinfo{author}{\bibfnamefont{S.~K.} \bibnamefont{Robinson}},
  \bibinfo{journal}{Annu. Re. Fluid Mech.} \textbf{\bibinfo{volume}{23}},
  \bibinfo{pages}{601} (\bibinfo{year}{1991}).

\bibitem[{\citenamefont{Adrian et~al.}(2000)\citenamefont{Adrian, Meinhart, and
  Tomkins}}]{AMT}
\bibinfo{author}{\bibfnamefont{R.~J.} \bibnamefont{Adrian}},
  \bibinfo{author}{\bibfnamefont{C.~D.} \bibnamefont{Meinhart}},
  \bibnamefont{and} \bibinfo{author}{\bibfnamefont{C.~D.}
  \bibnamefont{Tomkins}}, \bibinfo{journal}{J.~Fluid Mech.}
  \textbf{\bibinfo{volume}{422}}, \bibinfo{pages}{1} (\bibinfo{year}{2000}).

\bibitem[{\citenamefont{Ganapathisubramani
  et~al.}(2003)\citenamefont{Ganapathisubramani, Longmire, and Marusic}}]{gana}
\bibinfo{author}{\bibfnamefont{B.}~\bibnamefont{Ganapathisubramani}},
  \bibinfo{author}{\bibfnamefont{E.~K.} \bibnamefont{Longmire}},
  \bibnamefont{and} \bibinfo{author}{\bibfnamefont{I.}~\bibnamefont{Marusic}},
  \bibinfo{journal}{J.~Fluid Mech.} \textbf{\bibinfo{volume}{478}},
  \bibinfo{pages}{35} (\bibinfo{year}{2003}).

\bibitem[{\citenamefont{Tomkins and Adrian}(2003)}]{tomkins2}
\bibinfo{author}{\bibfnamefont{C.~D.} \bibnamefont{Tomkins}} \bibnamefont{and}
  \bibinfo{author}{\bibfnamefont{R.~J.} \bibnamefont{Adrian}},
  \bibinfo{journal}{J.~Fluid Mech.} \textbf{\bibinfo{volume}{490}},
  \bibinfo{pages}{37} (\bibinfo{year}{2003}).

\bibitem[{\citenamefont{K{\"a}hler}(2004)}]{kahler20}
\bibinfo{author}{\bibfnamefont{C.~J.} \bibnamefont{K{\"a}hler}}, Ph.D. thesis,
  \bibinfo{school}{DLR, Dt. Zentrum f{\"u}r Luft-und Raumfahrt}
  (\bibinfo{year}{2004}).

\bibitem[{\citenamefont{Adrian}(2007)}]{Adrian2007}
\bibinfo{author}{\bibfnamefont{R.~J.} \bibnamefont{Adrian}},
  \bibinfo{journal}{Phys. Fluids} \textbf{\bibinfo{volume}{19}},
  \bibinfo{pages}{1} (\bibinfo{year}{2007}).

\bibitem[{\citenamefont{Herpin et~al.}(2008)\citenamefont{Herpin, Wong,
  Stanislas, and Soria}}]{herpin2}
\bibinfo{author}{\bibfnamefont{S.}~\bibnamefont{Herpin}},
  \bibinfo{author}{\bibfnamefont{C.~Y.} \bibnamefont{Wong}},
  \bibinfo{author}{\bibfnamefont{M.}~\bibnamefont{Stanislas}},
  \bibnamefont{and} \bibinfo{author}{\bibfnamefont{J.}~\bibnamefont{Soria}},
  \bibinfo{journal}{Experiments in fluids} \textbf{\bibinfo{volume}{45}},
  \bibinfo{pages}{745} (\bibinfo{year}{2008}).

\bibitem[{\citenamefont{Schlatter et~al.}(2010)\citenamefont{Schlatter, Li,
  Brethouwer, Johansson, and Henningson}}]{schlatter}
\bibinfo{author}{\bibfnamefont{P.}~\bibnamefont{Schlatter}},
  \bibinfo{author}{\bibfnamefont{Q.}~\bibnamefont{Li}},
  \bibinfo{author}{\bibfnamefont{G.}~\bibnamefont{Brethouwer}},
  \bibinfo{author}{\bibfnamefont{A.~V.} \bibnamefont{Johansson}},
  \bibnamefont{and} \bibinfo{author}{\bibfnamefont{D.~S.}
  \bibnamefont{Henningson}}, \bibinfo{journal}{International Journal of Heat
  and Fluid Flow} \textbf{\bibinfo{volume}{31}}, \bibinfo{pages}{251}
  (\bibinfo{year}{2010}).

\bibitem[{\citenamefont{Marusic et~al.}(2010)\citenamefont{Marusic, McKeon,
  Monkewitz, Nagib, Smits, and Sreenivasan}}]{marusic2010}
\bibinfo{author}{\bibfnamefont{I.}~\bibnamefont{Marusic}},
  \bibinfo{author}{\bibfnamefont{B.}~\bibnamefont{McKeon}},
  \bibinfo{author}{\bibfnamefont{P.~A.} \bibnamefont{Monkewitz}},
  \bibinfo{author}{\bibfnamefont{H.~M.} \bibnamefont{Nagib}},
  \bibinfo{author}{\bibfnamefont{A.~J.} \bibnamefont{Smits}}, \bibnamefont{and}
  \bibinfo{author}{\bibfnamefont{K.~R.} \bibnamefont{Sreenivasan}},
  \bibinfo{journal}{Phys. Fluids} \textbf{\bibinfo{volume}{22}},
  \bibinfo{pages}{065103} (\bibinfo{year}{2010}).

\bibitem[{\citenamefont{Schr{\"o}der et~al.}(2011)\citenamefont{Schr{\"o}der,
  Geisler, {\'e}.~A.~A.~Staack, Elsinga, Scarano, Wieneke, Henning, Poelma, and
  Westerweel}}]{schroder}
\bibinfo{author}{\bibfnamefont{A.}~\bibnamefont{Schr{\"o}der}},
  \bibinfo{author}{\bibfnamefont{R.}~\bibnamefont{Geisler}},
  \bibinfo{author}{\bibfnamefont{K.}~\bibnamefont{{\'e}.~A.~A.~Staack}},
  \bibinfo{author}{\bibfnamefont{G.~E.} \bibnamefont{Elsinga}},
  \bibinfo{author}{\bibfnamefont{F.}~\bibnamefont{Scarano}},
  \bibinfo{author}{\bibfnamefont{B.}~\bibnamefont{Wieneke}},
  \bibinfo{author}{\bibfnamefont{A.}~\bibnamefont{Henning}},
  \bibinfo{author}{\bibfnamefont{C.}~\bibnamefont{Poelma}}, \bibnamefont{and}
  \bibinfo{author}{\bibfnamefont{J.}~\bibnamefont{Westerweel}},
  \bibinfo{journal}{Experiments in fluids} \textbf{\bibinfo{volume}{50}},
  \bibinfo{pages}{1071} (\bibinfo{year}{2011}).

\bibitem[{\citenamefont{Dennis and Nickels}(2011)}]{dennis201}
\bibinfo{author}{\bibfnamefont{D.~J.~C.} \bibnamefont{Dennis}}
  \bibnamefont{and} \bibinfo{author}{\bibfnamefont{T.~B.}
  \bibnamefont{Nickels}}, \bibinfo{journal}{J.~Fluid Mech.}
  \textbf{\bibinfo{volume}{673}}, \bibinfo{pages}{180} (\bibinfo{year}{2011}).

\bibitem[{\citenamefont{Jacobi and McKeon}(2013)}]{jacobi2013}
\bibinfo{author}{\bibfnamefont{I.}~\bibnamefont{Jacobi}} \bibnamefont{and}
  \bibinfo{author}{\bibfnamefont{B.~J.} \bibnamefont{McKeon}},
  \bibinfo{journal}{Experiments in fluids} \textbf{\bibinfo{volume}{54}},
  \bibinfo{pages}{1} (\bibinfo{year}{2013}).

\bibitem[{\citenamefont{Hunt et~al.}(1988)\citenamefont{Hunt, Wray, and
  Moin}}]{HWM}
\bibinfo{author}{\bibfnamefont{J.~C.~R.} \bibnamefont{Hunt}},
  \bibinfo{author}{\bibfnamefont{A.~A.} \bibnamefont{Wray}}, \bibnamefont{and}
  \bibinfo{author}{\bibfnamefont{P.}~\bibnamefont{Moin}}, \bibinfo{journal}{In
  its Studying Turbulence Using Numerical Simulation Databases, 2. Proceedings
  of the 1988 Summer Program} \textbf{\bibinfo{volume}{CTR-S88}},
  \bibinfo{pages}{193} (\bibinfo{year}{1988}).

\bibitem[{\citenamefont{Chong et~al.}(1990)\citenamefont{Chong, Perry, and
  Cantwell}}]{CPC}
\bibinfo{author}{\bibfnamefont{M.~S.} \bibnamefont{Chong}},
  \bibinfo{author}{\bibfnamefont{A.~E.} \bibnamefont{Perry}}, \bibnamefont{and}
  \bibinfo{author}{\bibfnamefont{B.~J.} \bibnamefont{Cantwell}},
  \bibinfo{journal}{Phys. Fluids A} \textbf{\bibinfo{volume}{2}},
  \bibinfo{pages}{765} (\bibinfo{year}{1990}).

\bibitem[{\citenamefont{Jeong and Hussein}(1995)}]{JH}
\bibinfo{author}{\bibfnamefont{J.}~\bibnamefont{Jeong}} \bibnamefont{and}
  \bibinfo{author}{\bibfnamefont{F.}~\bibnamefont{Hussein}},
  \bibinfo{journal}{J.~Fluid Mech.} \textbf{\bibinfo{volume}{285}},
  \bibinfo{pages}{69} (\bibinfo{year}{1995}).

\bibitem[{\citenamefont{Zhou et~al.}(1999)\citenamefont{Zhou, Adrian,
  Balachandar, and Kendall}}]{ZABK}
\bibinfo{author}{\bibfnamefont{J.}~\bibnamefont{Zhou}},
  \bibinfo{author}{\bibfnamefont{R.~J.} \bibnamefont{Adrian}},
  \bibinfo{author}{\bibfnamefont{S.}~\bibnamefont{Balachandar}},
  \bibnamefont{and} \bibinfo{author}{\bibfnamefont{T.~M.}
  \bibnamefont{Kendall}}, \bibinfo{journal}{J.~Fluid Mech.}
  \textbf{\bibinfo{volume}{387}}, \bibinfo{pages}{353} (\bibinfo{year}{1999}).

\bibitem[{\citenamefont{Haller}(2015)}]{haller2015}
\bibinfo{author}{\bibfnamefont{G.}~\bibnamefont{Haller}},
  \bibinfo{journal}{Annu. Re. Fluid Mech.} \textbf{\bibinfo{volume}{47}},
  \bibinfo{pages}{137} (\bibinfo{year}{2015}).

\bibitem[{\citenamefont{Lumley}(1967)}]{lumley1967}
\bibinfo{author}{\bibfnamefont{J.~L.} \bibnamefont{Lumley}},
  \bibinfo{journal}{Atmospheric turbulence and radio wave propagation} pp.
  \bibinfo{pages}{166--178} (\bibinfo{year}{1967}).

\bibitem[{\citenamefont{Sirovich}(1987)}]{sirovich1987turbulence}
\bibinfo{author}{\bibfnamefont{L.}~\bibnamefont{Sirovich}},
  \bibinfo{journal}{Quarterly of applied mathematics}
  \textbf{\bibinfo{volume}{45}}, \bibinfo{pages}{561} (\bibinfo{year}{1987}).

\bibitem[{\citenamefont{Herzog}(1986)}]{herzog1986l}
\bibinfo{author}{\bibfnamefont{S.}~\bibnamefont{Herzog}},
  \bibinfo{journal}{Dissertation Abstracts International Part B: Science and
  Engineering[DISS. ABST. INT. PT. B- SCI. \& ENG.],}
  \textbf{\bibinfo{volume}{47}} (\bibinfo{year}{1986}).

\bibitem[{\citenamefont{Moin and Moser}(1989)}]{moin1989}
\bibinfo{author}{\bibfnamefont{P.}~\bibnamefont{Moin}} \bibnamefont{and}
  \bibinfo{author}{\bibfnamefont{R.~D.} \bibnamefont{Moser}},
  \bibinfo{journal}{J.~Fluid Mech.} \textbf{\bibinfo{volume}{200}},
  \bibinfo{pages}{471} (\bibinfo{year}{1989}).

\bibitem[{\citenamefont{Liberzon et~al.}(2005)\citenamefont{Liberzon, Gurka,
  Tiselj, and Hetsroni}}]{liberzon200l}
\bibinfo{author}{\bibfnamefont{A.}~\bibnamefont{Liberzon}},
  \bibinfo{author}{\bibfnamefont{R.}~\bibnamefont{Gurka}},
  \bibinfo{author}{\bibfnamefont{I.}~\bibnamefont{Tiselj}}, \bibnamefont{and}
  \bibinfo{author}{\bibfnamefont{G.}~\bibnamefont{Hetsroni}},
  \bibinfo{journal}{Theoretical and Computational Fluid Dynamics}
  \textbf{\bibinfo{volume}{19}}, \bibinfo{pages}{115} (\bibinfo{year}{2005}).

\bibitem[{\citenamefont{Wu and Christensen}(2010)}]{wu2010spatial}
\bibinfo{author}{\bibfnamefont{Y.}~\bibnamefont{Wu}} \bibnamefont{and}
  \bibinfo{author}{\bibfnamefont{K.~T.} \bibnamefont{Christensen}},
  \bibinfo{journal}{J.~Fluid Mech.} \textbf{\bibinfo{volume}{655}},
  \bibinfo{pages}{380} (\bibinfo{year}{2010}).

\bibitem[{\citenamefont{Baltzer and Adrian}(2011)}]{baltzer2011st}
\bibinfo{author}{\bibfnamefont{J.~R.} \bibnamefont{Baltzer}} \bibnamefont{and}
  \bibinfo{author}{\bibfnamefont{R.~J.} \bibnamefont{Adrian}},
  \bibinfo{journal}{Phys. Fluids} \textbf{\bibinfo{volume}{23}},
  \bibinfo{pages}{015107} (\bibinfo{year}{2011}).

\bibitem[{\citenamefont{Shah and Bou-Zeid}(2014)}]{shah2014very}
\bibinfo{author}{\bibfnamefont{S.}~\bibnamefont{Shah}} \bibnamefont{and}
  \bibinfo{author}{\bibfnamefont{E.}~\bibnamefont{Bou-Zeid}},
  \bibinfo{journal}{Boundary-Layer Meteorology} \textbf{\bibinfo{volume}{153}},
  \bibinfo{pages}{355} (\bibinfo{year}{2014}).

\bibitem[{\citenamefont{Berkooz et~al.}(1993)\citenamefont{Berkooz, Holmes, and
  Lumley}}]{berkooz1993proper}
\bibinfo{author}{\bibfnamefont{G.}~\bibnamefont{Berkooz}},
  \bibinfo{author}{\bibfnamefont{P.}~\bibnamefont{Holmes}}, \bibnamefont{and}
  \bibinfo{author}{\bibfnamefont{J.~L.} \bibnamefont{Lumley}},
  \bibinfo{journal}{Annu. Re. Fluid Mech.} \textbf{\bibinfo{volume}{25}},
  \bibinfo{pages}{539} (\bibinfo{year}{1993}).

\bibitem[{\citenamefont{Schmid and Sesterhenn}(2008)}]{schmid2008de}
\bibinfo{author}{\bibfnamefont{P.~J.} \bibnamefont{Schmid}} \bibnamefont{and}
  \bibinfo{author}{\bibfnamefont{J.~L.} \bibnamefont{Sesterhenn}},
  \bibinfo{journal}{Bull. Amer. Phys. Soc}  (\bibinfo{year}{2008}).

\bibitem[{\citenamefont{Rowley et~al.}(2009)\citenamefont{Rowley, Mezi{\'c},
  Bagheri, Schlatter, and Henningson}}]{rowley2009spectral}
\bibinfo{author}{\bibfnamefont{C.~W.} \bibnamefont{Rowley}},
  \bibinfo{author}{\bibfnamefont{I.}~\bibnamefont{Mezi{\'c}}},
  \bibinfo{author}{\bibfnamefont{S.}~\bibnamefont{Bagheri}},
  \bibinfo{author}{\bibfnamefont{P.}~\bibnamefont{Schlatter}},
  \bibnamefont{and} \bibinfo{author}{\bibfnamefont{D.~S.}
  \bibnamefont{Henningson}}, \bibinfo{journal}{J.~Fluid Mech.}
  \textbf{\bibinfo{volume}{641}}, \bibinfo{pages}{115} (\bibinfo{year}{2009}).

\bibitem[{\citenamefont{Mezi{\'c}}(2005)}]{mezic2005spectral}
\bibinfo{author}{\bibfnamefont{I.}~\bibnamefont{Mezi{\'c}}},
  \bibinfo{journal}{Nonlinear Dynamics} \textbf{\bibinfo{volume}{41}},
  \bibinfo{pages}{309} (\bibinfo{year}{2005}).

\bibitem[{\citenamefont{Schmid}(2010)}]{schmid2010dynamic}
\bibinfo{author}{\bibfnamefont{P.~J.} \bibnamefont{Schmid}},
  \bibinfo{journal}{J.~Fluid Mech.} \textbf{\bibinfo{volume}{656}},
  \bibinfo{pages}{5} (\bibinfo{year}{2010}).

\bibitem[{\citenamefont{Schmid}(2011)}]{schmid2011application}
\bibinfo{author}{\bibfnamefont{P.~J.} \bibnamefont{Schmid}},
  \bibinfo{journal}{Experiments in fluids} \textbf{\bibinfo{volume}{50}},
  \bibinfo{pages}{1123} (\bibinfo{year}{2011}).

\bibitem[{\citenamefont{Muld et~al.}(2012)\citenamefont{Muld, Efraimsson, and
  Henningson}}]{muld2012flow}
\bibinfo{author}{\bibfnamefont{T.~W.} \bibnamefont{Muld}},
  \bibinfo{author}{\bibfnamefont{G.}~\bibnamefont{Efraimsson}},
  \bibnamefont{and} \bibinfo{author}{\bibfnamefont{D.~S.}
  \bibnamefont{Henningson}}, \bibinfo{journal}{Computers \& Fluids}
  \textbf{\bibinfo{volume}{57}}, \bibinfo{pages}{87} (\bibinfo{year}{2012}).

\bibitem[{\citenamefont{Pan et~al.}(2011)\citenamefont{Pan, Yu, and
  Wang}}]{pan2011dynamical}
\bibinfo{author}{\bibfnamefont{C.}~\bibnamefont{Pan}},
  \bibinfo{author}{\bibfnamefont{D.}~\bibnamefont{Yu}}, \bibnamefont{and}
  \bibinfo{author}{\bibfnamefont{J.}~\bibnamefont{Wang}},
  \bibinfo{journal}{Theoretical and Applied Mechanics Letters}
  \textbf{\bibinfo{volume}{1}}, \bibinfo{pages}{012002} (\bibinfo{year}{2011}).

\bibitem[{\citenamefont{Tang and Jiang}(2012)}]{tang2012dynamic}
\bibinfo{author}{\bibfnamefont{Z.}~\bibnamefont{Tang}} \bibnamefont{and}
  \bibinfo{author}{\bibfnamefont{N.}~\bibnamefont{Jiang}},
  \bibinfo{journal}{Science China Physics, Mechanics and Astronomy}
  \textbf{\bibinfo{volume}{55}}, \bibinfo{pages}{118} (\bibinfo{year}{2012}).

\bibitem[{\citenamefont{Zhang et~al.}(2014)\citenamefont{Zhang, Liu, and
  Wang}}]{zhang2014identification}
\bibinfo{author}{\bibfnamefont{Q.}~\bibnamefont{Zhang}},
  \bibinfo{author}{\bibfnamefont{Y.}~\bibnamefont{Liu}}, \bibnamefont{and}
  \bibinfo{author}{\bibfnamefont{S.}~\bibnamefont{Wang}},
  \bibinfo{journal}{Journal of Fluids and Structures}
  \textbf{\bibinfo{volume}{49}}, \bibinfo{pages}{53} (\bibinfo{year}{2014}).

\bibitem[{\citenamefont{Haller}(2000)}]{haller2000finding}
\bibinfo{author}{\bibfnamefont{G.}~\bibnamefont{Haller}},
  \bibinfo{journal}{Chaos: An Interdisciplinary Journal of Nonlinear Science}
  \textbf{\bibinfo{volume}{10}}, \bibinfo{pages}{99} (\bibinfo{year}{2000}).

\bibitem[{\citenamefont{Haller}(2001)}]{H}
\bibinfo{author}{\bibfnamefont{G.}~\bibnamefont{Haller}},
  \bibinfo{journal}{Physica D} \textbf{\bibinfo{volume}{149}},
  \bibinfo{pages}{248} (\bibinfo{year}{2001}).

\bibitem[{\citenamefont{Wang et~al.}(2003)\citenamefont{Wang, Haller, Banaszuk,
  and Tadmor}}]{wang2003closed}
\bibinfo{author}{\bibfnamefont{Y.}~\bibnamefont{Wang}},
  \bibinfo{author}{\bibfnamefont{G.}~\bibnamefont{Haller}},
  \bibinfo{author}{\bibfnamefont{A.}~\bibnamefont{Banaszuk}}, \bibnamefont{and}
  \bibinfo{author}{\bibfnamefont{G.}~\bibnamefont{Tadmor}},
  \bibinfo{journal}{Phys. Fluids} \textbf{\bibinfo{volume}{15}},
  \bibinfo{pages}{2251} (\bibinfo{year}{2003}).

\bibitem[{\citenamefont{Shadden et~al.}(2005)\citenamefont{Shadden, Lekien, and
  Marsden}}]{SLM}
\bibinfo{author}{\bibfnamefont{S.}~\bibnamefont{Shadden}},
  \bibinfo{author}{\bibfnamefont{F.}~\bibnamefont{Lekien}}, \bibnamefont{and}
  \bibinfo{author}{\bibfnamefont{J.}~\bibnamefont{Marsden}},
  \bibinfo{journal}{Physica D} \textbf{\bibinfo{volume}{212}},
  \bibinfo{pages}{271} (\bibinfo{year}{2005}).

\bibitem[{\citenamefont{Haller}(2011)}]{Ha}
\bibinfo{author}{\bibfnamefont{G.}~\bibnamefont{Haller}},
  \bibinfo{journal}{Physica D} \textbf{\bibinfo{volume}{240}},
  \bibinfo{pages}{574} (\bibinfo{year}{2011}).

\bibitem[{\citenamefont{Farazmand and Haller}(2012)}]{FH2}
\bibinfo{author}{\bibfnamefont{M.}~\bibnamefont{Farazmand}} \bibnamefont{and}
  \bibinfo{author}{\bibfnamefont{G.}~\bibnamefont{Haller}},
  \bibinfo{journal}{Chaos} \textbf{\bibinfo{volume}{22}},
  \bibinfo{pages}{013128} (\bibinfo{year}{2012}).

\bibitem[{\citenamefont{Farazmand and Haller}(2013)}]{farazmand2013attracting}
\bibinfo{author}{\bibfnamefont{M.}~\bibnamefont{Farazmand}} \bibnamefont{and}
  \bibinfo{author}{\bibfnamefont{G.}~\bibnamefont{Haller}},
  \bibinfo{journal}{Chaos: An Interdisciplinary Journal of Nonlinear Science}
  \textbf{\bibinfo{volume}{23}}, \bibinfo{pages}{023101}
  (\bibinfo{year}{2013}).

\bibitem[{\citenamefont{Farazmand}(2015)}]{farazmand2015hyperbolic}
\bibinfo{author}{\bibfnamefont{M.}~\bibnamefont{Farazmand}},
  \bibinfo{journal}{arXiv preprint arXiv:1501.05036}  (\bibinfo{year}{2015}).

\bibitem[{\citenamefont{Inanc et~al.}(2005)\citenamefont{Inanc, Shadden, and
  Marsden}}]{inanc2005optimal}
\bibinfo{author}{\bibfnamefont{T.}~\bibnamefont{Inanc}},
  \bibinfo{author}{\bibfnamefont{S.~C.} \bibnamefont{Shadden}},
  \bibnamefont{and} \bibinfo{author}{\bibfnamefont{J.~E.}
  \bibnamefont{Marsden}} (\bibinfo{year}{2005}).

\bibitem[{\citenamefont{Green et~al.}(2007)\citenamefont{Green, Rowley, and
  Haller}}]{GRH}
\bibinfo{author}{\bibfnamefont{M.~A.} \bibnamefont{Green}},
  \bibinfo{author}{\bibfnamefont{C.~W.} \bibnamefont{Rowley}},
  \bibnamefont{and} \bibinfo{author}{\bibfnamefont{G.}~\bibnamefont{Haller}},
  \bibinfo{journal}{J.~Fluid Mech.} \textbf{\bibinfo{volume}{572}},
  \bibinfo{pages}{111} (\bibinfo{year}{2007}).

\bibitem[{\citenamefont{Peng and Dabiri}(2008)}]{PD}
\bibinfo{author}{\bibfnamefont{J.}~\bibnamefont{Peng}} \bibnamefont{and}
  \bibinfo{author}{\bibfnamefont{J.}~\bibnamefont{Dabiri}},
  \bibinfo{journal}{J. Exp. Biol} \textbf{\bibinfo{volume}{211}},
  \bibinfo{pages}{280} (\bibinfo{year}{2008}).

\bibitem[{\citenamefont{Pan et~al.}(2009)\citenamefont{Pan, Wang, and
  Zhang}}]{pan2009identification}
\bibinfo{author}{\bibfnamefont{C.}~\bibnamefont{Pan}},
  \bibinfo{author}{\bibfnamefont{J.}~\bibnamefont{Wang}}, \bibnamefont{and}
  \bibinfo{author}{\bibfnamefont{C.}~\bibnamefont{Zhang}},
  \bibinfo{journal}{Science in China Series G: Physics, Mechanics and
  Astronomy} \textbf{\bibinfo{volume}{52}}, \bibinfo{pages}{248}
  (\bibinfo{year}{2009}).

\bibitem[{\citenamefont{Beron-Vera et~al.}(2010)\citenamefont{Beron-Vera,
  Olascoaga, Brown, Ko{\c{c}}ak, and Rypina}}]{beron2010invariant}
\bibinfo{author}{\bibfnamefont{F.~J.} \bibnamefont{Beron-Vera}},
  \bibinfo{author}{\bibfnamefont{M.~J.} \bibnamefont{Olascoaga}},
  \bibinfo{author}{\bibfnamefont{M.~G.} \bibnamefont{Brown}},
  \bibinfo{author}{\bibfnamefont{H.}~\bibnamefont{Ko{\c{c}}ak}},
  \bibnamefont{and} \bibinfo{author}{\bibfnamefont{I.~I.}
  \bibnamefont{Rypina}}, \bibinfo{journal}{Chaos: An Interdisciplinary Journal
  of Nonlinear Science} \textbf{\bibinfo{volume}{20}}, \bibinfo{pages}{017514}
  (\bibinfo{year}{2010}).

\bibitem[{\citenamefont{Beron-Vera et~al.}(2012)\citenamefont{Beron-Vera,
  Olascoaga, Brown, and Ko{\c{c}}ak}}]{beron2012zonal}
\bibinfo{author}{\bibfnamefont{F.~J.} \bibnamefont{Beron-Vera}},
  \bibinfo{author}{\bibfnamefont{M.~J.} \bibnamefont{Olascoaga}},
  \bibinfo{author}{\bibfnamefont{M.~G.} \bibnamefont{Brown}}, \bibnamefont{and}
  \bibinfo{author}{\bibfnamefont{H.}~\bibnamefont{Ko{\c{c}}ak}},
  \bibinfo{journal}{Journal of the Atmospheric Sciences}
  \textbf{\bibinfo{volume}{69}}, \bibinfo{pages}{753} (\bibinfo{year}{2012}).

\bibitem[{\citenamefont{Wilson et~al.}(2013)\citenamefont{Wilson, Tutkun, and
  Cal}}]{WTR}
\bibinfo{author}{\bibfnamefont{Z.~D.} \bibnamefont{Wilson}},
  \bibinfo{author}{\bibfnamefont{M.}~\bibnamefont{Tutkun}}, \bibnamefont{and}
  \bibinfo{author}{\bibfnamefont{R.~B.} \bibnamefont{Cal}},
  \bibinfo{journal}{Journal of Fluid Mechanics} \textbf{\bibinfo{volume}{728}},
  \bibinfo{pages}{396} (\bibinfo{year}{2013}).

\bibitem[{\citenamefont{Trefethen and Bau}(1997)}]{trefethen1997numerical}
\bibinfo{author}{\bibfnamefont{L.~N.} \bibnamefont{Trefethen}}
  \bibnamefont{and} \bibinfo{author}{\bibfnamefont{I.~D.} \bibnamefont{Bau}},
  \emph{\bibinfo{title}{Numerical linear algebra}}, vol.~\bibinfo{volume}{50}
  (\bibinfo{publisher}{Siam}, \bibinfo{year}{1997}).

\bibitem[{\citenamefont{Tirunagari et~al.}(2012)\citenamefont{Tirunagari,
  Vuorinen, Kaario, and Larmi}}]{tirunagari2012analysis}
\bibinfo{author}{\bibfnamefont{S.}~\bibnamefont{Tirunagari}},
  \bibinfo{author}{\bibfnamefont{V.}~\bibnamefont{Vuorinen}},
  \bibinfo{author}{\bibfnamefont{O.}~\bibnamefont{Kaario}}, \bibnamefont{and}
  \bibinfo{author}{\bibfnamefont{M.}~\bibnamefont{Larmi}},
  \bibinfo{journal}{CSI Journal of Computing} \textbf{\bibinfo{volume}{1}},
  \bibinfo{pages}{20} (\bibinfo{year}{2012}).

\bibitem[{\citenamefont{Mathur et~al.}(2007)\citenamefont{Mathur, Haller,
  Peacock, Ruppert-Felsot, and Swinney}}]{mathur2007uncovering}
\bibinfo{author}{\bibfnamefont{M.}~\bibnamefont{Mathur}},
  \bibinfo{author}{\bibfnamefont{G.}~\bibnamefont{Haller}},
  \bibinfo{author}{\bibfnamefont{T.}~\bibnamefont{Peacock}},
  \bibinfo{author}{\bibfnamefont{J.~E.} \bibnamefont{Ruppert-Felsot}},
  \bibnamefont{and} \bibinfo{author}{\bibfnamefont{H.~L.}
  \bibnamefont{Swinney}}, \bibinfo{journal}{Physical Review Letters}
  \textbf{\bibinfo{volume}{98}}, \bibinfo{pages}{144502}
  (\bibinfo{year}{2007}).

\bibitem[{\citenamefont{Foucaut et~al.}(2007)\citenamefont{Foucaut, Coudert,
  Kostas, Stanislas, Braud, Fourment, Delville, Tutkun, Mehdi, Johansson
  et~al.}}]{FOUCAUTetal2007}
\bibinfo{author}{\bibfnamefont{J.~M.} \bibnamefont{Foucaut}},
  \bibinfo{author}{\bibfnamefont{S.}~\bibnamefont{Coudert}},
  \bibinfo{author}{\bibfnamefont{J.}~\bibnamefont{Kostas}},
  \bibinfo{author}{\bibfnamefont{M.}~\bibnamefont{Stanislas}},
  \bibinfo{author}{\bibfnamefont{P.}~\bibnamefont{Braud}},
  \bibinfo{author}{\bibfnamefont{C.}~\bibnamefont{Fourment}},
  \bibinfo{author}{\bibfnamefont{J.}~\bibnamefont{Delville}},
  \bibinfo{author}{\bibfnamefont{M.}~\bibnamefont{Tutkun}},
  \bibinfo{author}{\bibfnamefont{F.}~\bibnamefont{Mehdi}},
  \bibinfo{author}{\bibfnamefont{P.}~\bibnamefont{Johansson}},
  \bibnamefont{et~al.}, in \emph{\bibinfo{booktitle}{7th {I}nternational
  {S}ymposium on {P}article {I}mage {V}elocimetry}} (\bibinfo{year}{2007}).

\bibitem[{\citenamefont{Green}(1995)}]{green1995fluid}
\bibinfo{author}{\bibfnamefont{S.~I.} \bibnamefont{Green}},
  \emph{\bibinfo{title}{Fluid vortices: fluid mechanics and its applications}},
  vol.~\bibinfo{volume}{30} (\bibinfo{publisher}{Springer Science \& Business
  Media}, \bibinfo{year}{1995}).

\bibitem[{\citenamefont{Smith et~al.}(1991)\citenamefont{Smith, Walker,
  Haidari, and Sobrun}}]{smith1991dynamics}
\bibinfo{author}{\bibfnamefont{C.~R.} \bibnamefont{Smith}},
  \bibinfo{author}{\bibfnamefont{J.~D.~A.} \bibnamefont{Walker}},
  \bibinfo{author}{\bibfnamefont{A.~H.} \bibnamefont{Haidari}},
  \bibnamefont{and} \bibinfo{author}{\bibfnamefont{U.}~\bibnamefont{Sobrun}},
  \bibinfo{journal}{Philosophical Transactions of the Royal Society of London
  A: Mathematical, Physical and Engineering Sciences}
  \textbf{\bibinfo{volume}{336}}, \bibinfo{pages}{131} (\bibinfo{year}{1991}).

\end{thebibliography}

\end{document}